\documentclass[11pt,a4paper]{article}
\pdfoutput=1
\usepackage{epsfig,graphicx}
\usepackage{cite}
\RequirePackage{mathrsfs}
\DeclareGraphicsExtensions{.pdf,.eps}

\newlength{\dinwidth}
\newlength{\dinmargin}
\def\fig#1{{Fig.~(\ref{#1})}}
\def\eq#1{{Eq.~(\ref{#1})}}
\newcommand{\Le}{\left(}
\newcommand{\Ra}{\right)}

\newcommand{\beq}{\begin{equation}}
\newcommand{\eeq}{\end{equation}}
\newcommand{\beqar}{\begin{eqnarray}}
\newcommand{\eeqar}{\end{eqnarray}}
\newcommand{\pd}[2]{\frac{\partial #1}{\partial #2}} 
\date{}
\begin{document}
%\input{header.tex}
%\begin{flushright}
%\vspace{-0.5cm}
%{\Large \bf DRAFt}\\
%\today
%\end{flushright}
%\thispagestyle{empty}

\title {{~}\\
{\Large \bf  Transport properties of a charged hot spot in an external electromagnetic field }\\}
\author{ 
{~}\\
{\large 
S.~Bondarenko$^{(1) },$
K.~Komoshvili$^{(1) }$,
A.~Prygarin$^{(1) }$}\\[7mm]
{\it\normalsize  $^{(1) }$Physics Department, Ariel University, Ariel 40700, Israel}\\}

\maketitle
\thispagestyle{empty}

\begin{abstract}
 We investigate adiabatic expansion of a charged and rotating fluid element consisting of weakly interacting particles,  which is initially perturbed by an external electromagnetic field.
A framework for the perturbative  calculation of the non-equilibrium distribution function of this fluid volume  is considered and the distribution function is calculated 
to the first order in the perturbative expansion. This distribution function, which describes the evolution of the element with constant entropy, allows to calculate
momentum flux tensor and viscosity coefficients of the expanding system. We show, that these viscosity coefficients  depend on the initial angular 
velocity of the spot and on the strength of its initial perturbation by the external field. Obtained results are applied to the phenomenology of the viscosity to the entropy ratio calculated in lattice models.

\end{abstract}

\newpage
\section{Introduction}

\,\,\,Collisions of relativistic nuclei in the RHIC and LHC experiments at very high energies led to the discovery of a new state of matter named quark-gluon plasma (QGP). At the 
initial stages of the scattering, this plasma resembles almost an ideal liquid 
 which microscopic structure is not yet well understood \cite{Shyr1,Shyr2,Berd,Koch,Liao,Nahrgang,Stein,Skokov,Rand}. The data obtained at the RHIC experiments is in a good agreement with the predictions of the 
ideal relativistic fluid dynamics, \cite{Fluid,Fluid1,Fluid2}, that establish fluid dynamics as the main theoretical tool
to describe collective flow in the collisions. As an input to the hydrodynamical evolution of the particles it
is assumed that after a very short time, $\tau\, <\, 1\,\, fm/c$ \cite{Fluid2},  the matter  reaches a thermal equilibrium and expands with a very small shear viscosity  \cite{Visc,Teaney}.
 
 Different models of QGP were proposed in order to describe its fluid behavior. 
Among other there are strongly coupled Quark Gluon Plasma (sQGP) model, see \cite{Shyr1} and references therein, different non-perturbative and lattice models \cite{Cass,Pajares, Lattice,Lattice1}. In our approach we will model a QGP system created in the process of high-energy scattering
as  a system of weakly interacting particles where the thermal equilibrium is achieved for a small elements 
of the colliding matter \cite{Berd,Skokov,Teaney,HotSp,HotSp1} separately at some initial moments of time.  
Namely, we assume, that the whole colliding system is not in a global equilibrium state, see \cite{Berd,Step}, and therefore, at the initial stages of the evolution,  only local equilibrium may be achieved for the small fluid elements of the matter, independently on each other in the first approximation, see \cite{Our1}. We can call these elements as hot spots, considering them as local initial fluctuations of the density, 
see for example \cite{New1,New2,New3,New4,New5}, and which can be related with the physics of saturation, 
\cite{GLR,MUQI,Sat1,Sat2,Sat3}, as well. 
Consequently, accordingly to \cite{Land1},  the expansion of the dense initial fluctuation
occurs with the constant entropy  which justifies the applicability of the hydrodynamical description of the process. This adiabatic expansion continues till the value of the particle's mean free path inside the expanding hot spot becomes comparable to the
size of the whole system. At this stage, instead of liquid, there appears a gas of interacting particles with rapidly decreasing density, see for example \cite{Eppel}.
 
  In kinetic approach the process of the dense system expansion, i.e. hydrodynamics flow, is described by the Vlasov's equation. This equation determines a microscopic distribution function for the fluid element which depends on time in the case of non-equilibrium processes. For the very dense system the applicability of Vlasov's equation is determined by the value of 
plasma parameter, the dynamical constraints on the value of this parameter coincide with the requests of the mean free path smallness introduced in \cite{Land1}. Additionally,  
an application of the fluid dynamics to the process of the hot spot expansion requires some initial conditions among which 
the most intriguing one is a condition of a small value of the shear viscosity/entropy ratio. For the collective flow, described by Vlasov's equation, usual perturbative mechanisms, see \cite{PerVis}, responsible for the shear viscosity value are absent. Indeed, in the Vlasov's equation framework, in the first approximation in the  plasma parameter, the collision term is zero implying an entropy conservation and collective flow of the particles without rescattering. 
To resolve this problems of the collective flow of an almost ideal fluid with very small viscosity coefficients some new mechanisms of the 
shear viscosity smallness were proposed, see \cite{Tuchin,AnVis,AnVis1,AnVis2,AnVis3}.
Additionally,  
for the non-isotropic initial conditions, the connection between stress-energy tensor and viscosity is more complicated than in the case of isotropic initial state, see \cite{Eppel} for example,  that means an existence of a family of different viscosity coefficients in the expanding system, for example see  \cite{Tuchin} or \cite{Landau3}.

 Referring to Vlasov's equation as to the equation which describes equilibrium state only, the viscosity coefficients will be equal zero by definition. Therefore some time-dependent fluctuations around the equilibrium must be considered
in the approach in order to generate the viscosity coefficients. The mechanisms behind these fluctuations may be, for example, some instabilities generated in the plasma, see \cite{AnVis,AnVis1,AnVis2,AnVis3} and references therein. The viscosity obtained by this way is called "anomalous" one, in contrast to viscosity coefficient created by the perturbative rescattering
mechanisms as in \cite{PerVis}.   
The proposed generation of "anomalous" viscosity and corresponding  solution of Vlasov's equation, in turn, requires an existence of some equilibrium state which is perturbed by fluctuations, but in the case of very fast expanding hot spot such set-up may be questionable. 

 In this paper we continue to develop a model proposed in \cite{Our1,Our2}. Namely, we assume that the local energy density fluctuations,  charged hot spots, see \cite{New1,New2,New3,New4,New5} and also \cite{TFire,HotSp}, are created at the very initial stage of interactions at  time $\tau\,<\,0.5\,fm/c\,$, see \cite{Fluid2,Teaney,Cass,HotSp,Our1,Cass11}. These fluctuations are very dense and small, whose size is much smaller than the proton size, see  \cite{HotSp,Our1}, with energy density is much larger than the density achieved  in high-energy interactions at the energies $\sqrt{s}\,<\,100\,GeV\,$. In our model we assume that the hot spots
consist of the particles with weak inter-particle interactions and have a non-zero charge.
The small coupling constant approximation makes our model useful for the electromagnetic type of interactions or for the modeling of the interactions of asymptotically free quarks and gluons  in QCD. In the later case, this expanding fluid element can be considered as local colored density fluctuation, see \cite{Cass11,Mish} for example.
Therefore, we consider our model as some "toy" model for the real QCD processes, which applicability is justified by the smallness of the strong coupling constant in the case 
of a fluctuation with $0.05\,-\,0.5\,fm$ size, see \cite{Our1}. In our model we assume also, that the charged hot spot is perturbated 
 by the external field at the initial moment of time, and only then the fluctuation is expanding  with relativistic velocity along z axis.
Thus, consequently, further hot spot's evolution is  
considered  as an adiabatic expansion of the very dense drop of matter, which kinetic  description is given by a non-stationary solution of Vlasov's equation
with some inhomogeneous initial conditions. In this model, for the simplicity, we consider only one kind of charge, nevertheless the generalization to the case of many charges is straightforward. As the initial condition for the expansion, we choose a 
"rigid body" condition, see \cite{Davidson,Our2}, where the whole fluid element is rotating with some angular velocity. 
The value of the angular velocity is a  parameter of the model which we can vary, it's emergence 
may be understood as consequence of the interaction of the charged fluctuation with crossed electrical and magnetic fields at the initial moment of time, see \cite{Davidson}.
Thus, in addition to the parameter which characterizes 
interaction of the hot spot with the external field, another degree of freedom is introduced in the model. 
As we will see further, those two new degrees of freedom determine leading values of the viscosity coefficients.

 In the next Section 2, we discuss the form of Vlasov's equation, its integral representation and perturbative framework for the solution as well as a form of the external electromagnetic field. In Section 3 we solve Vlasov's equation and calculate mean velocities of the expanding hot spot in the leading order of the perturbation.
Section 4 is dedicated to the calculation of the momentum flux tensor and viscosity coefficients of the system, whereas  
in the Section 5 the values of the shear viscosity coefficients at different values of the problem's parameters as functions of time are present.
In the Section 6 we determine the form  the viscosity coefficients dependence on temperature and present the calculation of the ratios of the viscosity coefficients to the entropy as functions of temperature in comparison with the lattice data.  
The Conclusions are presented in Section 7.

\section{Vlasov's equation for a charged hot spot in electromagnetic field}

\subsection{Vlasov's equations system}

 In this section we review main results of \cite{Our2} for Vlasov's equation. The usual form of Vlasov's equation for a 
system of charged particles in an external electromagnetic field is given by
\beq\label{Vlas11}
\frac{\partial\,f_{s}}{\partial\,t}\,+\,\vec{v}\,\frac{\partial\,f_{s}}{\partial\vec{r}}\,+
\,\Le\,q\vec{E}\,+\,\frac{q}{c}\,\vec{v}\times\vec{B}\Ra\,\frac{\partial\,f_{s}}{\partial\vec{p}}\,=\,0\,.
\eeq
In a case of non-stationary processes the equation must be solved together with   Maxwell's equations
\begin{eqnarray}\label{Vlas22}
&\,& \nabla\times\vec{E}\,=\,-\frac{1}{c}\,\frac{\partial \vec{B}}{\partial t}\,,\\
\label{Vlas33}
&\,& \nabla\times\vec{B}\,=\,\frac{1}{c}\,\frac{\partial \vec{E}}{\partial t}\,+\,\frac{4\pi\,}{c}\,\vec{j}\,=\,
\frac{1}{c}\,\frac{\partial \vec{E}}{\partial t}\,+\,\frac{4\pi\,q\,n\,}{c}\,\langle\,\vec{v}\,\rangle\,,\\
\label{Vlas44}
&\,& \nabla\cdot\,\vec{E}\,=\,4\,\pi\,q\,n\,\int\,f_{s}(\vec{r},\,\vec{p},\,t)\,d^{3}p\,,\\
&\,& \nabla\cdot\,\vec{B}\,=\,0\,,
\label{Vlas55}
\end{eqnarray}
where we normalize our distribution function as follows
\beq
\int\,f_{s}(\vec{r}=0,\,\vec{p},\,t=0)\,d^{3}p\,=\,1\,.
\eeq
Electric and magnetic fields in Maxwell equations \eq{Vlas22}-\eq{Vlas55} are a combination of the 
self-generated electric 
$\vec{E}^{s}$ and
 magnetic $\vec{B}^{s}$ fields of the matter element and external electromagnetic  field:
\beq\label{TotF}
 \vec{E}_{Total}\,=\,\vec{E}^{s}\,-\,\vec{E}^{ext}\,;\,\,
 \vec{B}_{Total}\,=\,\vec{B}^{s}\,+\,\vec{B}^{ext}\,. 
 \eeq
In the next subsection we discuss the configuration of these fields.

\subsection{Fields configuration}

 As an external electromagnetic field in the problem we consider the electromagnetic field  in the transverse plan 
which perturbates the density fluctuation at the initial moment of time only. In the rest frame of the hot spot of  interest, this field is given
by the following potentials 
\beq\label{Pot1}
\varphi(\vec{r},t)\,=\,A_{z}(\vec{r},t)\,=\,0\,,
\eeq
\beq\label{Pot2}
A_{x}(\vec{r},t)\,=\,-\,2\,q\,n_{0}\,V_{0}\,\theta(z\,-\,ct)\,\partial_{x}\,K_{0}(|r_{\bot}\,-\,b|/r_{0D})\,,
\eeq
\beq\label{Pot3}
A_{y}(\vec{r},t)\,=\,-\,2\,q\,n_{0}\,V_{0}\,\theta(z\,-\,ct)\,\partial_{y}\,K_{0}(|r_{\bot}\,-\,b|/r_{0D})\,,
\eeq 
see \cite{Our2}. Here $r_{0D}$ is a Debye length of the incident cloud of the particles which surrounds the hot spot, see \cite{Klim}-\cite{PScr}, $K_{0}$ is the Macdonald function and
$b$ is the transverse distance between incident drop  and hot spot at rest (impact parameter), in the following we consider the case
$b\,=\,0$.
The radial component of the 4-potential in the cylindrical coordinate system, has the following form:
\beq\label{Pot5}
A_{r}(\vec{r},t)\,=\,\,A_{x}(\vec{r},t)\cos\,\theta + A_{y}(\vec{r},t)\,\sin\,\theta\,,
\eeq
thus we obtain
\beq
A_{r}(\vec{r},t)\,=\,-\,
2\, q\, n_{0}\,V_{0}\,\theta(z\,-\,c t)\, 
\pd{K_{0}(|r|/r_{0D})}{r}\,=\,Q_{T}\,\theta(z\,-\,c t)\,K_{1}(\xi_{0})\,. 
\eeq
Here $\xi_{0}\,=\,\frac{r}{r_{0D}}$ and we defined an overall charge of the incident drop  as
\beq\label{Pot6}
Q_{T}\,=\,q\, n_{0}\,V_{0}\,.
\eeq
Therefore, the hot spot at rest is affected by the following external fields:
\beq\label{Pot4}
\vec{E}^{ext}_{r}\,=\,-\,\frac{1}{c}\frac{\partial\,A_{r}}{\partial\,t}\,\hat{e}_{r}\,=
\,\frac{Q_{T}}{r_{0D}}\,\delta(c t)\,K_{1}(\xi_{0})\,
\eeq
and
\beq\label{Pot7}
\vec{B}^{ext}_{\theta}\,=\frac{\partial\,A_{r}}{\partial\,z}\,\hat{e}_{\theta}\,
=\,\frac{Q_{T}}{r_{0D}}\,\delta(c t)\,K_{1}(\xi_{0})\,.
\eeq
Here we  took into account, that our fluid element is at $z\,=\,0$ 
at the time $t\,=\,0$. We also assume axially symmetry of the problem,  otherwise
some dependence of $Q_{T}$ on polar angle must be introduced.
 
The following configuration of the self-generated  fields suit our problem:
\begin{eqnarray}
&\,& \vec{E}^{s}_{r}\,=\,-\,\frac{1}{c}\frac{\partial\,A_{r}^{s}}{\partial\,t}\,\hat{e}_{r}\,,
\vec{E}^{s}_{\theta}\,=\,-\,\frac{1}{c}\frac{\partial\,A_{\theta}^{s}}{\partial\,t}\,\hat{e}_{\theta}\,;
\,\,\,\,\,\,\,\,\,\,\,\,\,\,\,\,\,\,\,\,\,\,\,\label{Fl1}\\
&\,& \vec{B}^{s}_{r}\,=\,-\,\frac{\partial\,A_{\theta}^{s}}{\partial\,z}\,\hat{e}_{r}\,,\,\,
\vec{B}^{s}_{\theta}\,=\,\frac{\partial\,A_{r}^{s}}{\partial\,z}\,\hat{e}_{\theta}\,,\,\,
\vec{B}^{s}_{z}\,=\,\frac{1}{r}\frac{\partial\,r\,A_{\theta}^{s}}{\partial\,r}\,\hat{e}_{z}\,\label{Fl2},
\end{eqnarray}
see \cite{Our2,Davidson}.
%In terms of electromagnetic field Maxwell equations \eq{Vlas33}-\eq{Vlas44} have the following form:
%\begin{eqnarray}
%&\,& \label{Pot11}
%-\frac{\partial B_{\theta}}{\partial z}\, = \,\frac{1}{c}\,\frac{\partial E_{r}}{\partial %t}\,+\,\frac{4\pi\,}{c}\,j_{r}\,
%;\\
%&\,& \label{Pot22}
%\frac{1}{r}\,\frac{\partial ( r\,E_{r})}{\partial r}\, = \,4\pi\,q\,n\,\,;\\
%&\,& \label{Pot33}
%\frac{\partial B_{r}}{\partial z}\,-\frac{\partial B_{z}}{\partial r}\, = \,\frac{1}{c}\,\frac{\partial %E_{\theta}}{\partial t}\,+\,\frac{4\pi\,}{c}\,j_{\theta}\,;\\
%&\,& \label{Pot55}
%\frac{\partial E_{\theta}}{\partial z}\, = \,\frac{1}{c}\frac{\partial B_{r}}{\partial t}\,\,;\\
%&\,& \label{Pot66}
%\frac{1}{r}\,\frac{\partial ( r\,E_{\theta})}{\partial r}\, = \,
%-\,\frac{1}{c}\frac{\partial B_{z}}{\partial t}\,\,;\\
%&\,& \label{Pot44}
%\frac{1}{r}\,\frac{\partial ( r\,B_{\theta})}{\partial r}\, = \,\frac{4\pi\,}{c}\,j_{z}\,\,;\\
%&\,& \label{Pot77}
%\frac{\partial E_{r}}{\partial z}\, = \,-\,\frac{1}{c}\frac{\partial B_{\theta}}{\partial t}\,.
%\end{eqnarray}
In general, when we consider a non-stationary problem, a  
formulation of Maxwell equations in terms of potentials $A_{r}\,,A_{\theta}$\footnote{We use the potentials only for the self-consistent fields calculations, in the following we drop the $s$ subscript from the
potentials symbols.} is more useful.
So, the following equations hold for the potentials:
\begin{eqnarray}
&\,& \label{Pot8}
\frac{1}{c^2}\frac{\partial^2 A_{r}}{\partial t^2}\,-\,\frac{\partial^2 A_{r}}{\partial z^2}\,=\,
\frac{4\pi\,}{c}\,j_{r}\,;\\
&\,& \label{Pot9}
\frac{1}{c^2}\frac{\partial^2 A_{\theta}}{\partial t^2}\,-\,\frac{\partial^2 A_{\theta}}{\partial z^2}
\,-\,\frac{1}{r}\frac{\partial}{\partial r}
\Le\,r\frac{\partial A_{\theta}}{\partial r}\Ra\,+\,\frac{A_{\theta}}{r^2}\,=\,
\frac{4\pi\,}{c}\,j_{\theta}\,;\\
&\,& \label{Pot10}
-\frac{1}{r\,c}\frac{\partial^2 \Le r\,A_{r}\Ra}{\partial t\,\partial r}\,=\,4\pi\,q\,n\,;\\
&\,& \label{Pot11}
\frac{1}{r}\frac{\partial^2 \Le r\,A_{r}\Ra}{\partial r\,\partial z}\,=\,\frac{4\pi\,}{c}\,j_{z}\,.
\end{eqnarray}
We obtained, that equation  \eq{Pot8} is a one-dimensional  wave equations for the
potential $A_{r}\,$ with a source which may depend on the potentials of the problem. 
In turn, equation  \eq{Pot9} is a kind of non-homogeneous Bessel  equation for the potential $A_{\theta}\,$,  where
the source  may depend on the potentials as well.
Equations \eq{Pot10}-\eq{Pot11} have  different meaning.
Namely, we can rewrite them in the following form:
\beq\label{Pot55}
q\,\frac{\partial n}{\partial z}\,+\,\frac{1}{c^2}\,\frac{\partial j_{z}}{\partial t}\,=\,0\,.
\eeq
The problem is formulated for the relativistic motion of the drop along $z$ axis when
\beq
j_{z}\,=\,q\,n\,<v_{z}>\,\approx\,\,q\,n\,c\,
\eeq
implying for \eq{Pot55}:
\beq\label{Pot66}
\frac{\partial n}{\partial z}\,+\,\frac{1}{c}\,\frac{\partial n}{\partial t}\,=\,0\,,
\eeq
which is homogeneous wave equation for the wave propagation in positive $z$ direction.
We see, that both \eq{Pot10}-\eq{Pot11} are simply provide a condition that to a first approximation
our fluid element expands along $z$ axis with $<v_{z}>\,\approx\,c\,$ velocity  .

\subsection{Vlasov's equation in  integral form}

 In order to simplify the solution of our equation, the first step we make is a 
reformulation of  Vlasov's equation in terms of dimensionless variables.
We introduce:
\beq\label{Dimen}
t\,\rightarrow\,\frac{t}{t_{0}}\,\,,\vec{r}\,\rightarrow\,\frac{\vec{r}}{l_{0}}\,\,,l_{0}\,=\,t_{0}\,c\,\,,
\vec{p}\,\rightarrow\,\frac{\vec{p}}{mc}\,\,,\vec{v}\,\rightarrow\,\frac{\vec{v}}{c}\,\,,
\eeq
where $t_{0}$ is a time of  applicability of  Vlasov's equation and $l_{0}$ corresponds to this time length.
So, in the following, we will understand as variables the dimensionless quantities from \eq{Dimen}. 
We rewrite \eq{Vlas11}, obtaining the following equation:
\beq\label{VlasDL}
\frac{\partial\,f_{s}}{\partial\,t}\,+\,\vec{v}\,\frac{\partial\,f_{s}}{\partial\vec{r}}\,+
\,\frac{t_{0}}{m c}\,\vec{F}\,\frac{\partial\,f_{s}}{\partial\vec{p}}\,=\,0\,.
\eeq
Further, we assume the  parameter
\beq\label{PertPar}
\frac{t_{0} |\vec{F}|}{m c}\,\simeq\,\frac{l_{0} |\vec{F}|}{m c^2}\,,
\eeq
as a  perturbative parameter of the problem.
In both cases, either this parameter is large or small, it can serve as the parameter 
for a perturbative expansion of functions. In our further calculations we take this parameter as a small one, that implies
a smallness of the work of dissipative forces during the hot spot expansion relatively to its kinetic energy .

 Following to  \cite{Our1} we rewrite the differential equation \eq{VlasDL} in the integral form.
First of all we rewrite Vlasov's equation as 
\beq
\label{Vl1}
\,\frac{\partial\,f_{s}}{\partial\,t}\,+\,\vec{v}\,\frac{
\partial\,f_{s}}{\partial \vec{r}}\,=\,-\,\frac{t_{0} \vec{F}}{mc}\,
\frac{\partial\,f_{s}}{\partial \vec{p}}\,.
\eeq
The l.h.s.  of this equation we can write as the well-known transport equation, see \cite{Vlad},
whose fundamental solution $\mathscr{E}$ satisfies
\beq\label{Vl2}
\,\frac{\partial\,\mathscr{E}}{\partial\,t}\,+\,\vec{v}\,\frac{
\partial\,\mathscr{E}}{\partial \vec{r}}\,=\,\delta(t)\,\delta^{2}(\vec{r})\,,
\eeq 
here $\vec{r}$ is the 2-dimensional $\vec{r}\,=\,(r,z)$ vector.
The fundamental solution of \eq{Vl2} is well known:
\beq\label{Vl3}
\mathscr{E}\,=\,\theta(t)\,\delta(\,r\,-\,v_{r}\,t\,)\,\delta(\,z\,-\,v_{z}\,t\,).
\eeq 
With the help of \eq{Vl2}, we rewrite \eq{Vl1} as an integral equation:
\beq\label{Vl4}
f_{s}(\vec{r},\vec{v},t)=-\frac{t_{0} }{m c}\,\int d r^{'} d z^{'} dt^{'}
\mathscr{E}(t-t^{'},\vec{r}-\vec{r}^{'})\,\vec{F}(t^{'},\vec{r}^{'})\,
\frac{\partial\,f_{s}(\vec{r}^{'},\vec{v},t^{'})}{\partial \vec{p}}+
f_{0}(\vec{r}-\vec{v}\,t\, ,\vec{v})\,,
\eeq 
where the function $f_{0}(r,\vec{\zeta})$ will be determined later. 
Inserting \eq{Vl3} into \eq{Vl4}, we obtain:
\beq\label{Vl5}
f_{s}(\vec{r},\vec{v},t)\, = \,
-\,\frac{t_{0}}{mc}\int_{0}^{t}d t^{'}
\vec{F}(t^{'},\vec{r}-\vec{v}(t-t^{'}))
\frac{\partial\,f_{s}(\vec{r}-\vec{v}(t-t^{'}),\vec{v},t^{'})}{\partial \vec{p}}
+\,
 f_{0}(\vec{r}-\vec{v}t ,\vec{v}).
\eeq 
Now, our Lorentz force $\vec{F}$ consists of two parts, external one and part which corresponds to self-generated fields:
\beq
\vec{F}\,=\,\vec{F}^{ext}\,+\,\vec{F}^{s}\,,
\eeq  
where we have for the external part of the force in the terms of dimensionless variables:
\begin{eqnarray}
&\,& \label{Vl6}
F_{r}^{ext}\,=\,q\,\Le -\,E_{r}^{ext}\,-\,v_{z}\,B_{\theta}^{ext} \Ra\,=\,
-\,\frac{q\,Q_{T}}{l_{0}^{2} \hat{r}_{0D}}\,K_{1}(\xi_{0})\,\Le\,1\,+\,v_z\Ra\,\delta(t)
\,\\
&\,& \label{Vl7}
F_{\theta}^{ext}\,=\,0\,,\\
&\,& \label{Vl8}
F_{z}^{ext}\,=\,q\,\,v_{r}\,B_{\theta}^{ext}\,=\,
\frac{q\,Q_{T}\,}{l_{0}^{2} \hat{r}_{0D}}\,v_{r}\,K_{1}(\xi_{0})\,\delta(t)\,,
\end{eqnarray}  
with redefined $\xi_{0}$
\beq\label{Vl88}
\xi_{0}\,=\,\frac{|r\,-\,\hat{r}_{0D}|}{\hat{r}_{0D}}
\eeq
and with $\hat{r}_{0D}\,=\,r_{0D}/l_{0}$.
Thus using \eq{Vl6}-\eq{Vl8}, we obtain for \eq{Vl5}:
\begin{eqnarray}\label{Vl9}
&\,& f_{s}(\vec{r},\vec{v},t)\, = \,
-\,\frac{t_{0}}{mc}\int_{0}^{t}d t^{'}
\vec{F}^{s}(t^{'},\vec{r}-\vec{v}(t-t^{'}))
\frac{\partial\,f_{s}(\vec{r}-\vec{v}(t-t^{'}),\vec{v},t^{'})}{\partial \vec{p}}+\,\nonumber\\
&\,& + \frac{t_{0}}{mc}\frac{q Q_{T}}{l_{0}^{2} \hat{r}_{0D}} K_{1}(\xi_{0}(v_{r} t))
\Le\frac{\partial f_{s0}(\vec{r}-\vec{v}t,\vec{v})}{\partial p_{r}}\Le 1+v_z\Ra
-v_{r}\frac{\partial\,f_{s0}(\vec{r}-\vec{v}t,\vec{v})}{\partial p_{z}}\Ra
+f_{0}(\vec{r}-\vec{v}t ,\vec{v})\,,
\end{eqnarray}
here we denoted  $f_{s0}\,=\,f_{s}(\vec{r},\vec{v},t=0)$.
Initial condition of the problem is provided by  \eq{Vl9} by taking $t\,=\,0$. We have
\beq\label{Vl10}
f_{s0}(\vec{r},\vec{v})\,=\,\Lambda_{0}\,\Le\frac{\partial f_{s0}(\vec{r},\vec{v})}{\partial p_{r}}
\Le 1+v_z\Ra
-v_{r}\frac{\partial\,f_{s0}(\vec{r},\vec{v})}{\partial p_{z}}\Ra
+f_{0}(\vec{r},\vec{v})\,,
\eeq 
with
additional small parameter of the problem  
\beq\label{Vl11}
\Lambda_{0}\,=\,\frac{t_{0}}{mc}\frac{q Q_{T}}{l_{0}^{2} \hat{r}_{0D}} K_{1}(\xi_{0})\,=\,
\frac{q Q_{T}}{m c^2 r_{0D}} K_{1}(\xi_{0})\,<<\,1\,,
\eeq 
that value characterizes a smallness of the interaction of the charged fluctuation with the external field at the initial moment of time relative to the hot spot's kinetic energy.
Here the function $f_{0}(\vec{r},\vec{v})$ provides an initial condition for the problem in absence of external fields. 
An analytical solution of  \eq{Vl10} for the $f_{s0}$  function is 
unknown, further we will consider only perturbative solutions of this equation.

\subsection{Initial conditions for Vlasov's equation}

There are two functions which we need as initial conditions in order to solve \eq{Vl9}, they are $f_{s0}$ and $f_{0}$.
We can solve perturbatively \eq{Vl10} considering $\Lambda_{0}$ as a small parameter:
\beq
f_{s0}(\vec{r},\vec{v})\,=\,\sum_{n\,=\,0}^{\infty}\,{\Lambda_{0}^{n}}\,f_{s0n}
\eeq
At the first order on this parameter the solution of \eq{Vl9} is:
\beq\label{DFun1}
f_{s0}(\vec{r},\vec{v})\,=\,\Lambda_{0}\,\Le\frac{\partial f_{0}(\vec{r},\vec{v})}{\partial p_{r}}
\Le 1+v_z\Ra
-v_{r}\frac{\partial\,f_{0}(\vec{r},\vec{v})}{\partial p_{z}}\Ra
+f_{0}(\vec{r},\vec{v})\,,
\eeq
see \cite{Our2}. Therefore, an important step of the problem is the choice of the initial condition
for the Vlasov's equation without an external field. So, first of all, we define the following
numericall parameters\footnote{In the dimensionless form.}:
\beq
\beta\,=\,|v_{z}|\,\propto\,1,\,\,
\gamma\,=\,\sqrt{1-v^2}\,\approx\,\sqrt{1-\beta^2}\,<<\,1.
\eeq
The energy of a relativistic particle is given by the following Hamiltonian:
\beq
H\,=\,m c^2\,\sqrt{1\, +\,p^{2}_{\bot}\,+\,p_{z}^2}\,\approx\,m c^2\,\sqrt{1\, +\,p_{z}^2}\,\Le
1\,+\,\frac{p^{2}_{\bot}}{2\,\Le 1\, +\,p_{z}^2 \Ra}\,\Ra.
\eeq
In our notations we have:
\beq\label{DFunc4}
p_{z}\,=\,\frac{\beta}{\gamma}\,=\,\frac{ v_{z}}{\gamma}\,,\,\,\,\,p_{r}\,=\,v_{r}\,,\,\,\,\,
p_{\theta}\,=\,v_{\theta}\,.
\eeq
Thus we obtain:
\beq\label{DFunc41}
H\,\approx\,m c^2 \frac{\beta}{\gamma}\,+\,m c^2\,\frac{\gamma}{\beta}\frac{p^{2}_{\bot}}{2}\,=\,H_{\|}\,+\,H_{\bot}\,.
\eeq
Now, following by \cite{Davidson}, we define our initial distribution function as a kind of "rigid-rotor" equilibrium
distribution function with factorization of longitudinal and transverse velocities:
\beq\label{DFun2}
f_{0}(\vec{r},\vec{v})\,=\,\frac{\gamma}{\beta}\Le \frac{m^2 c^3}{2 \pi}\Ra\, \delta \Le H_{\bot}\,-\,P_{\theta}\omega_{r}\,-\,k T_{\bot}\Ra\,
\delta \Le m c p_{z}\,-\,\frac{m c \beta}{\gamma}\Ra\,\phi(z)\,,
\eeq
where we adopt
\beq
P_{\theta}\,=\,\frac{\gamma}{\beta}\,r\,l_{0}\,m c\Le p_{\theta}\,-\,\omega_{c}\,r\,l_{0}\,/\,2 c\,\Ra\,,
\eeq
with
\beq\label{MagF}
\omega_{c}\,=\,\frac{|q|\,B_{z0}}{m\,c}\,.
\eeq
Here
\beq
A^{0}_{\theta}\,=\,\,B_{z0}\,r\,/\,2
\eeq
is the solution of homogeneous \eq{Pot9} and
function $\phi(z)\,$ determines the position of the drop in the laboratory coordinate frame on the $z$ axis. 
We note also, that the form of \eq{DFun2} function is consistent with requests on microcanonical ensemble distribution function which describes an adiabatic states of the matter.

  The normalization of \eq{DFun2} function is usual:
\beq\label{InCond1}
\int d^3 p\,f_{0}(\vec{r},\vec{v})\,=\,\frac{\gamma}{\beta}\Le \frac{m^ 2 c^3}{2  \pi}\Ra\,\int d^2 p_{\bot}\,dp_{z}\,
\delta \Le H_{\bot}\,-\,P_{\theta}\omega_{r}\,-\,k T_{\bot}\Ra\,
\delta \Le m c p_{z}\,-\,\frac{m c \beta}{\gamma}\Ra\,\phi(z)\,=\,\phi(z)\,,
\eeq
with the request $\phi(z)_{z\,=\,0}\,=\,1$. We notice also, that 
the integral of the delta-function in \eq{InCond1} is not zero only for some  values of 
$r$. Indeed, rewriting 
\beq\label{DFunc3}
H_{\bot}-\omega_{r}\,P_{\theta}\,=\,m c^2\,
\frac{\gamma}{2\,\beta}\Le\,p_{r}^{2}\,+\,(p_{\theta}\,-\,\frac{r l_{0}}{c}\omega_{r} )^{2}\,\Ra\,+\,\psi(r)\,,
\eeq
with an effective potential $\psi(r)$ 
\beq
\psi(r)\,=\,\frac{m\,\gamma}{2\,\beta}\,\omega_{r}\,\Le \omega_{c}\,-\,\omega_{r}\Ra\,r^{2}\,l_{0}^{2}\,,
\eeq
we obtain that 
for the distribution function \eq{DFun1} the density profile is non-zero if only
\beq\label{In14}
n(r) = \left\{
\begin{array}{rl}
& 1\,=\,const.,\,\,0\,\leq\,r\,<\,r_{b} \\
& 0 \,,\,\,\,\,\,\,\,\,\,\,\,\,\,\,\,\,\,\,\,\,\,\,\,\,\,\,\,\,\,  r\,\,>\,r_{b}\,,\\
\end{array} \right.
\eeq
where
\beq\label{In15}
r_{b}^{2}\,=\,\frac{\beta}{\gamma}\frac{2\,k\,T_{\bot}\,}{\omega_{r}\,m\,l_{0}^{2}\,\Le \omega_{c}\,-\,\omega_{r}\,\Ra\,}\,.
\eeq

\section{Averaged velocities in the first order approximation}

 The parameter  \eq{PertPar} is the parameter which determines a perturbative expansion of any function in our framework. As we mentioned above, for the adiabatic expansion of the hot spot, this  parameter is small:
\beq
\Lambda\,=\,\frac{t_{0} |\vec{F}|}{m c}\,\,<\,1\,,
\eeq 
and we can construct the following series 
\beq\label{Pert1}
f_{s}(\vec{r},\,\vec{v},\,t)\,=\,\sum_{i=0}^{\infty}\,f_{si}(\vec{r},\,\vec{v},\,t)\,\Lambda^{i}\,,\,\,\,\,\,
\vec{E}^{s}(\vec{r}\,,\,t)\,=\,\sum_{i=0}^{\infty}\,\vec{E}^{s}_{i}(\vec{r}\,,\,t)\,\Lambda^{i}\,,\,\,\,\,\,\,
\vec{B}^{s}(\vec{r}\,,\,t)\,=\,\sum_{i=0}^{\infty}\,\vec{B}^{s}_{i}(\vec{r}\,,\,t)\,\Lambda^{i}\,.
\eeq
In the first order the distribution function is given by \eq{DFun1} with only $\vec{r}$ argument shift:
\beq\label{DistrF1}
f_{s0}(\vec{r},\vec{v},t)\,=\,f_{0}(\vec{r}-\vec{v} t,\vec{v})\,+\,
\,\Lambda_{0}\,\Le\frac{\partial f_{0}(\vec{r}-\vec{v} t,\vec{v})}{\partial p_{r}}
\Le 1+v_z\Ra
-v_{r}\frac{\partial\,f_{0}(\vec{r}-\vec{v} t,\vec{v})}{\partial p_{z}}\Ra\,,
\eeq 
whereas for the calculation of the $f_{s1}(\vec{r},\vec{v},t)\,$ term we need to calculate 
$\vec{E}^{s}_{0}$ and $\vec{B}^{s}_{0}$ fields using \eq{Pot8}-\eq{Pot9}. Further, the only zero order of $\Lambda$
parameter is considered, next order contribution we will investigate in a separate publication.

  In turn, the calculation of the potentials  and transport properties of the system requires
a knowledge of the averaged velocities:
\beq
\langle\,v_{r}\,\rangle_{s}\,=\,\sum_{i=0}^{\infty}\,\langle\,v_{r}\,\rangle_{s}^{i}\,\Lambda^{i}\,,\,\,\,\,
\langle\,v_{\theta}\,\rangle_{s}\,=\,\sum_{i=0}^{\infty}\,\langle\,v_{\theta}\,\rangle_{s}^{i}\,\Lambda^{i}\,\,\,\,\,
\langle\,v_{z}\,\rangle_{s}\,=\,\sum_{i=0}^{\infty}\,\langle\,v_{z}\,\rangle_{s}^{i}\,\Lambda^{i}\,.
\eeq
The values of $\langle\,v_{r}\,\rangle_{s}^{0}$, $\langle\,v_{\theta}\,\rangle_{s}^{0}$ and
$\langle\,v_{z}\,\rangle_{s}^{0}$ are calculated in the next subsection.

\subsection{Radial velocity in the first order approximation}

 An averaged radial velocity to the first perturbative order in sense of \eq{Pert1},  is defined as
\beq\label{RadVel1}
\langle\,v_{r}\,\rangle_{s}^{0}\,=\,\frac{\int\,d^{3}p\,v_{r}\,f_{s}(\vec{r},\,\vec{v},\,t)\,}
{\,\int\,d^{3}p\,f_{s}(\vec{r},\,\vec{v},t)}\,=\,
\frac{\int\,d^{3}p\,v_{r}\,f_{s0}(\vec{r},\,\vec{v},\,t)\,}
{\,\int\,d^{3}p\,f_{s0}(\vec{r},\,\vec{v},t)}\,
\eeq 
with the distribution function of \eq{DistrF1} which satisfies condition \eq{In14} at $t\,=\,0$.
The calculation of the normalization factor in this equation is simple.
For the denominator we have:
\beq\label{RadVel11} 
N\,=\,\int\,d^{3}p\,f_{s0}(\vec{r},\,\vec{v},t)\,=\,\frac{\phi(z\,-\,\beta t)}{\sqrt{1\,+\,t^2\,\alpha/\rho}}\,,
\eeq
see 
\eq{InCond1} and calculations below.
In the numerator, due to the argument shift in \eq{DistrF1},
we rewrite \eq{DFunc3}:
\beq
H_{\bot}-\omega_{r}\,P_{\theta}\,=\,m c^2\,
\frac{\gamma}{2\,\beta}\Le\,p_{r}^{2}\,+\,(p_{\theta}\,-\,\frac{(r\,-\,v_{r}\,t) l_{0}}{c}\omega_{r} )^{2}\,\Ra\,+\,
\frac{m\,\gamma}{2\,\beta}\,\omega_{r}\,\Le \omega_{c}\,-\,\omega_{r}\Ra\,(r\,-\,v_{r}\,t)^{2}\,l_{0}^{2}
\,,
\eeq
or
\beq\label{RadVel2}
H_{\bot}-\omega_{r}P_{\theta}=\Le\,\alpha t^2 +\rho\Ra\Le p_{r}-
\frac{\,\alpha r t}{\alpha t^2 +\rho}\Ra^{2}+
\rho\,(p_{\theta}\,-\,\frac{(r\,-\,v_{r}\,t) l_{0}}{c}\omega_{r} )^{2}+
\alpha\,r^{2}\,\Le 1-\frac{\,\alpha  t^{2}}{\alpha t^2 +\rho}\Ra=
H_{\bot}+\psi(r)\,,
\eeq
where
\beq\label{Al}
\alpha\,=\,\frac{m\,\gamma}{2\,\beta}\,\omega_{r}\,l_{0}^{2}\Le \omega_{c}\,-\,\omega_{r}\Ra\,,\,\,\,
\rho\,=\,m c^2\,\frac{\gamma}{2\,\beta}\,,
\,\,\,\psi(r)=\alpha\,r^{2}\,\Le 1-\frac{\,\alpha  t^{2}}{\alpha t^2 +\rho}\Ra\,
\eeq
and where  $v_{r}\,=\,p_{r}\,$ identity was used.

 Now consider the contribution in the numerator from the first term of \eq{DistrF1} in \eq{RadVel1}:
\beq\label{RadVel11Add1}
\langle\,v_{r}\,\rangle_{s}^{01}\,=\,\frac{\int\,d^{3}p\,v_{r}\,f_{0}\,}{\,\int\,d^{3}p\,f_{s0} }\,.
\eeq 
The integral over $p_{z}$ in \eq{RadVel1} is simple:
\beq
m c\int\,dp_{z}\,\delta \Le m c p_{z}\,-\,\frac{m c \beta}{\gamma}\Ra\,\phi(z-v_{z} t)\,=\,\phi(z\,-\,\beta t)\,,
\eeq
and therefore
\beq
\langle\,v_{r}\,\rangle_{s}^{01}\,=\,N^{-1}\,\frac{\gamma}{\beta}\Le \frac{m c^2}{2  \pi}\Ra\,\int d^2 p_{\bot}\,v_{r}\,
\delta \Le H_{\bot}\,-\,P_{\theta}\omega_{r}\,-\,k T_{\bot}\Ra\,\phi(z\,-\,\beta t)\,.
\eeq
Shifting $p_{r}$ variable in the even $H$ function we obtain:
\beq\label{Shift1}
\langle\,v_{r}\,\rangle_{s}^{01}\,=\,\,N^{-1}\,\frac{\gamma}{\beta}\Le \frac{m c^2}{2  \pi}\Ra
\frac{\,\alpha r t}{\alpha t^2 +\rho}
\int d^2 p_{\bot}\,
\delta \Le H_{\bot}\,-\,P_{\theta}\omega_{r}\,-\,k T_{\bot}\Ra\,\phi(z\,-\,\beta t)\,.
\eeq
Performing additional variables change
\beq\label{Shift2}
p_{r}\,\rightarrow\,\frac{p_{r}}{\sqrt{\alpha t^2 +\rho}}\,,\,\,\,\,
p_{\theta}\,\rightarrow\,\frac{p_{\theta}}{\sqrt{\rho}}\,
\eeq
and taking into account that in terms of new variable $p^2\,=\,p_{r}^2\,+\,p_{\theta}^{2}$ we have $d^2\,p_{\bot}\,=\,\pi\,d\,p^2$
we get the final answer for this term:
\beq\label{RadVel33}
\langle\,v_{r}\,\rangle_{s}^{01}\,=\,\,N^{-1}\,r\,t\,
\frac{\omega_{r}\,l_{0}^{2}\Le \omega_{c}\,-\,\omega_{r}\Ra\,/\,c^2}
{\Le\,1\,+\,t^2\,\omega_{r}\,l_{0}^{2}\Le \omega_{c}\,-\,\omega_{r}\Ra\,/\,c^2\Ra^{3/2}}\,\phi(z\,-\,\beta t)\,=\,
\,r\,t\,\frac{\alpha/\rho}{ 1+t^{2}\alpha/\rho}\,.
\eeq
In the case when
\beq
\,\frac{t^2\,\alpha}{\rho}\,\ll\,1,
\eeq
we obtain for the radial velocity 
\beq\label{RadVel3}
\langle\,v_{r}\,\rangle_{s}^{01}\,\approx\,
r\,t\,\frac{\alpha}{\rho}
\,=\,\frac{r\,t\,}{ c^2}
\omega_{r}\,l_{0}^{2}\Le \omega_{c}\,-\,\omega_{r}\Ra\,\ll\,1\,,
\eeq
whereas at large $t$ when
\beq
\,\frac{t^2\,\alpha}{\rho}\,\gg\,1,
\eeq
the answer for the radial velocity is the following:
\beq
\langle\,v_{r}\,\rangle_{s}^{01}\,\approx\,\frac{r}{ t}\,.
\eeq
In the following we consider only the limit $|\alpha|/\rho\,\ll\,1$ for any $t$ of interest.

 The second term which contributes in \eq{RadVel1} is the following:
\beq\label{RadVel6}
\langle\,v_{r}\,\rangle_{s}^{02}\,=\,\,N^{-1}\,
\int\,d^2 p_{\bot}\,dp_{z}\,v_{r}\,\Lambda_{0}\,\frac{\partial f_{0}(\vec{r}-\vec{v} t,\vec{v})}{\partial p_{r}}
\Le 1+v_z\Ra\,\phi(z\,-\,v_{z}\,t)\,.
\eeq
In the integration of this integral there is a term which contains $\frac{\partial \Lambda_{0}}{\partial p_{r}}$ derivative.
This derivative has an additional small parameter $1/\hat{r}_{0D}$, therefore further we will consider
$\Lambda_{0}$ as a constant with $\xi_{0}$ defined as in \eq{Vl88}.Thereby we have:
\beq\label{RadVel61}
\langle\,v_{r}\,\rangle_{s}^{02}\,=\,-\,N^{-1}\,\Lambda_{0}\,
\int\,d^2 p_{\bot}\,dp_{z}\,f_{0}(\vec{r}-\vec{v} t,\vec{v})\,
\Le 1+v_z\Ra\,\phi(z\,-\,v_{z}\,t)\,.
\eeq
The integration gives:
\beq\label{RadVel62}
\langle\,v_{r}\,\rangle_{s}^{02}\,=\,-\,N^{-1}\,\Lambda_{0}\,\Le\,
\frac{\phi(z\,-\,\beta t)}{\sqrt{1\,+\,t^2\,\alpha/\rho}}\,
+\,\gamma\,\int\,d^2 p_{\bot}\,dp_{z}\,
f_{0}(\vec{r}-\vec{v} t,\vec{v})\,p_{z}\phi(z\,-\,v_{z} t)\,\Ra\,
\eeq
or finally
\beq\label{RadVel7}
\langle\,v_{r}\,\rangle_{s}^{02}\,=\,-\,\Lambda_{0}\,\Le\,1\,+\,\beta\,\Ra\,.
\eeq

  The contribution of the third term in \eq{RadVel1} with \eq{DistrF1} distribution function is zero 
due the derivative of the product of Delta function and $\phi$ function in \eq{DistrF1}.  
Therefore, taking all terms together we obtain finally:
\beq\label{RadVel44}
\langle\,v_{r}\,\rangle_{s}^{0}=\Le\,r\,t\,
\frac{\alpha/\rho}{ 1+t^{2}\alpha/\rho}\,\,
\,-\,
\Lambda_{0}\,\Le\,1\,+\,\beta\,\Ra\,
\Ra\,,
\eeq  
which in liner in $t$ approximation is
\beq\label{RadVel4}
\langle\,v_{r}\,\rangle_{s}^{0}=\Le\,r\,t\,
\frac{\omega_{r}\,l_{0}^{2}\Le \omega_{c}-\omega_{r}\Ra}{ c^2}\,-\,
\Lambda_{0}\,\Le\,1\,+\,\beta\,\Ra\,
\Ra\,.
\eeq

\subsection{Azimuthal velocity in the first order approximation}

 The averaged azimuthal velocity is defined similarly to \eq{RadVel1}:
\beq\label{AzVel1}
\langle\,v_{\theta}\,\rangle_{s}^{0}\,=\,\frac{\int\,d^{3}p\,v_{\theta}\,f_{s0}(\vec{r},\,\vec{v},\,t)\,}
{\,\int\,d^{3}p\,f_{s0}(\vec{r},\,\vec{v},\,t)}\,,
\eeq  
where the only contribution in this integral comes from the first part of the distribution function \eq{DistrF1}.
Performing shift in the delta function of \eq{RadVel2}
\beq\label{TTrans1}
p_{\theta}\,\rightarrow\,\frac{(r\,-\,v_{r}\,t) l_{0}}{c}\omega_{r}\,,
\eeq
we obtain:
\beq\label{AzVel2}
\langle\,v_{\theta}\,\rangle_{s}^{0}\,=\,N^{-1}\,\frac{l_{0}\,\omega_{r}}{c}\,\int\,d^{3}p\,\Le\,r\,-\,v_{r}\,t\Ra\,
f_{s0}(\vec{r},\,\vec{v},\,t)\,.
\eeq
Inserting function \eq{DistrF1} into \eq{AzVel2} we obtain:
\beq\label{AzVel22}
\langle\,v_{\theta}\,\rangle_{s}^{0}\,=\,N^{-1}\,\frac{l_{0}\omega_{r}}{c} 
\frac{\phi(z-\beta t)}{\Le 1\,+\,t^2\,\alpha/\rho\Ra^{1/2}}\,
\Le r - \langle\,v_{r}\,\rangle_{s}^{0} t\Ra\,=\,\frac{l_{0}\omega_{r}}{c} 
\,
\Le r \,-\, \langle\,v_{r}\,\rangle_{s}^{0} t\Ra\,.
\eeq

\subsection{Longitudinal velocity in the first order approximation}

 The last, averaged $z$-axis velocity, is defined by the following integral:
\beq\label{ZVel1}
\langle\,v_{z}\,\rangle_{s}^{0}\,=\,N^{-1}\Le\int\,d^{3}p\,f_{0}\,v_{z}\,+\,
\,\int\,d^{3}p\,\Lambda_{0}\,v_{z}\,\frac{\partial f_{0}}{\partial p_{r}}\,\Le 1+v_{z}\Ra\,-\,
\,\int\,d^{3}p\,\Lambda_{0}\,v_{z}\,v_{r}\frac{\partial f_{0}}{\partial p_{z}}\Ra\,.
\eeq 
Let's consider these terms one by one.

 The first term in \eq{ZVel1} is simple:
\beq\label{ZVel2}
\langle\,v_{z}\,\rangle_{s}^{01}\,=\,N^{-1}\,\int\,d^{3}p\,f_{0}\,v_{z}\,= \beta\,,
\eeq
see calculations in the previous sections.

 The second term in \eq{ZVel1} does not contribute and taking into account \eq{DFunc4} we have for the third term:
\beq\label{ZVel3}
\langle\,v_{z}\,\rangle_{s}^{02}\,=\,N^{-1}\,\gamma\,\Lambda_{0}\,
\int\,d^{3}p\,v_{r}\,f_{0}\,=\,\gamma\,\Lambda_{0}\,\langle\,v_{r}\,\rangle_{s}^{01}\,.
\eeq
Taking both terms together we obtain:
\beq\label{ZVel4}
\langle\,v_{z}\,\rangle_{s}^{0}\,=\Le\beta\,+\,\gamma\Lambda_{0}\,\frac{\alpha}{\rho}\frac{r\,t\,}{\Le 1\,+\,t^2\,\alpha/\rho\Ra}\Ra\,,
\eeq
which the linear on $t$  order gives
\beq\label{ZVel5}
\langle\,v_{z}\,\rangle_{s}^{0}\,=\Le\beta\,+\,\gamma\Lambda_{0}\,r\,t\,\frac{\alpha}{\rho}\Ra\,.
\eeq

\section{ Viscosity coefficients  of the system}

 For the non-isotropic plasma system in an external field the shear viscosity coefficients can be found from the 
general momentum flux tensor expression\footnote{The expression presents the tensor in the dimensionless form. In order to obtain a dimensional tensor the expression must be multiplied on the $c^{2}$ factor.}:
\beq
\sigma_{\alpha \beta}\,=\,\,n\,m\,\int\,d^3\,p\,
\Le\,v_{\alpha}\,-\,\langle\,v_{\alpha}\,\rangle\,\Ra\,
\Le\,v_{\beta}\,-\,\langle\,v_{\beta}\,\rangle\,\Ra\,
f_{s}(\vec{r},\vec{v},t)\,=\,P_{\alpha \beta}\,+\,n\,m\,
\langle\,v_{\alpha}\,\rangle\,\langle\,v_{\beta}\,\rangle\,+\,\sigma^{'}_{\alpha \beta}\,
%=\,
%\eta_{\alpha \beta}^{\rho \gamma}\,\frac{\partial \langle\,v_{\rho}\,\rangle}{\partial q_{\gamma}}\,
\eeq 
where
\beq
\sigma^{'}_{\alpha \beta}\,\propto\,\eta_{\alpha \beta}^{\rho \gamma}\,\frac{\partial \langle\,v_{\rho}\,\rangle}{\partial q_{\gamma}}\,
\eeq
is a tensor of viscosity stress 
with viscosity coefficients $\eta_{\alpha \beta}^{\rho \gamma}\,$  symmetrical for the pair of indexes $\alpha,\beta$.
There are diagonal and non-diagonal terms in this expression, further we will calculate these terms separately.

\subsection{Diagonal parts of the momentum flux tensor }
\subsubsection{$\sigma_{r r}$ part of the momentum flux tensor}

 The $\sigma_{r r}$ part of the momentum flux tensor has the following form
\beq\label{RR1}
\sigma_{r r}\,=\,n\,m\,\int\,d^3\,p\,
\Le\,v_{r}\,-\,\langle\,v_{r}\,\rangle^{0}_{s}\,\Ra\,
\Le\,v_{r}\,-\,\langle\,v_{r}\,\rangle^{0}_{s}\,\Ra\,
f_{s0}(\vec{r},\vec{v},t)\,.
\eeq 
Performing the variables change
\beq\label{RR11}
v_{r}\,\rightarrow\,v_{r}\,-\,\langle\,v_{r}\,\rangle^{01}_{s}\,
\eeq
we obtain:
\beq\label{RR14}
\sigma_{r r}=n m\int\,d^3\,p
\Le v_{r}-\langle v_{r}\rangle^{02}_{s}\Ra^{2}
f_{s0}(\vec{r},\vec{v},t)=
n m \int\,d^3\,p\,
v_{r}^{2} f_{s0}(\vec{r},\vec{v},t)+
n m\,\Le\langle v_{r} \rangle^{02}_{s}\Ra^{2}
\int\,d^3\,p\,
f_{s0}(\vec{r},\vec{v},t)\,
\eeq 
with 
\beq\label{RR131}
H_{\bot}-\omega_{r}\,P_{\theta}\,=\,\Le\,\alpha t^2 +\rho\Ra p_{r}^{2}\,+\,
\,\rho\,p_{\theta}^{2}\,+\,
\alpha\,r^{2}\,\Le\,1-\,\frac{\,\alpha  t^{2}}{\alpha t^2 +\rho}\Ra\,.
\eeq
The only unknown integral here is the integral on $v_{r}^{2}$ variable:
\beq\label{RR5}
\sigma_{r r}^{1}\,=\,n m\,\int\,\,d^3 p\,v_{r}^{2} f_{s0}(\vec{r},\vec{v},t)\,.
\eeq  
This integral has two non-zero contributions, see \eq{DistrF1}\,. The first one is
\beq\label{RR6}
\sigma_{r r}^{11}\,=\,n m\,\int\,\,d^3 p\,v_{r}^{2}\,f_{0}(\vec{r}-\vec{v} t,\vec{v})\,
\eeq
which we can write as
\beq\label{RR7}
\sigma_{r r}^{11}=\frac{n m\gamma}{\beta}\Le \frac{m c^2}{2  \pi}\Ra
\frac{1}{\Le\alpha t^2 +\rho\Ra^{3/2}}\frac{1}{\sqrt{\rho}}
\int d^2 \,p_{\bot}\, p_{r}^{2}\,
\delta \Le H_{\bot}+\psi(r)-k T_{\bot}\Ra\phi(z-\beta t)\,,
\eeq
with
\beq\label{RR8}
H_{\bot}\,+\,\psi(r)\,\,=\, p_{r}^{2}\,+\,
p_{\theta}^{2}\,+\,\psi(r)\,=\,\, p_{r}^{2}\,+\,
p_{\theta}^{2}\,+\,
\alpha\,r^{2}\,\Le\,1-\,\frac{\,\alpha  t^{2}}{\alpha t^2 +\rho}\Ra\,.
\eeq
Using $d^2 \,p_{\bot}\,=\,\pi\,d\,H_{\bot}$ we obtain:
\beq\label{RR10}
\sigma_{r r}^{11}\,=\,n m\,\frac{ \beta}{\gamma}\,
\frac{1}{1+t^2\alpha\,/\,\rho}\,
\Le
\frac{k T_{\bot}}{m c^2}\,-\,\frac{\psi(r)}{m c^2}
\Ra\frac{\phi(z-\beta t)}{\sqrt{ 1+t^2\alpha\,/\,\rho}}\,.
\eeq

 The second non-zero contribution in \eq{RR5} is
\beq\label{RR12}
\sigma_{r r}^{12}= n m\,\Lambda_{0}\int\,d^3 p\,v_{r}^{2}\,
\frac{\partial f_{0}(\vec{r}-\vec{v} t,\vec{v})}{\partial p_{r}}
\Le\,1+\,v_{z}\,\Ra
\,=\,-2\,n m\,\Lambda_{0}\,\int\,d^3 p\,v_{r}\,f_{0}(\vec{r}-\vec{v} t,\vec{v})\Le\,1+\,v_{z}\,\Ra\,.
\eeq
An integration gives:
\beq\label{RR121}
\sigma_{r r}^{12}\,=\, 
-2\,n m\,\Lambda_{0}\langle v_{r} \rangle^{01}_{s}\,\Le\,1\,+\,\beta\,\Ra\,
\frac{\phi(z-\beta t)}{\sqrt{1+t^2 \alpha/\rho}}\,.
\eeq

 Taking all terms  together we obtain:
\beqar\nonumber\label{RR13}
\sigma_{r r}&=&n m\,\Le
\frac{ \beta}{\gamma}
\frac{1}{1+t^2\alpha\,/\,\rho}
\Le
\frac{k T_{\bot}}{m c^2}-\frac{\psi(r)}{m c^2}
\Ra
+\Le\langle v_{r} \rangle^{0}_{s}\Ra^{2}\Ra\frac{\phi(z-\beta t)}{\sqrt{1+t^2 \alpha/\rho}}\,+\,\\
&+&
n m\,\Le
2\,\langle v_{r} \rangle^{01}_{s}\,\langle v_{r} \rangle^{02}_{s}\,+
\Le\langle v_{r} \rangle^{02}_{s}\Ra^{2}-\Le\langle v_{r} \rangle^{0}_{s}\Ra^{2}
\Ra
\frac{\phi(z-\beta t)}{\sqrt{1+t^2 \alpha/\rho}}\,.
\eeqar
Now, using the identity
\beq\label{RR14Add1}
\,\langle v_{r} \rangle^{01}_{s}\,=\,
r\,\frac{\partial \langle v_{r} \rangle^{0}_{s}}{\partial r}\,
\eeq
we obtain the following shear viscosity coefficient:
\beq
\eta_{rr}^{rr}\,\simeq\,n m\,\,r\,\langle v_{r} \rangle^{01}_{s}\,
\frac{\phi(z-\beta t)}{\sqrt{1+t^2 \alpha/\rho}}\,=
\,\tilde{n}(t)\,r\,\langle v_{r} \rangle^{01}_{s}\,,
\eeq
with
\beq\label{RR15}
\tilde{n}(t)\,=\,n m\,\frac{\phi(z-\beta t)}{\sqrt{1+t^2 \alpha/\rho}}\,.
\eeq

\subsubsection{$\sigma_{\theta\theta}$ part of the momentum flux tensor}

 The $\sigma_{\theta\theta}$ part of the momentum flux tensor has the following form
\beq\label{TT1}
\sigma_{\theta\theta}\,=\,n\,m\,\int\,d^3\,p\,
\Le\,v_{\theta}\,-\,\langle\,v_{\theta}\,\rangle^{0}_{s}\,\Ra\,
\Le\,v_{\theta}\,-\,\langle\,v_{\theta}\,\rangle^{0}_{s}\,\Ra\,
f_{s0}(\vec{r},\vec{v},t)\,
\eeq 
or 
\beq\label{TT2}
\sigma_{\theta\theta}=n m\,\int\,d^3 p\, v_{\theta}^{2} f_{s0}(\vec{r},\vec{v},t) -
2 n m \langle\,v_{\theta}\,\rangle^{0}_{s} \int\, d^3 p\, v_{\theta} f_{s0}(\vec{r},\vec{v},t) +
n m \Le\langle\,v_{\theta}\,\rangle^{0}_{s}\Ra^{2}\int\,d^3 p \,f_{s0}(\vec{r},\vec{v},t)\,.
\eeq 
Again, the only unknown integral in \eq{TT2} is the integral on $v_{\theta}^{2}$ variable:
\beq\label{TT3}
\sigma_{\theta\theta}^{1}\,=\,n m\,\int\,\,d^3 p\,v_{\theta}^{2} f_{s0}(\vec{r},\vec{v},t)\,.
\eeq
Performing the variable shift 
\beq\label{TT4}
p_{\theta}\,\rightarrow\,p_{\theta}\,+\,\frac{l_{0}\omega_{r}}{c}\Le r - v_{r}t\Ra
\eeq
we obtain: 
\beq\label{TT31}
\sigma_{\theta\theta}^{1}=n m\,
\int d^3 p\,\Le p_{\theta}\,+\,\frac{l_{0}\omega_{r}}{c}\Le r - v_{r}t\Ra
\Ra^{2}\,f_{s0}(\vec{r},\vec{v},\vec{v} )\,
\eeq
with $H_{\bot}$ given by \eq{RR131}.

 The first contribution in this integral is
\beq\label{TT32}
\sigma_{\theta\theta}^{11}=\frac{n m\gamma}{\beta}\Le \frac{m c^2}{2  \pi}\Ra\,\frac{1}{\rho^{3/2}}
\frac{1}{\sqrt{\alpha t^2 +\rho}}
\int d^2 p_{\bot} p_{\theta}^{2}\,
\delta \Le H_{\bot}+\psi(r)-k T_{\bot}\Ra\phi(z-\beta t)\,
\eeq
with $H_{\bot}$ from \eq{RR8}.
Similarly to \eq{RR7} we have:
\beq\label{TT33}
\sigma_{\theta\theta}^{11}=\frac{n m \beta}{\gamma m c^2 }
\frac{1}{\sqrt{1+t^2\alpha\,/\,\rho}}
\int dH_{\bot} H_{\bot}\,
\delta \Le H_{\bot}+\psi(r)-k T_{\bot}\Ra\phi(z-\beta t)\,,
\eeq
which gives:
\beq\label{TT34}
\sigma_{\theta\theta}^{11}=
n m\,\frac{ \beta}{\gamma}\,
\Le
\frac{k T_{\bot}}{m c^2}-\frac{\psi(r)}{m c^2}\,\Ra\,\frac{\phi(z-\beta t)}{\sqrt{1+t^2\alpha\,/\,\rho}}\,.
\eeq
The linear on $p_{\theta}$ term in  \eq{TT31} is zero, therefore for the next non-vanishing contribution in 
 \eq{TT31} we obtain:
\beq\label{TT35}
\sigma_{\theta\theta}^{12}=n m\,\frac{l_{0}^{2}\omega_{r}^{2}}{c^{2}}
\int d^3 p\,\Le \, r - v_{r}\,t\,\Ra^{2}\,f_{s0}(\vec{r},\vec{v},\vec{v} )\,,
\eeq 
that, after  $v_{r}\,\rightarrow\,v_{r}\,-\,\langle v_{r}\rangle^{01}_{s}$ variable change, gives:
\beq\label{TT36}
\sigma_{\theta\theta}^{12}=n m\,\frac{l_{0}^{2}\omega_{r}^{2}}{c^{2}}
\Le \, r^{2}\, -\,2\,r\,t\,\langle v_{r} \rangle^{0}_{s}\,+\,t^{2}\,
\Le\,\langle v_{r}^{2} \rangle^{0}_{s}\,+\,\Le\langle v_{r}\rangle^{01}_{s}\Ra^{2}\,\Ra\,
\Ra\,\frac{\phi(z-\beta t)}{\sqrt{1+t^2\alpha\,/\,\rho}} \,,
\eeq 
where
\beq\label{TT37}
n m\,\langle v_{r}^{2} \rangle^{0}_{s}\,=\,N^{-1}\,n m\,\int d^3 p\,v_{r}^{2}\,f_{s0}(\vec{r},\vec{v} ,\vec{v} )\,=\,
\,N^{-1}\,\sigma_{rr}^{1}\,,
\eeq
see \eq{RR6}-\eq{RR121}.

 Now, taking all the terms of \eq{TT2} together at the linear on $\Lambda_{0}$ order, we obtain:
\beqar\label{TT5}\nonumber 
\sigma_{\theta\theta}& =& 
n m\,\Le\frac{ \beta}{\gamma}\,
\Le
\frac{k T_{\bot}}{m c^2}-\frac{\psi(r)}{m c^2}\Ra
\Le 1 + \frac{l_{0}^{2}\omega_{r}^{2}t^{2}}{c^{2}\Le 1 + t^{2} \alpha/\rho\Ra}\Ra+
\Le\langle\,v_{\theta}\,\rangle^{0}_{s}\Ra^{2}
\Ra\frac{\phi(z-\beta t)}{\sqrt{1+t^2\alpha/\rho}}\,-\,\\
&-& n m\,\frac{l_{0}^{2}\omega_{r}^{2}}{c^{2}}\,
r\,
\Le 1 - \frac{t}{r} \langle v_{r} \rangle^{01}_{s}\Ra
\Le r - 2 t \langle v_{r} \rangle^{0}_{s}\, +\,
 t \langle v_{r} \rangle^{01}_{s}
\Ra\frac{\phi(z-\beta t)}{\sqrt{1+t^2\alpha/\rho}}\,.
\eeqar
Using  identity \eq{RR14Add1} with
\beq\label{TT6}
\frac{\partial \langle v_{\theta} \rangle^{0}_{s} }{\partial r}\,=\,\frac{l_{0}\omega_{r}}{c}\,
\Le 1 - \frac{t}{r} \langle v_{r} \rangle^{01}_{s}\Ra
\eeq
we obtain the following expressions for the viscosity coefficient:
\beq\label{TT8}
\eta_{\theta\theta}^{r\theta}\,\simeq\,\tilde{n}(t)\,\frac{l_{0}\omega_{r}}{c}\,r\,
\Le r - 2 t \langle v_{r} \rangle^{0}_{s} + t \langle v_{r} \rangle^{01}_{s}\Ra\,.
\eeq

\subsubsection{$\sigma_{z z}$ part of the momentum flux tensor} 

  Similarly to the previous subsections we have:
\beq\label{ZZ1}
\sigma_{z z}=n m\,\int\,d^3 p\, v_{z}^{2} f_{s0}(\vec{r},\vec{v},t) -
2 n m \langle\,v_{z}\,\rangle^{0}_{s} \int\, d^3 p\, v_{z} f_{s0}(\vec{r},\vec{v},t) +
n m \Le\langle\,v_{z}\,\rangle^{0}_{s}\Ra^{2}\int\,d^3 p \,f_{s0}(\vec{r},\vec{v},t)\,.
\eeq 
Our integral of interest is
\beq\label{ZZ2}
\sigma_{z z}^{1}=n m\,\int\,d^3 p\, v_{z}^{2}\, f_{s0}(\vec{r},\vec{v},t) \,.
\eeq 
There are two contributions in this integral, see \eq{DistrF1}. The first one is:
\beq\label{ZZ2Add1}
\sigma_{z z}^{11}=n m\,\int\,d^3 p\, v_{z}^{2} \, f_{0}(\vec{r}-\vec{v} t, \vec{v}) \,=\,n m\,\beta^{2}\,
\frac{\phi(z-\beta t)}{\sqrt{1+t^2\alpha/\rho}}\,.
\eeq 
The second contribution is
\beq\label{ZZ3}
\sigma_{z z}^{12}=\,-\,n m\,\Lambda_{0}\,\int\,d^3 p\, v_{z}^{2} \,v_{r}\,
\frac{\partial f_{0}(\vec{r}-\vec{v} t, \vec{v}) }{\partial p_{z}}\,=\,
2\,\gamma^{2}\,\,n m\,\Lambda_{0}\,\int\,d^3 p\, p_{z} \,v_{r}\,
f_{0}(\vec{r}-\vec{v} t, \vec{v})\,.
\eeq
Simple calculations gives:
\beq\label{ZZ3Add1}
\sigma_{z z}^{12}=\,2\,n m\,\gamma\,\beta\,\,\Lambda_{0}\,\langle v_{r} \rangle^{01}_{s}\,
\frac{\phi(z-\beta t)}{\sqrt{1+t^2\alpha/\rho}}\,.
\eeq
Now, taking into account \eq{ZVel3} result and summing up all the terms of \eq{ZZ1}, we obtain in linear on $\Lambda_{0}$ order: 
\beq\label{ZZ4}
\sigma_{z z}\simeq\,0\,.
\eeq

\subsection{Non-diagonal parts of the momentum flux tensor}

\subsubsection{$\sigma_{r\theta}$ part of the momentum flux tensor}

 The $\sigma_{r \theta}$ part of the momentum flux tensor reads as
\beq\label{RT1}
\sigma_{r \theta}\,=\,n\,m\,\int\,d^3\,p\,
\Le\,v_{r}\,-\,\langle\,v_{r}\,\rangle^{0}_{s}\,\Ra\,
\Le\,v_{\theta}\,-\,\langle\,v_{\theta}\,\rangle^{0}_{s}\,\Ra\,
f_{s0}(\vec{r},\vec{v},t)\,
\eeq 
or 
\beqar\label{RT2}
\sigma_{r \theta}& = &n m\,\int\,d^3 p\, v_{r} \, v_{\theta} f_{s0}(\vec{r},\vec{v},t) +
n m\, \langle\,v_{r}\,\rangle^{0}_{s} \langle\,v_{\theta}\,\rangle^{0}_{s}
\int\,d^3 p \,f_{s0}(\vec{r},\vec{v},t)\,-\,\nonumber\\
& - & \,n m \,\langle\,v_{r}\,\rangle^{0}_{s} \int\, d^3 p\, v_{\theta} f_{s0}(\vec{r},\vec{v},t) \,
- \,n m \langle\,v_{\theta}\,\rangle^{0}_{s} \int\, d^3 p\, v_{r} f_{s0}(\vec{r},\vec{v},t)  \,.
\eeqar
The only unknown contribution in \eq{RT2} is
\beq\label{RT3}
\sigma_{r \theta}^{1}\,=\,n\,m\,\int\,d^3\,p\,v_{r} \, v_{\theta} f_{s0}(\vec{r},\vec{v},t)\,.
\eeq
This integral, after the variable shifting \eq{TT4}, has the following form:
\beq\label{RT3Add1}
\sigma_{r \theta}^{1}\,=\,n\,m\,\int\,d^3\,p\,v_{r} \,
\Le v_{\theta} + \frac{l_{0}\omega_{r}}{c}
\Le
r - v_{r} t 
\Ra\Ra
f_{s0}(\vec{r},\vec{v},t)\,=\,\,n\,m\,\int\,d^3\,p\,
 \frac{l_{0}\omega_{r}}{c}\,v_{r} \,
\Le
r - v_{r} t
\Ra
f_{s0}(\vec{r},\vec{v},t)\,.
\eeq
The integration gives:
\beq\label{RT4}
\sigma_{r \theta}^{1}\,=\,n\,m\Le\,r\,\frac{l_{0}\omega_{r}}{c}\,\langle\,v_{r}\,\rangle^{0}_{s}\,-\,
\,n\,m\,t\,\frac{l_{0}\omega_{r}}{c}\,
\Le\,\langle\,v_{r}^{2}\,\rangle^{0}_{s}+\Le\langle\,v_{r}\,\rangle^{01}_{s}\Ra^{2}\,\Ra
\Ra\,
\frac{\phi(z-\beta t)}{\sqrt{1+t^2\alpha/\rho}}\,.
\eeq
Taking all terms of \eq{RT1} together we obtain:
\beqar\label{RT5}\nonumber
\sigma_{r \theta}& = & n m\,\frac{l_{0}\omega_{r}}{c}\,
\Le
-t\,\frac{ \beta}{\gamma}\,
\frac{1}{1+t^2\alpha\,/\,\rho}\,
\Le
\frac{k T_{\bot}}{m c^2}\,-\,\frac{\psi(r)}{m c^2}
\Ra\,+\,\langle\,v_{r}\,\rangle^{0}_{s}\,\langle\,v_{\theta}\,\rangle^{0}_{s}\,
\Ra\frac{\phi(z-\beta t)}{\sqrt{1+t^2\alpha/\rho}}\,+\,\\
& +& \,
n m\,\frac{l_{0}\omega_{r}}{c}\,
\Le 
-r \langle\,v_{r}\,\rangle^{0}_{s}\Le 1 - \frac{t}{r} \langle\,v_{r}\,\rangle^{01}_{s}  \Ra
+
t \langle\,v_{r}\,\rangle^{01}_{s}\,\langle\,v_{r}\,\rangle^{02}_{s}
\Ra
\frac{\phi(z-\beta t)}{\sqrt{1+t^2\alpha/\rho}}\,.
\eeqar
From \eq{RR14Add1} and \eq{TT6}  we obtain for the viscosity coefficients:
\beq\label{RT6}
\eta_{r\theta}^{r\theta}\,\simeq\,\tilde{n}(t)\,r\,
\langle v_{r} \rangle^{0}_{s}\,
\eeq
and
\beq\label{RT7}
\eta_{r\theta}^{rr}\,\simeq\,
\tilde{n}(t)\,\frac{l_{0}\omega_{r}}{c}\,r\,t\,
\langle v_{r} \rangle^{02}_{s}\,.
\eeq

\subsubsection{$\sigma_{r z}$ part of the momentum flux tensor}

 The $\sigma_{r z}$ part of the momentum flux tensor reads as
\beq\label{RZ1}
\sigma_{r z}\,=\,n\,m\,\int\,d^3\,p\,
\Le\,v_{r}\,-\,\langle\,v_{r}\,\rangle^{0}_{s}\,\Ra\,
\Le\,v_{z}\,-\,\langle\,v_{z}\,\rangle^{0}_{s}\,\Ra\,
f_{s0}(\vec{r},\vec{v},t)\,
\eeq 
or, again after the variable shift
\beq\label{RZ11}
v_{r}\,\rightarrow\,v_{r}\,-\,\langle\,v_{r}\,\rangle^{01}_{s}\,,
\eeq
the $\sigma_{r z}$ acquires the following form:
\beq\label{RZ12}
\sigma_{r z}\,=\,n\,m\,\int\,d^3\,p\,
\Le\,v_{r}\,-\,\langle\,v_{r}\,\rangle^{02}_{s}\,\Ra\,
\Le\,v_{z}\,-\,\langle\,v_{z}\,\rangle^{0}_{s}\,\Ra\,
f_{s0}(\vec{r},\vec{v},t)\,
\eeq 
with argument of $f_{s0}$ given by \eq{RR131}. 
This integral has four contributions which are calculated further.

 We have:
\beq\label{RZ3}
\sigma_{r z}^{1}\,=\,n\,m\,\int\,d^3\,p\,v_{r} \, v_{z} \,f_{s0}(\vec{r},\vec{v},t)\,.
\eeq
In turn there are two contribution in \eq{RZ3}. The first one is
\beq\label{RZ6}
\sigma_{r z}^{11}\,=\, n\,m\,\Lambda_{0}\,\int\,d^3\,p\,v_{r}
\, v_{z} \,\frac{\partial f_{0}(\vec{r} - \vec{v}t)\,}{\partial p_{r}}\,\Le 1 + v_{z}\Ra\,=\,
\,- n\,m\,\Lambda_{0}\,\int\,d^3\,p\,
\, v_{z} \,f_{0}(\vec{r} - \vec{v}t)\,\,\Le 1 + v_{z}\Ra\,,
\eeq
or 
\beq\label{RZ7}
\sigma_{r z}^{11}\,=\,\,- N^{-1}\, n\,m\,\Lambda_{0}\,\Le\,\beta\,+\,\beta^{2}\,\Ra\,=\,n\,m\,\beta\,
\langle\,v_{r}\,\rangle^{02}_{s}\,\frac{\phi(z-\beta t)}{\sqrt{ 1+t^2\alpha\,/\,\rho}}\,=\,\,n\,m\,
\langle\,v_{r}\,\rangle^{02}_{s}\,\langle\,v_{z}\,\rangle^{0}_{s}\,
\frac{\phi(z-\beta t)}{\sqrt{ 1+t^2\alpha\,/\,\rho}}
\,,
\eeq
with liner on $\Lambda_{0}$ precision. The second contribution in \eq{RZ3} is
\beq\label{RZ8}
\sigma_{r z}^{12}\,=\,- n\,m\,\Lambda_{0}\,\int\,d^3\,p\,v_{r}^{2}\,
v_{z} \,\frac{\partial f_{0}(\vec{r} - \vec{v}t)\,}{\partial p_{z}}\,=\,n\,m\,\Lambda_{0}\,\gamma\,
\int\,d^3\,p\,v_{r}^{2}\,f_{0}(\vec{r} - \vec{v}t)\,.
\eeq
Using \eq{RR6}  we obtain for this integral:
\beq\label{RZ9}
\sigma_{r z}^{12}\,=\,
\,n m\,\beta\,\Lambda_{0}
\frac{1}{1+t^2\alpha\,/\,\rho}\,
\Le
\frac{k T_{\bot}}{m c^2}\,-\,\frac{\psi(r)}{m c^2}
\Ra\frac{\phi(z-\beta t)}{\sqrt{ 1+t^2\alpha\,/\,\rho}}\,.
\eeq

  The second contribution in \eq{RZ12} is the following:
\beq\label{RZ10}
\sigma_{r z}^{2}\,=\,
\,- n\,m\,\langle\,v_{z}\,\rangle^{0}_{s}\,\int\,d^3\,p\,
\, v_{r} \, f_{s0}(\vec{r},\vec{v},t)\,=\,-\,\, n\,m\,
\langle\,v_{r}\,\rangle^{02}_{s}\,\langle\,v_{z}\,\rangle^{0}_{s}\,\frac{\phi(z-\beta t)}{\sqrt{ 1+t^2\alpha\,/\,\rho}}\,,
\eeq
see \eq{RadVel6}.

 The third and fourth contributions in \eq{RZ12} are simple. We have consequently:
\beq\label{RZ111}
\sigma_{r z}^{3}\,=\,
\,- n\,m\,\langle\,v_{r}\,\rangle^{02}_{s}\,\int\,d^3\,p\,
\, v_{z} \, f_{s0}(\vec{r},\vec{v},t)\,=\,-\,\, n\,m\,
\langle\,v_{r}\,\rangle^{02}_{s}\,\langle\,v_{z}\,\rangle^{0}_{s}\,
\frac{\phi(z-\beta t)}{\sqrt{ 1+t^2\alpha\,/\,\rho}}\,
\eeq 
and
\beq\label{RZ121}
\sigma_{r z}^{4}\,=\,
\, n\,m\,\langle\,v_{r}\,\rangle^{02}_{s}\,\langle\,v_{z}\,\rangle^{0}_{s}\,
\int\,d^3\,p\, f_{s0}(\vec{r},\vec{v},t)\,=\,\, n\,m\,
\langle\,v_{r}\,\rangle^{02}_{s}\,\langle\,v_{z}\,\rangle^{0}_{s}\,
\frac{\phi(z-\beta t)}{\sqrt{ 1+t^2\alpha\,/\,\rho}}\,.
\eeq 

 Taking all these contributions together we obtain:
\beq\label{RZ13} 
\sigma_{r z}\,=\,\tilde{n}(t)\,
\beta\,\Lambda_{0}
\frac{1}{1+t^2\alpha\,/\,\rho}\,
\Le
\frac{k T_{\bot}}{m c^2}\,-\,\frac{\psi(r)}{m c^2}
\Ra\,
\eeq
that means $\eta_{r z}\,\propto\,\Lambda^{2}_{0}$ in our perturbative scheme.

\subsubsection{$\sigma_{\theta z}$ part of the momentum flux tensor}

 The last $\sigma_{\theta z}$ part of the momentum flux tensor reads as
\beq\label{TZ1}
\sigma_{\theta z}\,=\,n\,m\,\int\,d^3\,p\,
\Le\,v_{\theta}\,-\,\langle\,v_{\theta}\,\rangle^{0}_{s}\,\Ra\,
\Le\,v_{z}\,-\,\langle\,v_{z}\,\rangle^{0}_{s}\,\Ra\,
f_{s0}(\vec{r},\vec{v},t)\,
\eeq 
or 
\beqar\label{TZT2}
\sigma_{\theta z}& = &n m\,\int\,d^3 p\, v_{z} \, v_{\theta} f_{s0}(\vec{r},\vec{v},t) +
n m\, \langle\,v_{z}\,\rangle^{0}_{s} \langle\,v_{\theta}\,\rangle^{0}_{s}
\int\,d^3 p \,f_{s0}(\vec{r},\vec{v},t)\,-\,\nonumber\\
& - & \,n m \,\langle\,v_{z}\,\rangle^{0}_{s} \int\, d^3 p\, v_{\theta} f_{s0}(\vec{r},\vec{v},t) \,
- \,n m \langle\,v_{\theta}\,\rangle^{0}_{s} \int\, d^3 p\, v_{z} f_{s0}(\vec{r},\vec{v},t)  \,.
\eeqar
The only non-known contribution in \eq{RT2} is
\beq\label{TZ3}
\sigma_{\theta z}^{1}\,=\,n\,m\,\int\,d^3\,p\,v_{z} \, v_{\theta} f_{s0}(\vec{r},\vec{v},t)\,.
\eeq
This integral, after the variable's shift \eq{TT4}, has the following form:
\beq\label{TZ4}
\sigma_{\theta z}^{1}\,=\,n\,m\,\int\,d^3\,p\,v_{z} \,
\Le v_{\theta} + \frac{l_{0}\omega_{r}}{c}
\Le
r - v_{r} t 
\Ra\Ra
f_{s0}(\vec{r},\vec{v},t)\,=\,\,n\,m\,\int\,d^3\,p\,
 \frac{l_{0}\omega_{r}}{c}\,v_{z} \,
\Le
r - v_{r} t
\Ra
f_{s0}(\vec{r},\vec{v},t)\,.
\eeq
The first part of integral \eq{TZ4} is simple:
\beq\label{TZ5}
\sigma_{\theta z}^{11}\,=\,n\,m\,r\,\frac{l_{0}\omega_{r}}{c}\,\langle\,v_{z}\,\rangle^{0}_{s}\,\frac{\phi(z-\beta t)}{\sqrt{ 1+t^2\alpha\,/\,\rho}}\,.
\eeq
In the second part of \eq{TZ4} we perform \eq{RZ11} variable change obtaining
\beq\label{TZ6}
\sigma_{\theta z}^{12}\,=\,- n\,m\,t\,\frac{l_{0}\omega_{r}}{c}\,
\int\,d^3\,p\,v_{z} \,
\Le v_{r} + \langle\,v_{r}\,\rangle^{01}_{s} \Ra\,f_{s0}(\vec{r},\vec{v},t)\,. 
\eeq
Using \eq{RZ3}-\eq{RZ9} results we obtain for the \eq{TZ6} integral:
\beq\label{TZ7}
\sigma_{\theta z}^{12}\,=\,- n\,m\,t\,\frac{l_{0}\omega_{r}}{c}\,
\Le \langle\,v_{z}\,\rangle^{0}_{s} \langle\,v_{r}\,\rangle^{01}_{s} +
\langle\,v_{z}\,\rangle^{0}_{s} \langle\,v_{r}\,\rangle^{02}_{s} +
\beta\,\Lambda_{0}
\frac{1}{1+t^2\alpha\,/\,\rho}\,
\Le
\frac{k T_{\bot}}{m c^2}\,-\,\frac{\psi(r)}{m c^2}
\Ra
\Ra
\frac{\phi(z-\beta t)}{\sqrt{ 1+t^2\alpha\,/\,\rho}}\,.
\eeq
Now, taking all contributions of \eq{TZ1} together, we obtain:
\beq\label{TZ8}
\sigma_{\theta z}\,=\,-n\,m\,t\,\frac{l_{0}\omega_{r}}{c}\,
\beta\,\Lambda_{0}
\frac{1}{1+t^2\alpha\,/\,\rho}\,
\Le
\frac{k T_{\bot}}{m c^2}\,-\,\frac{\psi(r)}{m c^2}
\Ra
\frac{\phi(z-\beta t)}{\sqrt{ 1+t^2\alpha\,/\,\rho}}\,,
\eeq
that again means $\eta_{\theta z}\,\propto\,\Lambda^{2}_{0}$ in our perturbative scheme.

\section{Time dependence of the viscosity coefficients}

 The following viscosity coefficients are calculated in the previous section\footnote{The calculated coefficients are dimensionless. In order to write these coefficients in the dimensional form we have to multiply the expressions on the $c\,l_{0}$ factor.}:
\beqar\label{VC1} 
\eta_{rr}^{rr}\,& \simeq & \,\tilde{n}(t)\,r\,\langle v_{r} \rangle^{01}_{s}\,,\\
\eta_{\theta\theta}^{r\theta}\,& \simeq & \,\tilde{n}(t)\,\frac{l_{0}\omega_{r}}{c}\,r\,
\Le r - 2 t \langle v_{r} \rangle^{0}_{s} + t \langle v_{r} \rangle^{01}_{s}\Ra\,,\label{VC11}\\
\eta_{r\theta}^{r\theta}\,& \simeq & \,\tilde{n}(t)\,r\,
\langle v_{r} \rangle^{0}_{s}\,,\label{VC12}\\
\eta_{r\theta}^{rr}\,& \simeq & \,
\tilde{n}(t)\,\frac{l_{0}\omega_{r}}{c}\,r\,t\,
\langle v_{r} \rangle^{02}_{s}\,\label{VC13}.
\eeqar
All of these coefficients are functions of dimensionless $r$ variables therefore in the further calculations we will take 
$r\,=\,1$ value of $r$, see \eq{Dimen}.
Also, we define the following parameters of the problem:
\beqar\label{VC2} 
A\,& = &\,\frac{l_{0}\omega_{r}}{c}\,,\,\,\,0\,<\,A\,<\,1\,, \\
B\,& = &\,\frac{\alpha}{\rho}\,=\,\frac{l_{0}^{2}\omega_{r}\,\omega_{c}}{c^2}\,-\,\frac{l_{0}^{2}\omega_{r}^{2}\,}{c^2}\,=\,
A\,\Le\,C\,-\,A\,\Ra\,,\,\,\,0\,<\,B\,<\,1\,,\,\,\,A\,<\,C\,,\label{VC21}  \\
\langle\,v_{r}\,\rangle^{01}_{s}\,& = & \,\frac{t\,\alpha/\rho}{1+t^2\,\alpha/\rho}\,=\,\frac{t\,B}{1+t^{2}\,B}\,,\,\,\,
0\,<\,t\,<\,1\,,\label{VC22} \\
\langle v_{r} \rangle^{0}_{s}\,& = & 
\langle\,v_{r}\,\rangle^{01}_{s}\,+\,\langle\,v_{r}\,\rangle^{02}_{s}\,=\,
\frac{t\,B}{1+t^{2}\,B}\,-\,\Lambda_{0}\,\Le\,1+\,\beta\,\Ra\,,\,\,\,
\Lambda_{0}\,\ll\,1\,,\,\,\,\beta\,\approx\,1\,\label{VC23} .
\eeqar
Taking $\phi(z-\beta t)\,\approx\,\phi(0)\,=\,1$ and normalizing the expressions on the mutual $n\,m$ factor, we rewrite these coefficients in the following form:
\beqar\label{VC3} 
\eta_{rr}^{rr}\,& \simeq & \,\frac{t\,B}{\Le 1+t^{2}\,B \Ra^{3/2}}\,,\label{VC35}\\
\eta_{\theta\theta}^{r\theta}\,& \simeq & \,\frac{A}{\Le 1+t^{2}\,B \Ra^{1/2}}\,\Le
1\,-\,\frac{t^{2}\,B}{1\,+\,t^{2}\,B}\,+\,2\,t\,\Lambda_{0}\Le\,1+\,\beta\,\Ra\,
\Ra\,,\label{VC33}\\
\eta_{r\theta}^{r\theta}\,& \simeq & \,
\frac{1}{\Le 1+t^{2}\,B \Ra^{1/2}}\,\left|\,\Lambda_{0}\Le\,1+\,\beta\,\Ra
\,-\,\frac{t\,B}{1\,+\,t^{2}\,B}\,
\right|
\,,\label{VC31} \\
\eta_{r\theta}^{rr}\,& \simeq & \,
\frac{A\,t\,}{\Le 1+t^{2}\,B \Ra^{1/2}}\,\left|\,\Lambda_{0}\Le\,1+\,\beta\,\Ra
\,\right|
\,.\label{VC32} 
\eeqar
The sign of $\Lambda_{0}$ parameter can be different depending on the sign of the external field in  \eq{Vl11}.
In the case when $\Lambda_{0}\,>\,0$ an external field is compressing the initial fluctuation, 
when  $\Lambda_{0}\,<\,0$  
the field expands the fluctuation, the case $\Lambda_{0}\,=\,0$ corresponds to the absence of the external field.
In the viscosity coefficients \eq{VC31}-\eq{VC32} we will take the absolute values of the expressions assuming that at $t\,=\,0$ the coefficients are positive.

\subsection{$\Lambda_{0}\,\geq\,0\,$ case }

\begin{figure}[thbp]
\begin{center}
\psfig{file=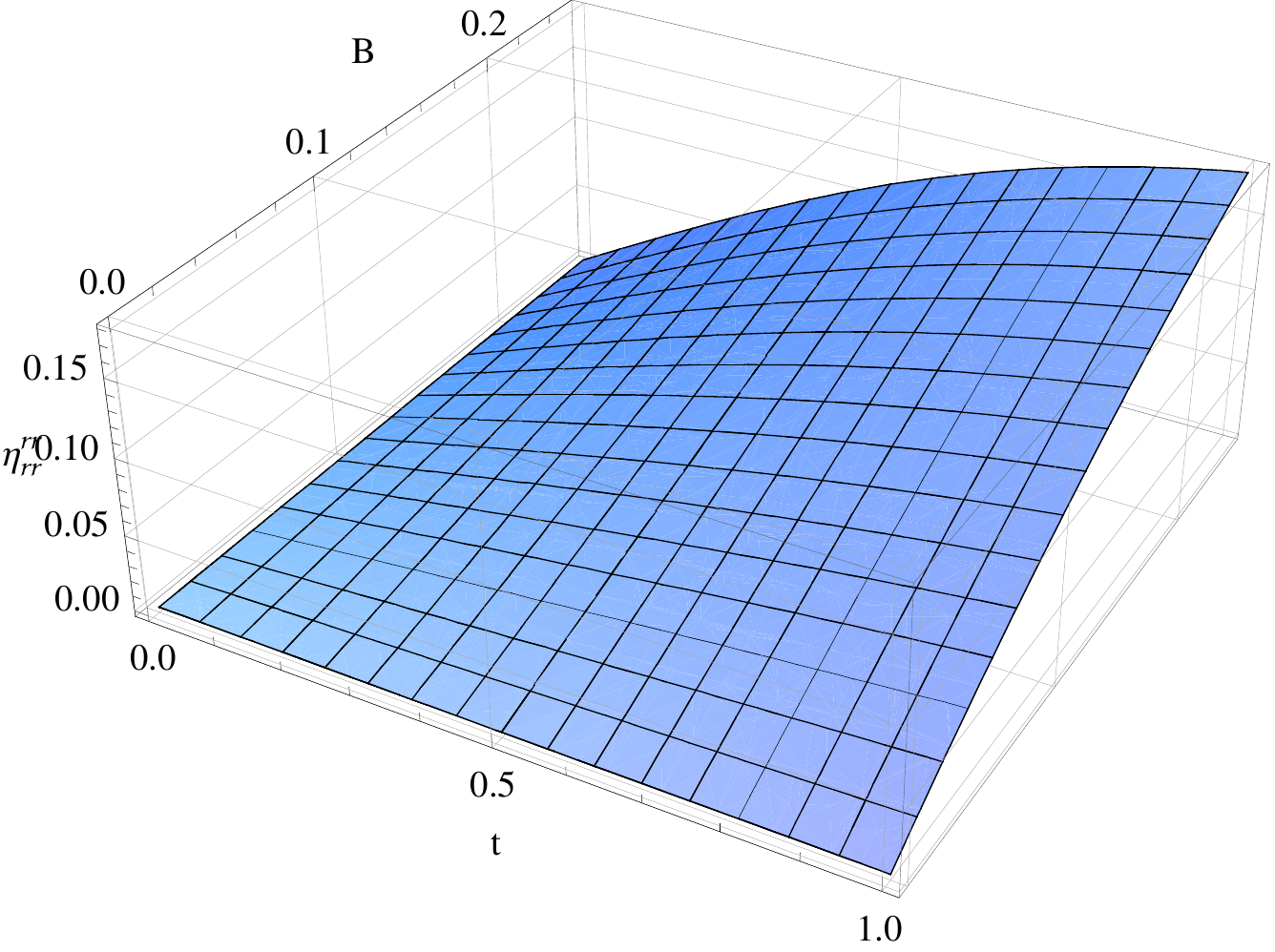 ,width=90mm} 
%\psfig{file=Pos_Lambda_r_t_t_t_Coeff_B0_01_L0_5.eps ,width=60mm} & 
%\psfig{file=Pos_Lambda_r_t_t_t_Coeff_B0_L0_5.eps,width=60mm}\\
% &  \\ 
%\fig{Vis_rttt_1}-c & \fig{Vis_rttt_1}-d \\
% &  \\
\end{center}
\caption{\it The $\eta^{r r}_{r r}$ viscosity coefficient as function of $B$ and $t$.}
\label{Vis_rrrr}
\end{figure}

\begin{figure}[thbp]
\begin{center}
\begin{tabular}{c c}
\psfig{file=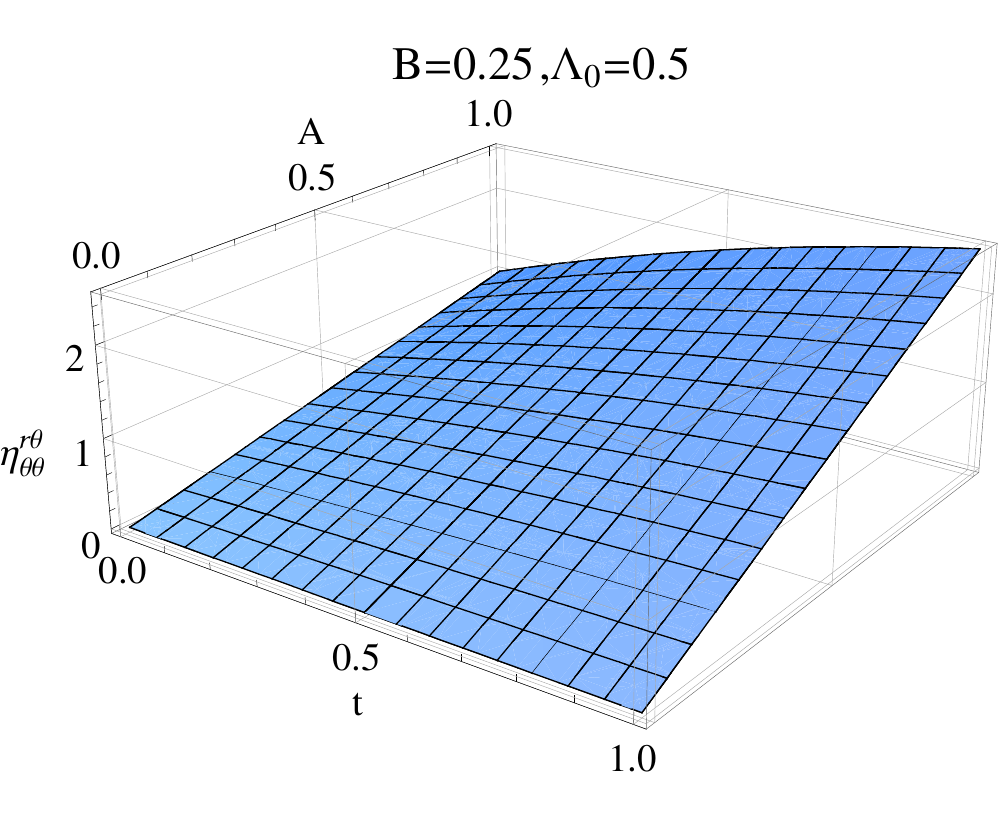 ,width=70mm} & 
\psfig{file=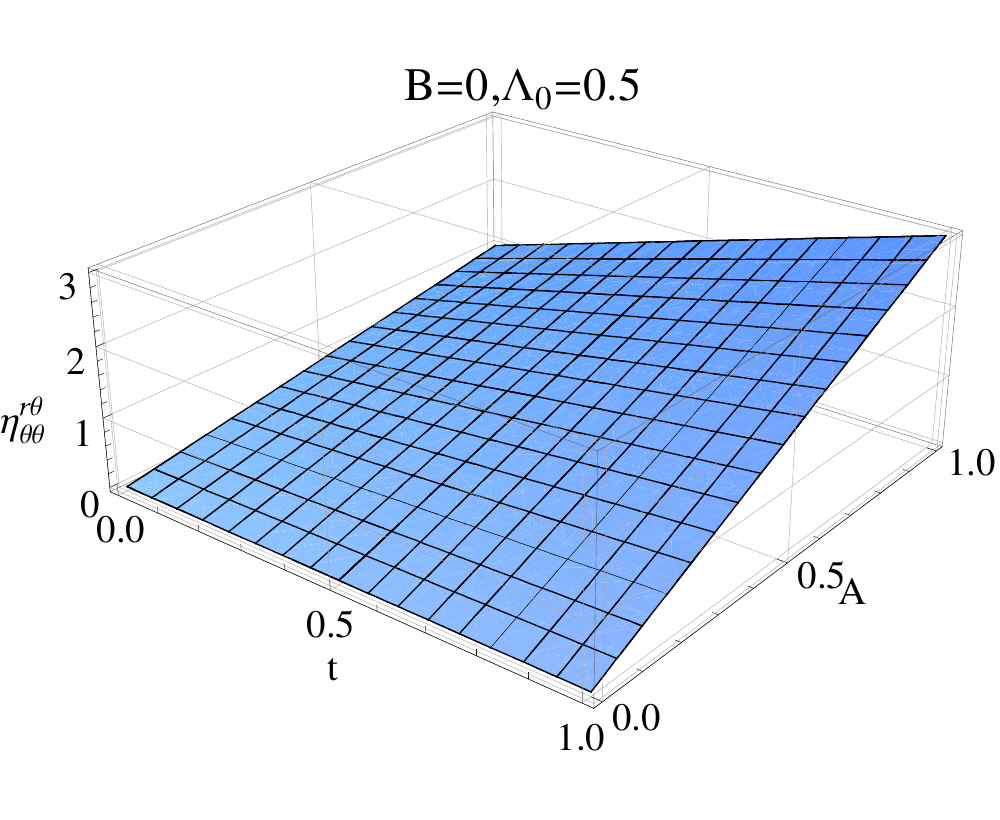,width=70mm}\\
 &  \\ 
\fig{Vis_rttt_1}-a & \fig{Vis_rttt_1}-b \\
 &  \\
%\psfig{file=Pos_Lambda_r_t_t_t_Coeff_B0_01_L0_5.eps ,width=60mm} & 
%\psfig{file=Pos_Lambda_r_t_t_t_Coeff_B0_L0_5.eps,width=60mm}\\
% &  \\ 
%\fig{Vis_rttt_1}-c & \fig{Vis_rttt_1}-d \\
% &  \\
\end{tabular}
\end{center}
\caption{\it The $\eta^{r\theta}_{\theta \theta}$ viscosity coefficient as function of $A$ and $t$ at $\Lambda_{0}=0.5$
 and different values of $B$.}
\label{Vis_rttt_1}
\end{figure}

\begin{figure}[thbp]
\begin{center}
\begin{tabular}{c c}
\psfig{file=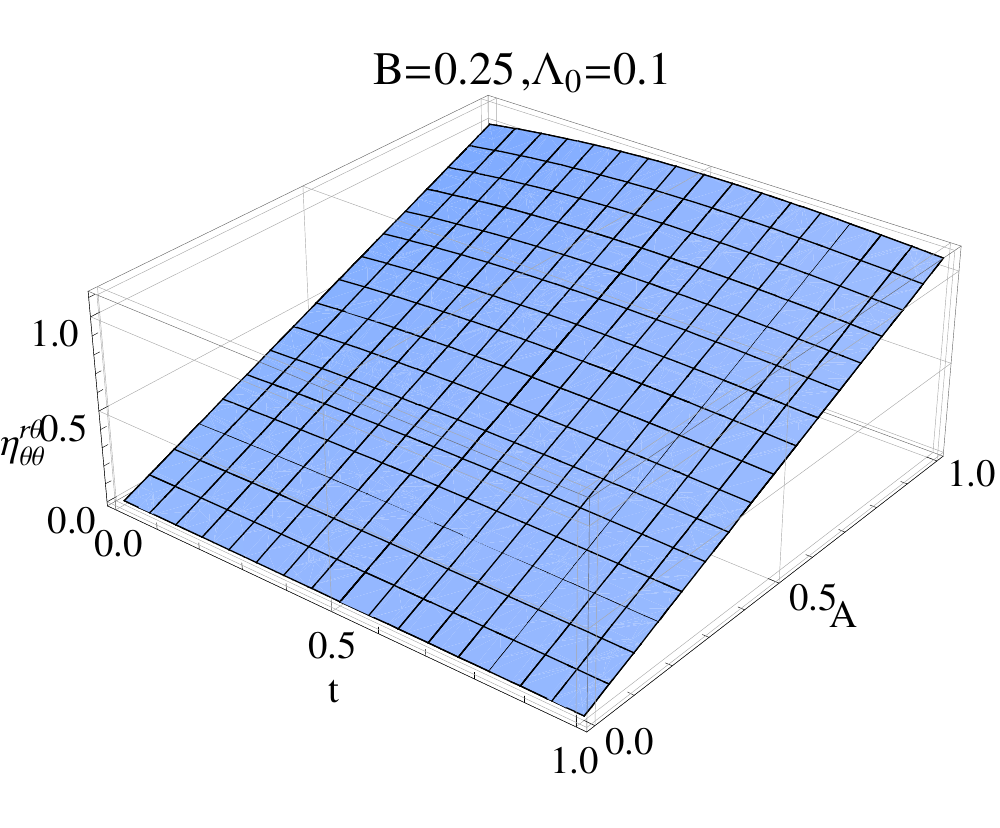 ,width=70mm} & 
\psfig{file=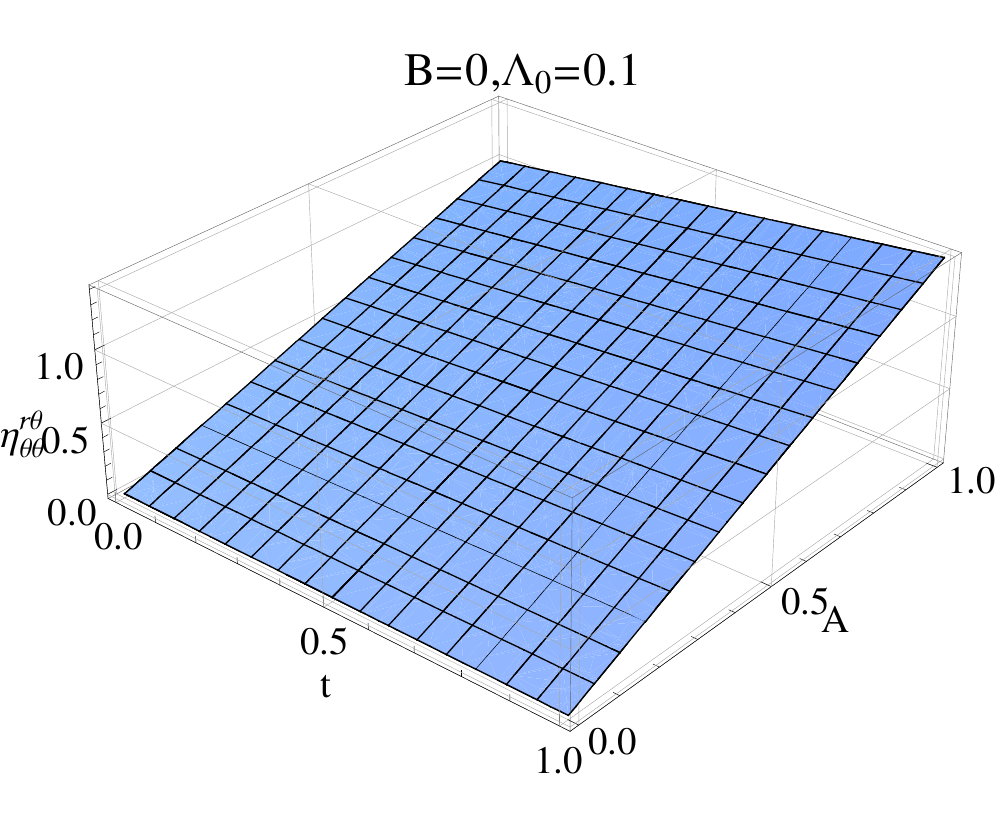,width=70mm}\\
 &  \\ 
\fig{Vis_rttt_2}-a & \fig{Vis_rttt_2}-b \\
 &  \\
%\psfig{file=Pos_Lambda_r_t_t_t_Coeff_B0_01_L0_5.eps ,width=60mm} & 
%\psfig{file=Pos_Lambda_r_t_t_t_Coeff_B0_L0_5.eps,width=60mm}\\
% &  \\ 
%\fig{Vis_rttt_1}-c & \fig{Vis_rttt_1}-d \\
% &  \\
\end{tabular}
\end{center}
\caption{\it The $\eta^{r\theta}_{\theta \theta}$ viscosity coefficient as function of $A$ and $t$ at $\Lambda_{0}=0.1$
 and different values of $B$.}
\label{Vis_rttt_2}
\end{figure}

\begin{figure}[thbp]
\begin{center}
\begin{tabular}{c c}
\psfig{file=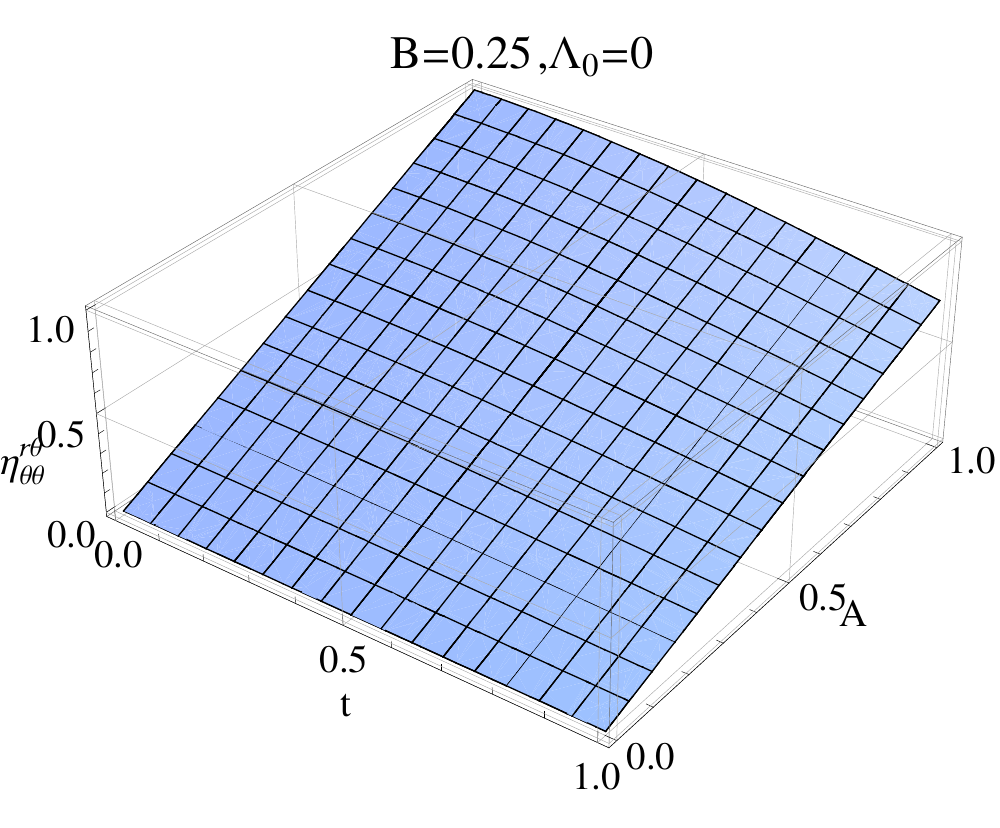 ,width=70mm} & 
\psfig{file=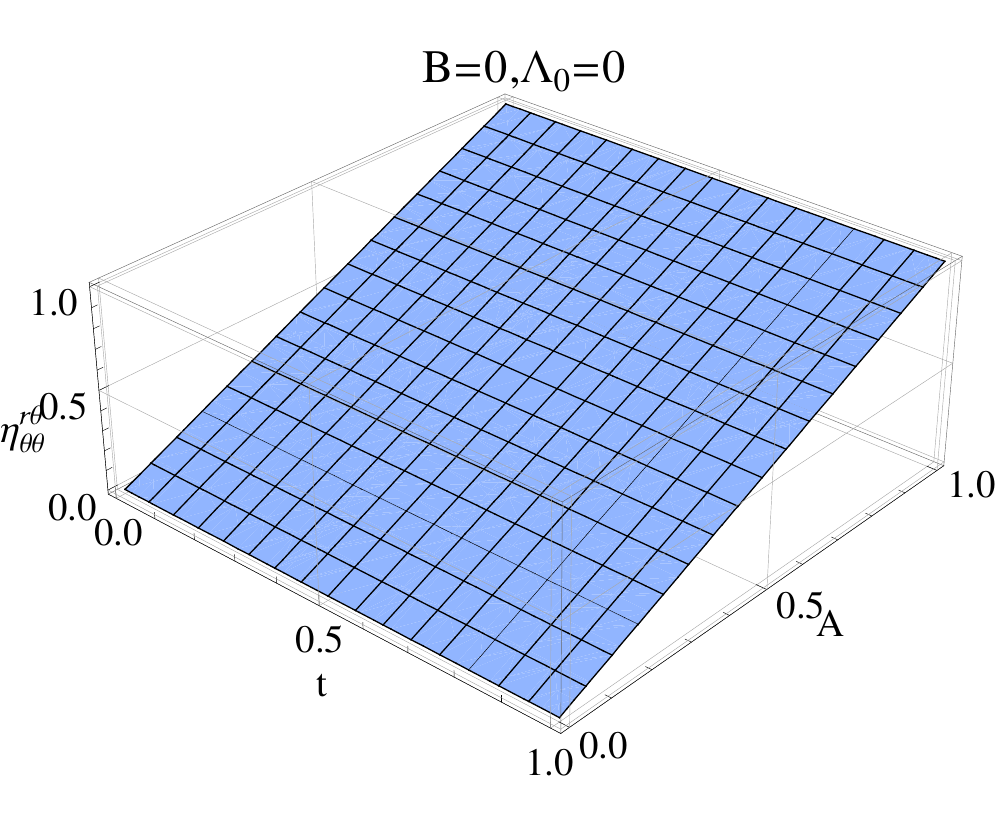,width=70mm}\\
 &  \\ 
\fig{Vis_rttt_3}-a & \fig{Vis_rttt_3}-b \\
 &  \\
%\psfig{file=Pos_Lambda_r_t_t_t_Coeff_B0_01_L0_5.eps ,width=60mm} & 
%\psfig{file=Pos_Lambda_r_t_t_t_Coeff_B0_L0_5.eps,width=60mm}\\
% &  \\ 
%\fig{Vis_rttt_1}-c & \fig{Vis_rttt_1}-d \\
% &  \\
\end{tabular}
\end{center}
\caption{\it The $\eta^{r\theta}_{\theta \theta}$ viscosity coefficient as function of $A$ and $t$ at $\Lambda_{0}=0$
 and different values of $B$.}
\label{Vis_rttt_3}
\end{figure}

 The first viscosity coefficient \eq{VC35} we plot as function of parameters $B$ and $t$, see \fig{Vis_rrrr}. 
Due \eq{VC21} restrictions, the parameter $B$ is varied from $0$ till $0.25$, whereas $0\,<\,t\,<\,1$. 
This coefficient does not depend on the $\Lambda_{0}$ parameter, so this plot is valid for any value of $\Lambda_{0}$.
Plots of the $\eta^{r\theta}_{\theta \theta}$ viscosity coefficient are presented in 
\fig{Vis_rttt_1}-\fig{Vis_rttt_3} at different values of $B$ and $\Lambda_{0}$ as functions of $A$ and $t$ as well.

\begin{figure}[thbp]
\begin{center}
\begin{tabular}{c c}
\psfig{file=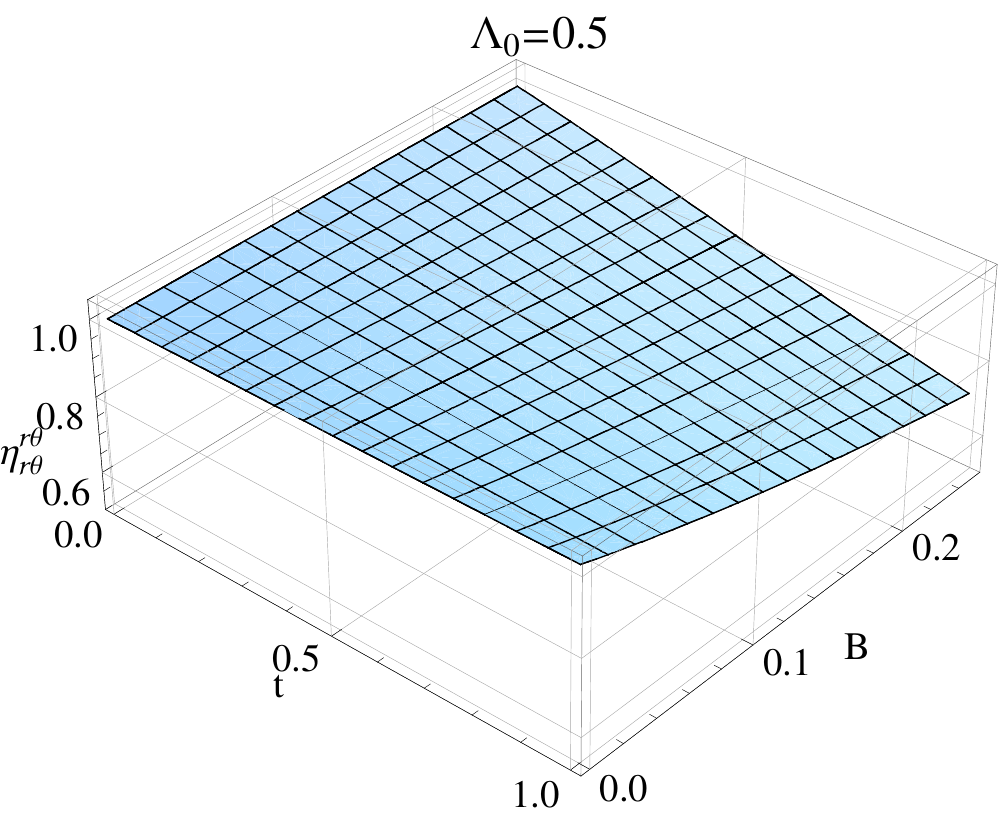 ,width=80mm} & 
\psfig{file=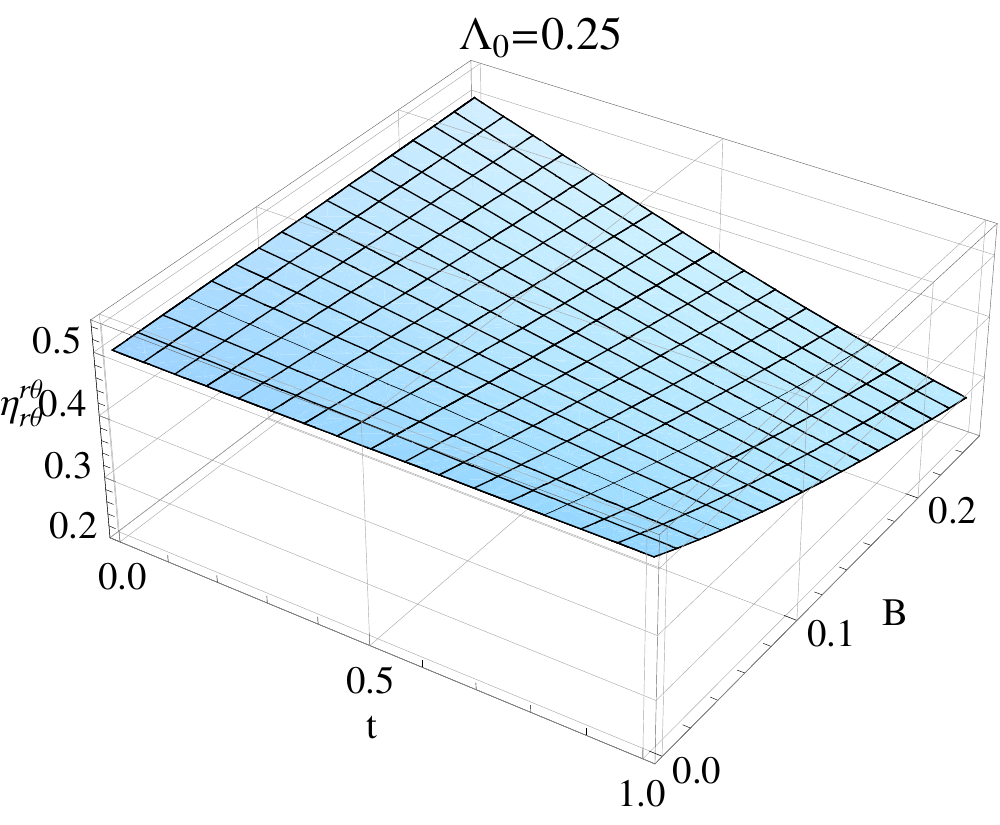,width=80mm}\\
 &  \\ 
\fig{Vis_rtrt}-a & \fig{Vis_rtrt}-b \\
 &  \\
\psfig{file=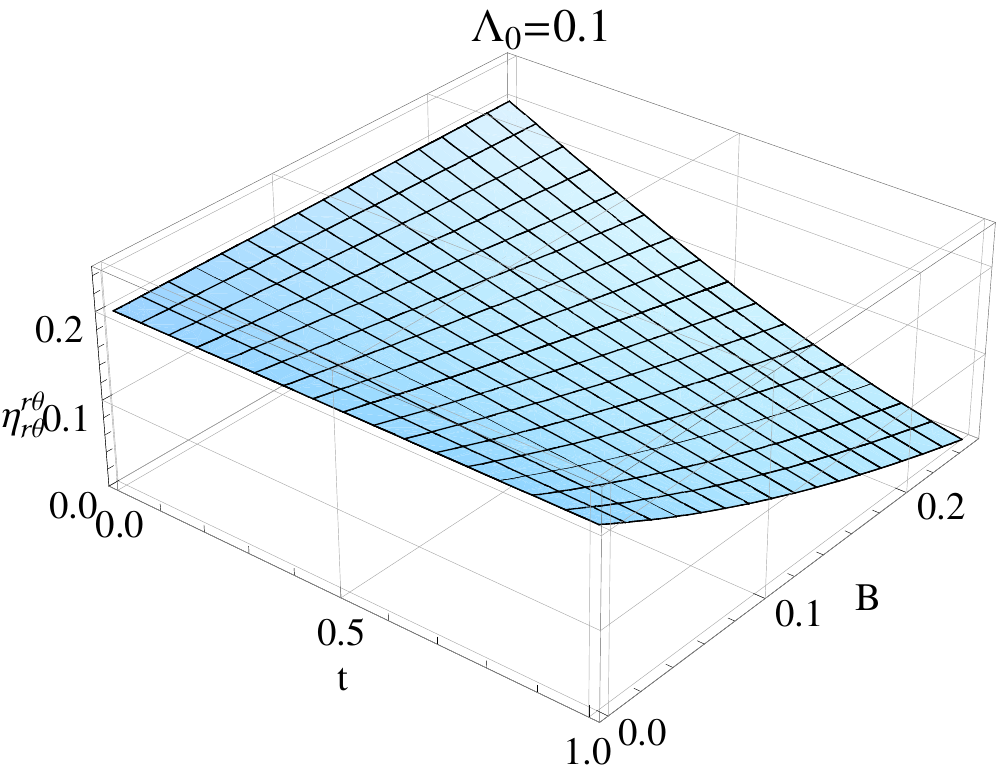 ,width=80mm} & 
\psfig{file=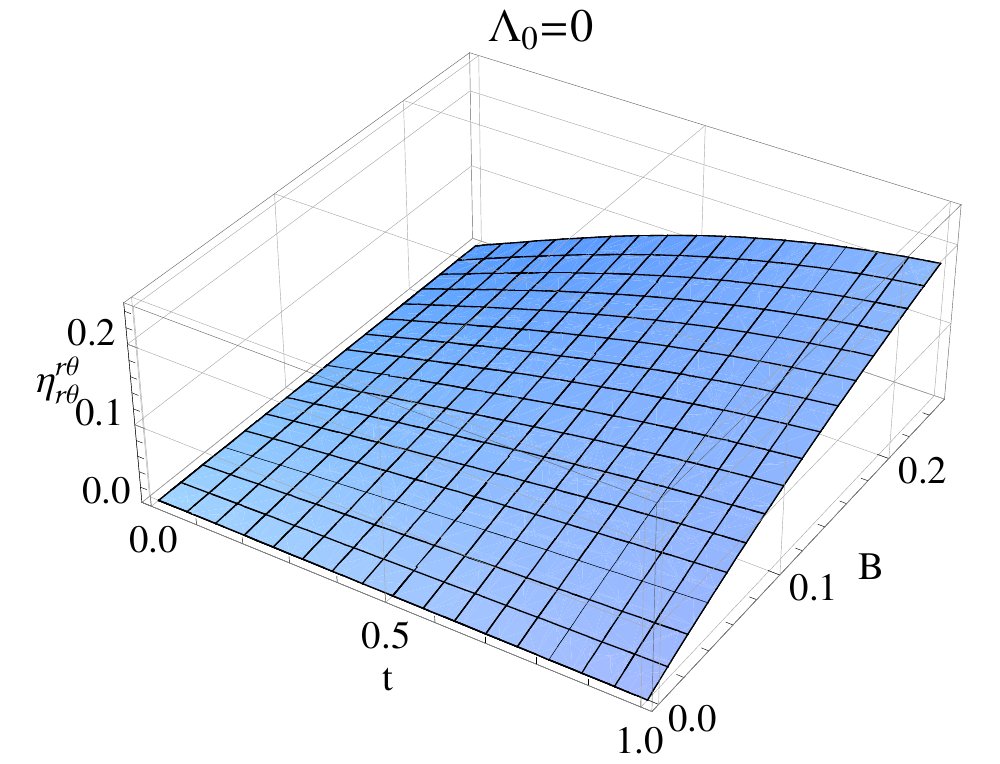,width=80mm}\\
 &  \\ 
\fig{Vis_rtrt}-c & \fig{Vis_rtrt}-d \\
 &  \\
\end{tabular}
\end{center}
\caption{\it The $\eta^{r\theta}_{r \theta}$ viscosity coefficient as function of $B$ and $t$ at 
different values of $\Lambda_{0}$.}
\label{Vis_rtrt}
\end{figure}

\begin{figure}[thbp]
\begin{center}
\begin{tabular}{c c}
\psfig{file=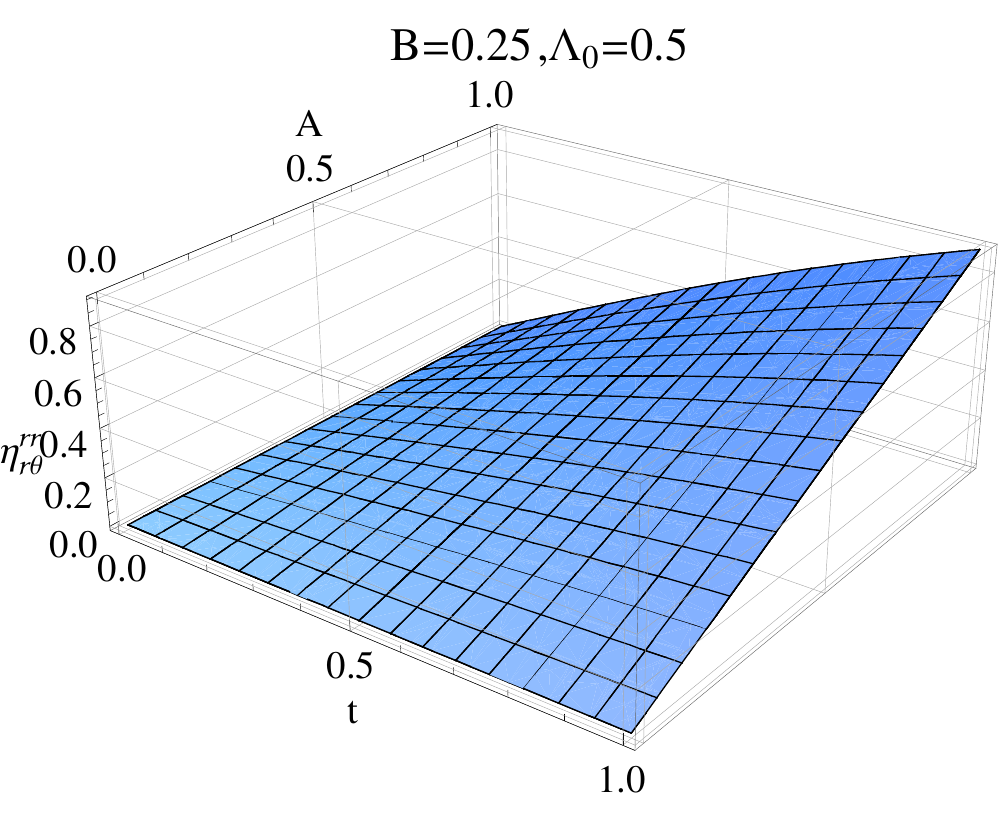 ,width=70mm} & 
\psfig{file=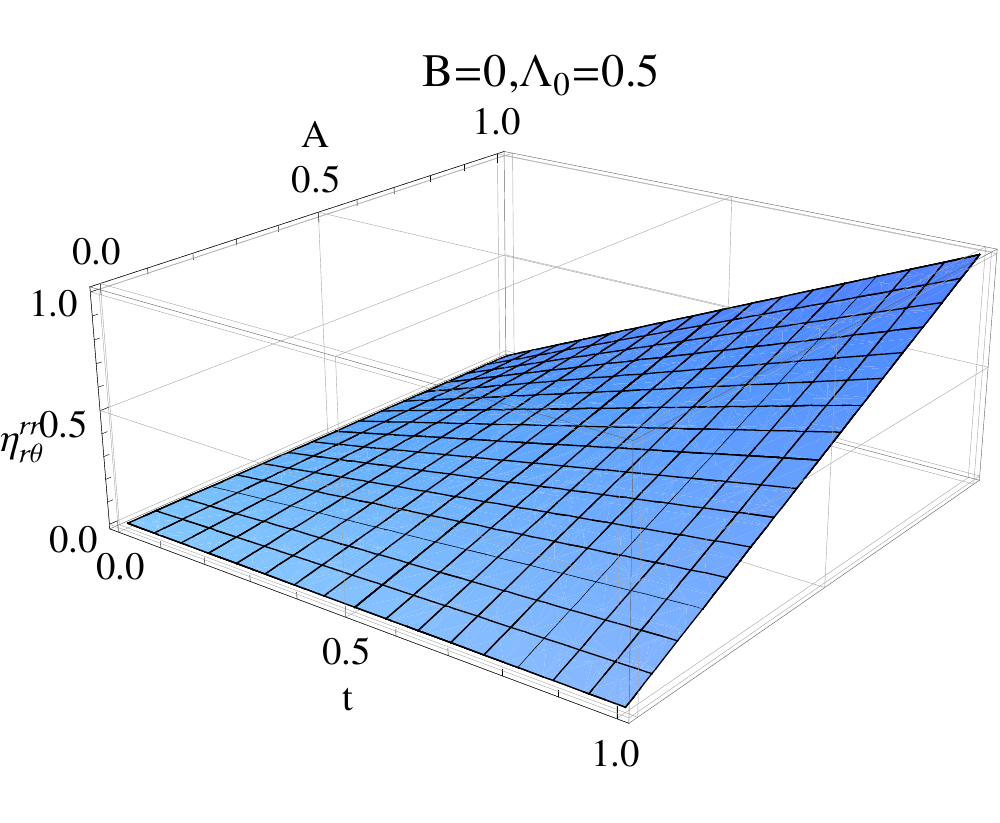,width=70mm}\\
 &  \\ 
\fig{Vis_rrrt_1}-a & \fig{Vis_rrrt_1}-b \\
 &  \\
%\psfig{file=Pos_Lambda_r_t_t_t_Coeff_B0_01_L0_5.eps ,width=60mm} & 
%\psfig{file=Pos_Lambda_r_t_t_t_Coeff_B0_L0_5.eps,width=60mm}\\
% &  \\ 
%\fig{Vis_rttt_1}-c & \fig{Vis_rttt_1}-d \\
% &  \\
\end{tabular}
\end{center}
\caption{\it The $\eta^{r r}_{r \theta}$ viscosity coefficient as function of $A$ and $t$ at $\Lambda_{0}=0.5$
 and different values of $B$.}
\label{Vis_rrrt_1}
\end{figure}

\begin{figure}[thbp]
\begin{center}
\begin{tabular}{c c}
\psfig{file=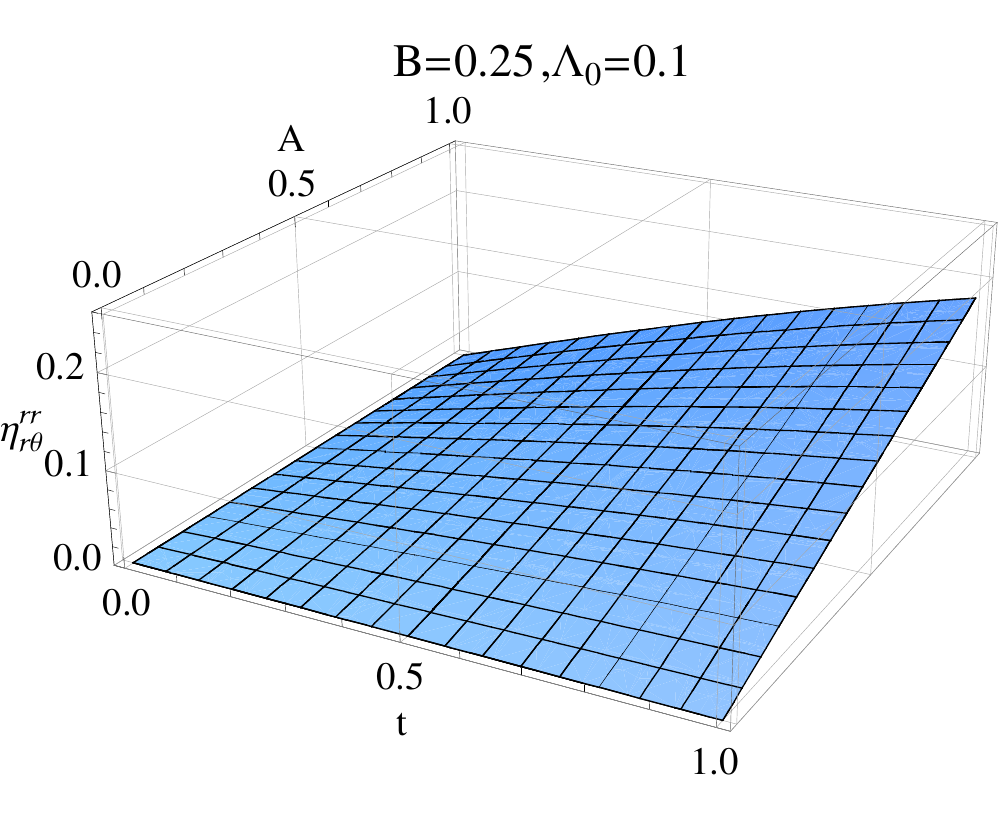 ,width=70mm} & 
\psfig{file=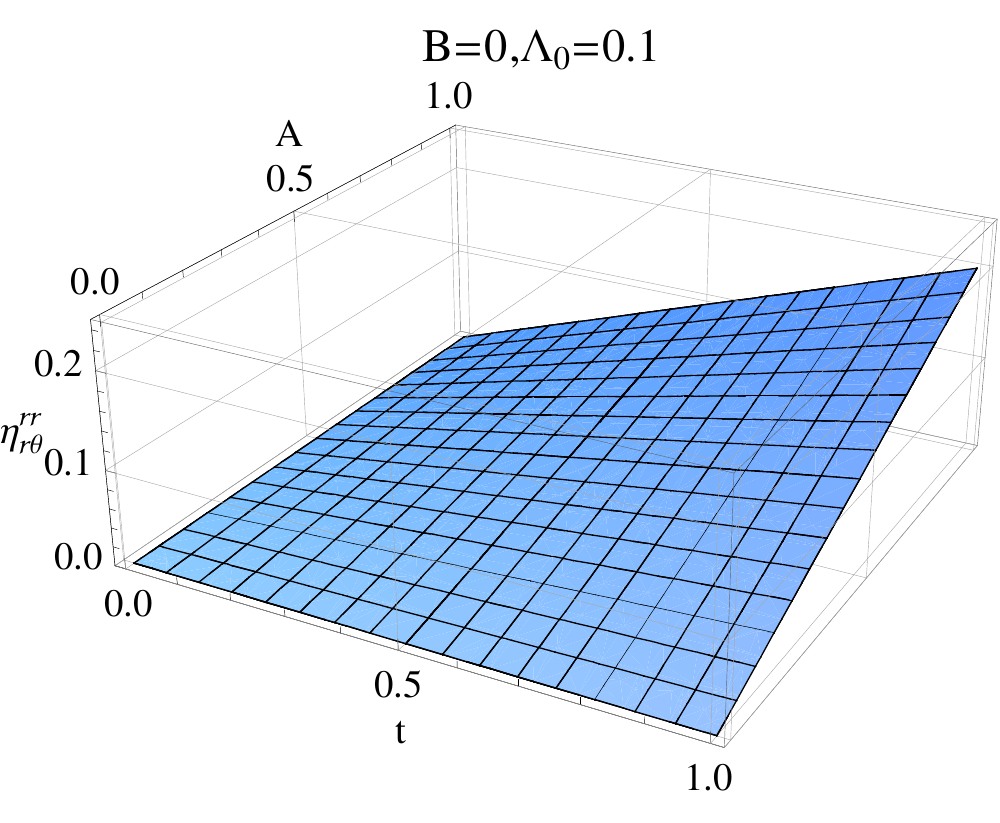,width=70mm}\\
 &  \\ 
\fig{Vis_rrrt_2}-a & \fig{Vis_rrrt_2}-b \\
 &  \\
%\psfig{file=Pos_Lambda_r_t_t_t_Coeff_B0_01_L0_5.eps ,width=60mm} & 
%\psfig{file=Pos_Lambda_r_t_t_t_Coeff_B0_L0_5.eps,width=60mm}\\
% &  \\ 
%\fig{Vis_rttt_1}-c & \fig{Vis_rttt_1}-d \\
% &  \\
\end{tabular}
\end{center}
\caption{\it The $\eta^{r r}_{r \theta}$ viscosity coefficient as function of $A$ and $t$ at $\Lambda_{0}=0.1$
 and different values of $B$.}
\label{Vis_rrrt_2}
\end{figure}

\begin{figure}[thbp]
\begin{center}
\begin{tabular}{c c}
\psfig{file=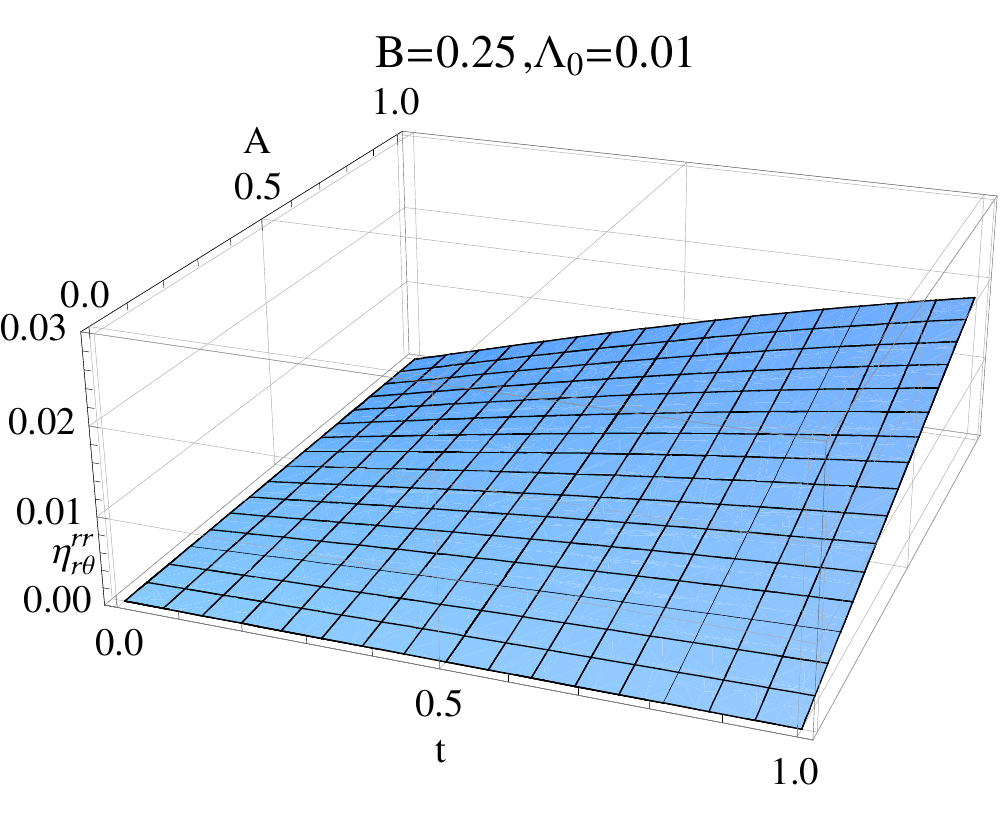 ,width=70mm} & 
\psfig{file=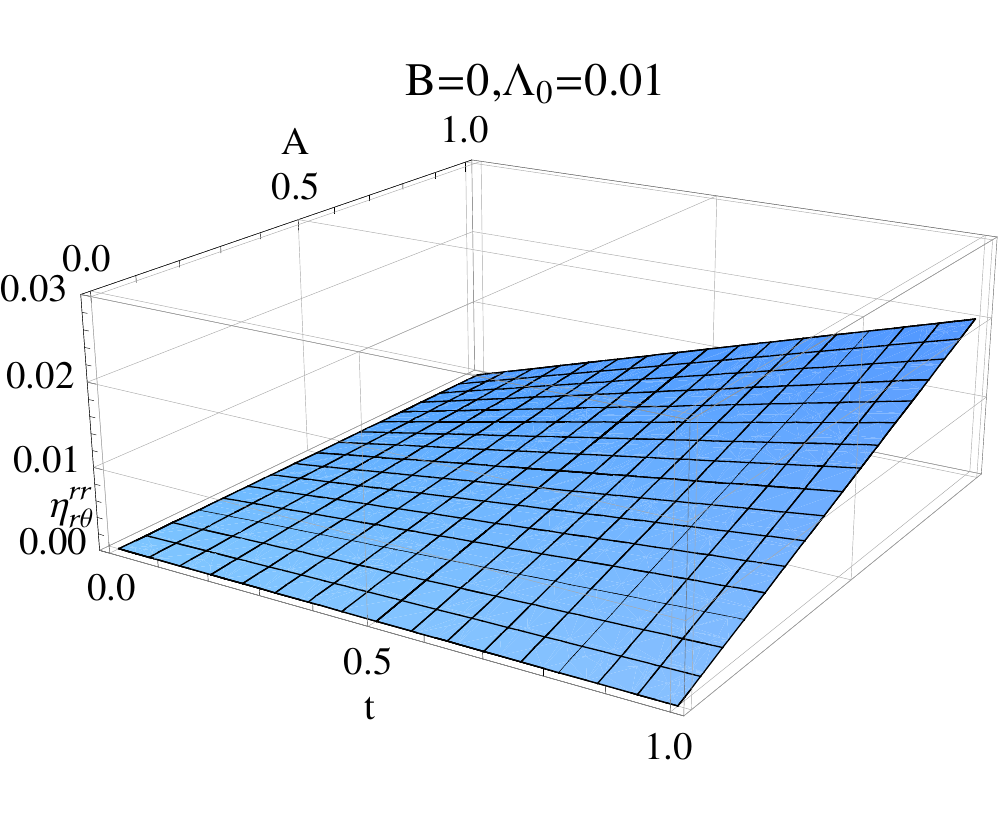,width=70mm}\\
 &  \\ 
\fig{Vis_rrrt_3}-a & \fig{Vis_rrrt_3}-b \\
 &  \\
%\psfig{file=Pos_Lambda_r_t_t_t_Coeff_B0_01_L0_5.eps ,width=60mm} & 
%\psfig{file=Pos_Lambda_r_t_t_t_Coeff_B0_L0_5.eps,width=60mm}\\
% &  \\ 
%\fig{Vis_rttt_1}-c & \fig{Vis_rttt_1}-d \\
% &  \\
\end{tabular}
\end{center}
\caption{\it The $\eta^{r r}_{r \theta}$ viscosity coefficient as function of $A$ and $t$ at $\Lambda_{0}=0$
 and different values of $B$.}
\label{Vis_rrrt_3}
\end{figure}

 The viscosity coefficient $\eta^{r\theta}_{r \theta}$ does not depend on $A$ and plots of this coefficient are presented as functions of $B$ at different values of $\Lambda_{0}$ in \fig{Vis_rtrt}. For the $\Lambda_{0}\,=\,0$ case the positive value of the coefficient is taken. Parameters $\Lambda_{0}$ and $B$ are combined it such way that in the given order of the approximation the expression in the brackets in \eq{VC31}-\eq{VC32} remains positive for any $t$.
Values of the last viscosity coefficient $\eta_{r \theta}^{r r}$ are presented in  \fig{Vis_rrrt_1}-\fig{Vis_rrrt_3}
as functions of $A$ and $t$ at different values of $B$ and $\Lambda_{0}$.

\subsection{$\Lambda_{0}\,<\,0\,$ case }

\begin{figure}[thbp]
\begin{center}
\begin{tabular}{c c}
\psfig{file=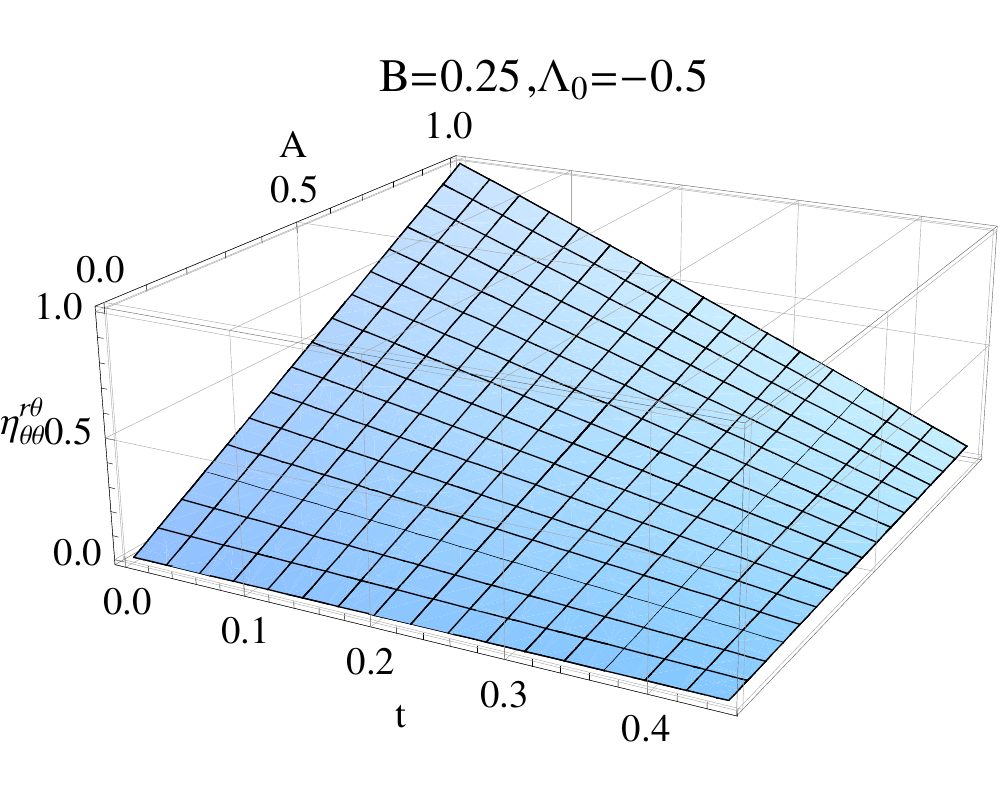 ,width=70mm} & 
\psfig{file=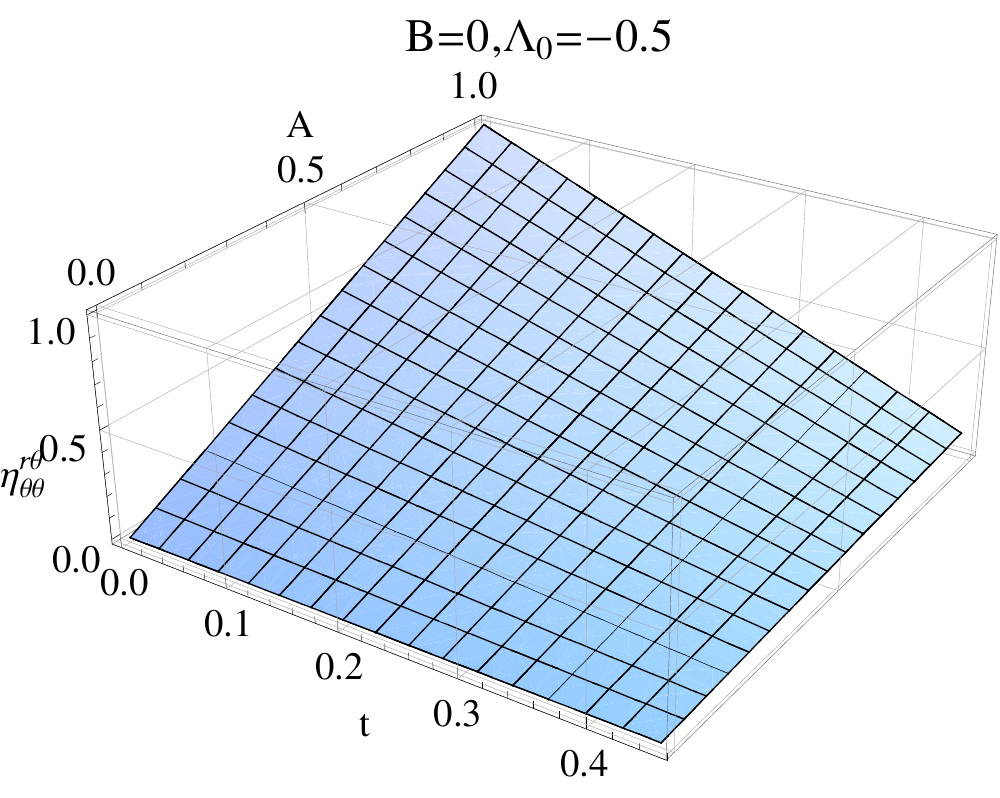,width=70mm}\\
 &  \\ 
\fig{Neg_Vis_rttt_1}-a & \fig{Neg_Vis_rttt_1}-b \\
 &  \\
%\psfig{file=Pos_Lambda_r_t_t_t_Coeff_B0_01_L0_5.eps ,width=60mm} & 
%\psfig{file=Pos_Lambda_r_t_t_t_Coeff_B0_L0_5.eps,width=60mm}\\
% &  \\ 
%\fig{Vis_rttt_1}-c & \fig{Vis_rttt_1}-d \\
% &  \\
\end{tabular}
\end{center}
\caption{\it The $\eta^{r \theta}_{\theta \theta}$ viscosity coefficient as function of $A$ and $t$ at $\Lambda_{0}=-0.5$
 and different values of $B$.}
\label{Neg_Vis_rttt_1}
\end{figure}
\begin{figure}[thbp]
\begin{center}
\begin{tabular}{c c}
\psfig{file=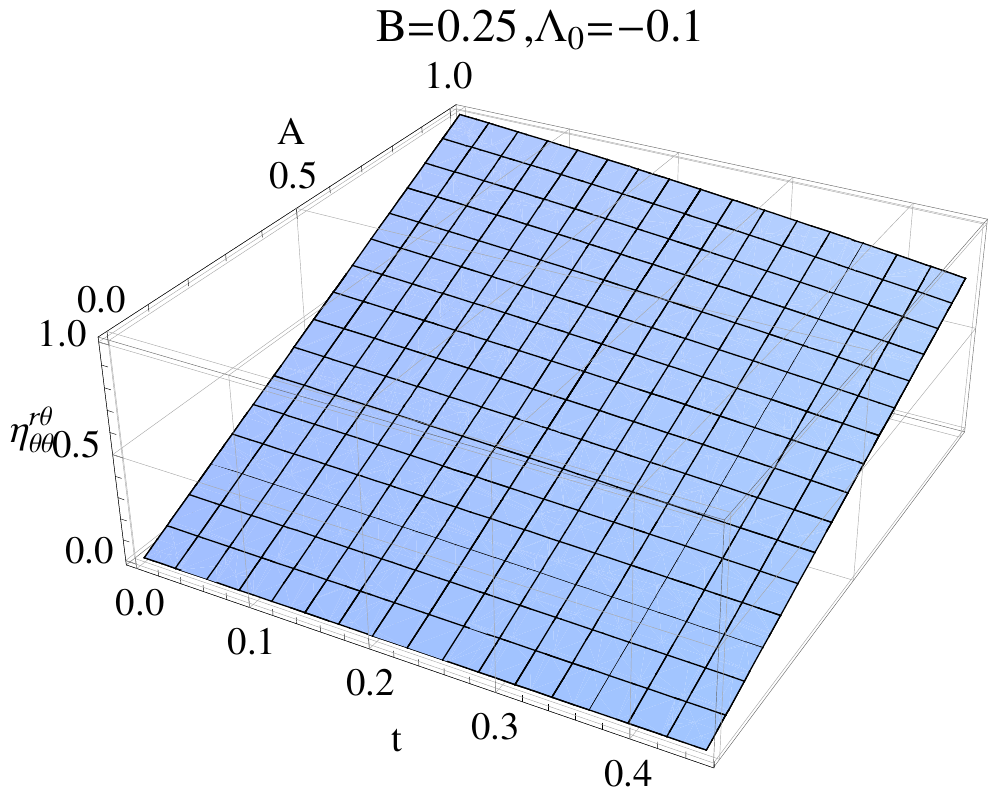 ,width=70mm} & 
\psfig{file=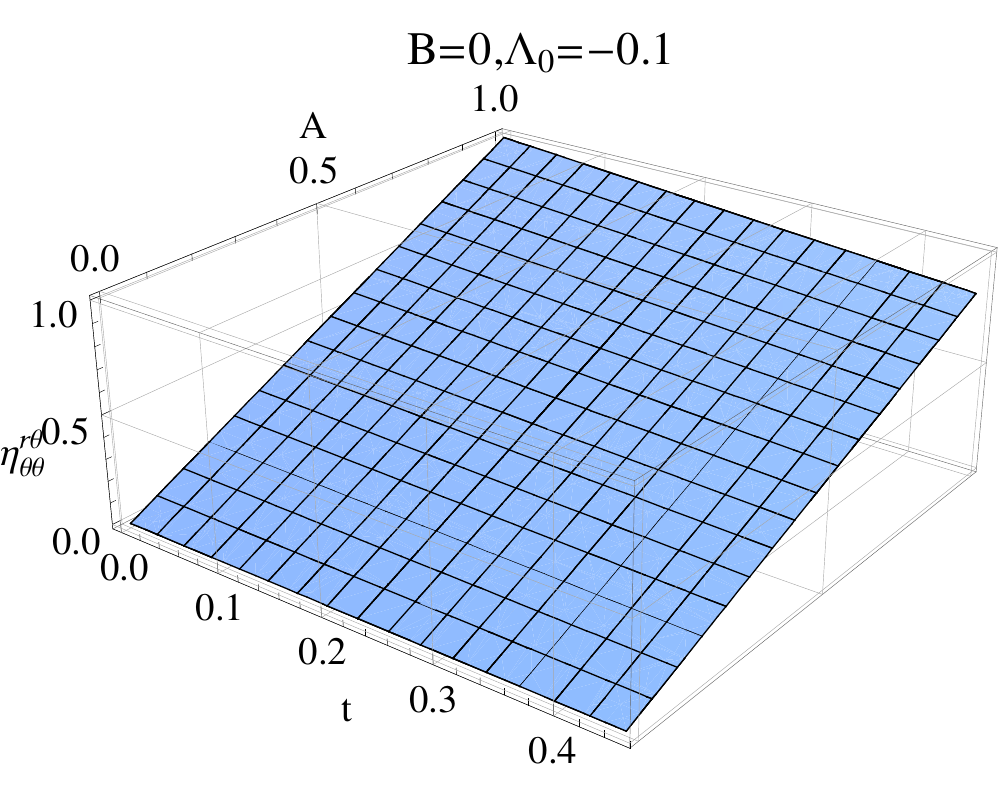,width=70mm}\\
 &  \\ 
\fig{Neg_Vis_rttt_2}-a & \fig{Neg_Vis_rttt_2}-b \\
 &  \\
%\psfig{file=Pos_Lambda_r_t_t_t_Coeff_B0_01_L0_5.eps ,width=60mm} & 
%\psfig{file=Pos_Lambda_r_t_t_t_Coeff_B0_L0_5.eps,width=60mm}\\
% &  \\ 
%\fig{Vis_rttt_1}-c & \fig{Vis_rttt_1}-d \\
% &  \\
\end{tabular}
\end{center}
\caption{\it The $\eta^{r \theta}_{\theta \theta}$ viscosity coefficient as function of $A$ and $t$ at $\Lambda_{0}=-0.1$
 and different values of $B$.}
\label{Neg_Vis_rttt_2}
\end{figure}
\begin{figure}[thbp]
\begin{center}
\begin{tabular}{c c}
\psfig{file=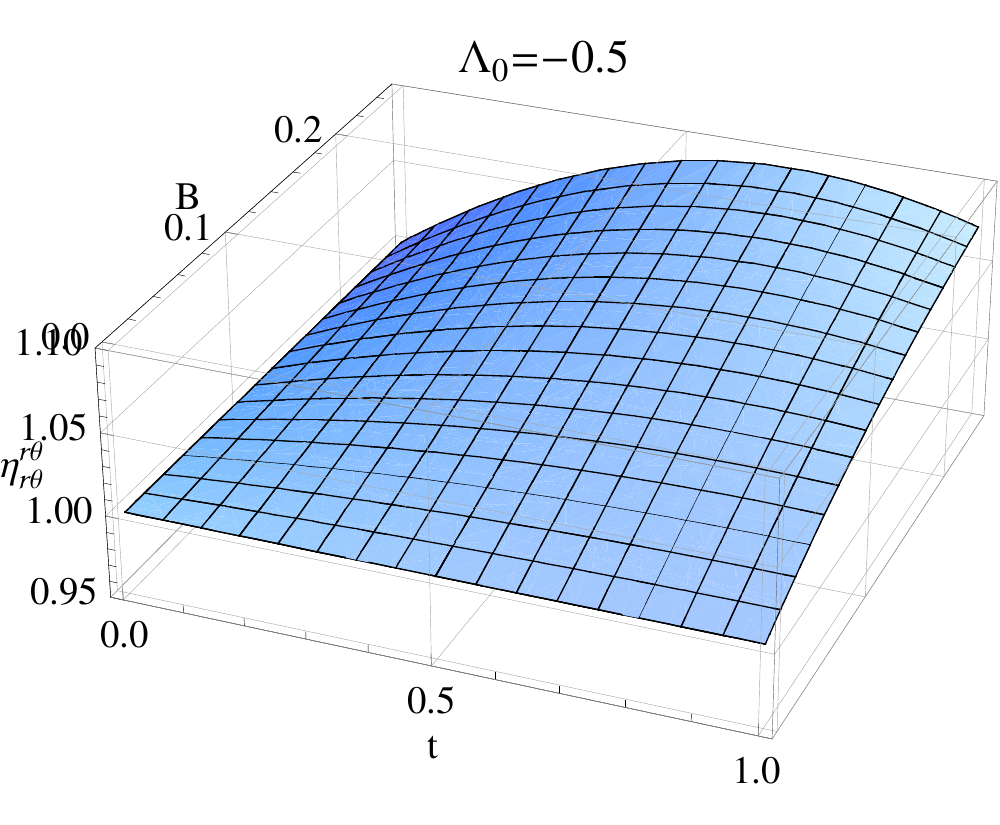 ,width=80mm} & 
\psfig{file=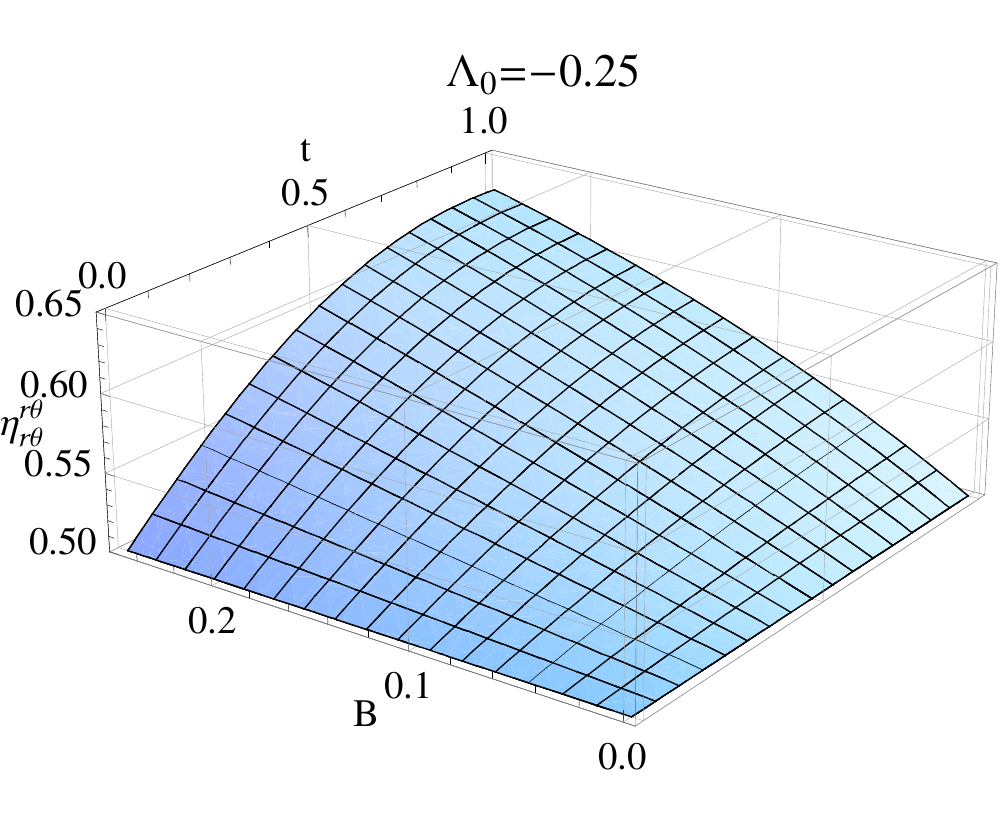,width=80mm}\\
 &  \\ 
\fig{Vis_rtrt}-a & \fig{Vis_rtrt}-b \\
 &  \\
\psfig{file=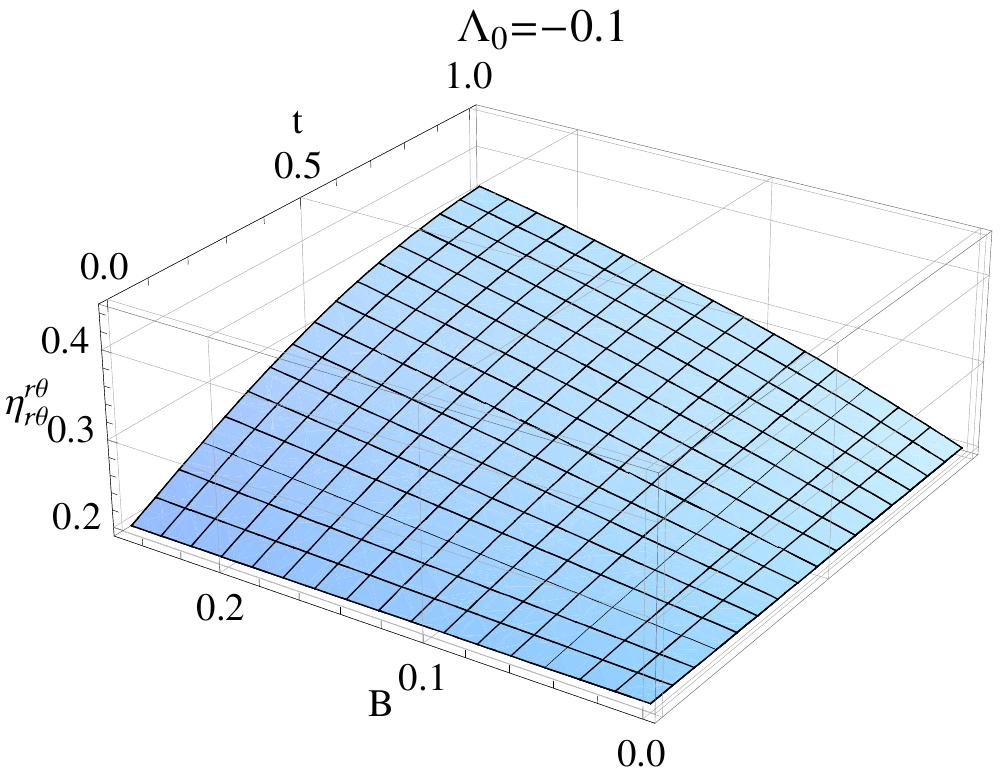 ,width=80mm} & 
\psfig{file=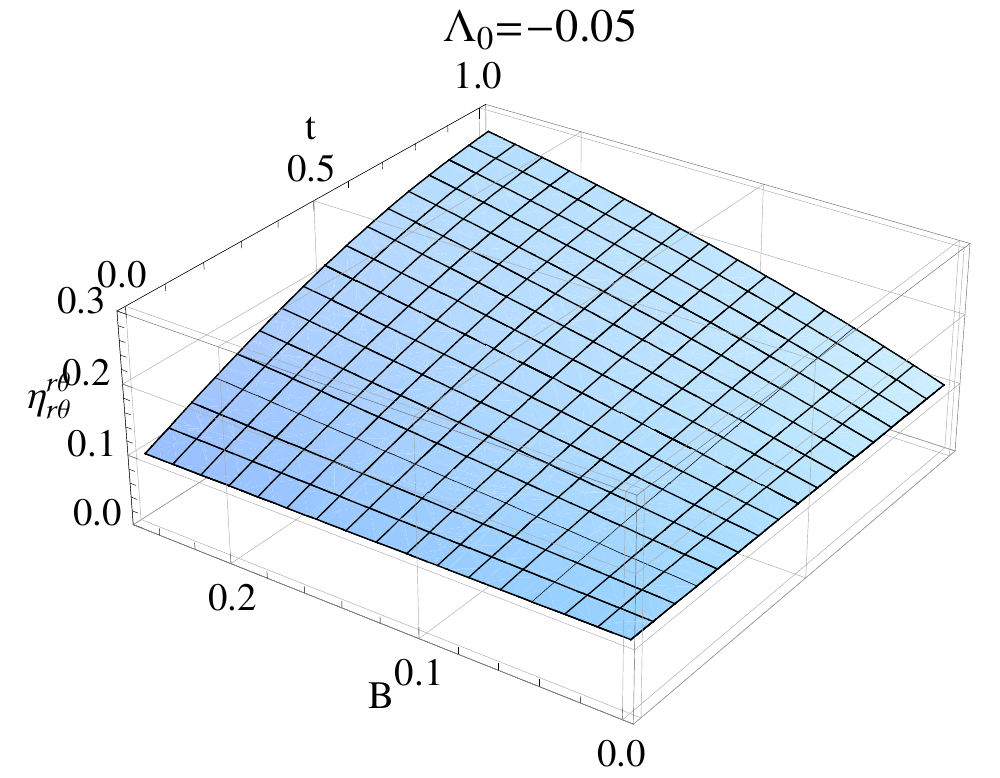,width=80mm}\\
 &  \\ 
\fig{Neg_Vis_rtrt}-c & \fig{Neg_Vis_rtrt}-d \\
 &  \\
\end{tabular}
\end{center}
\caption{\it The $\eta^{r\theta}_{r \theta}$ viscosity coefficient as function of $B$ and $t$ at 
different negative values of $\Lambda_{0}$.}
\label{Neg_Vis_rtrt}
\end{figure}
At the case of $\Lambda_{0}\,<\,0\,$ the only coefficients \eq{VC33}-\eq{VC31}  are changed. 
Due the negative sign of $\Lambda_{0}$ in \eq{VC33}, time evolution of this coefficient is limited by smaller value
of $t$, namely for this coefficient we have
$0\,<\,t\,<\,0.45$ approximately. The plots which represent values of $\eta_{r \theta}^{\theta \theta}$ are given by  \fig{Neg_Vis_rttt_1}-\fig{Neg_Vis_rttt_2}. Finally, plots of the $\eta_{r \theta}^{r \theta}$   viscosity coefficient are presented in \fig{Neg_Vis_rtrt} .

\section{Temperature dependence of the viscosity coefficients and quark-gluon plasma phenomenology}

 In order to rewrite transport coefficients \eq{VC1}-\eq{VC13} as functions of temperature, we introduce an effective temperature of the expanding fluid element as
\beq\label{Tt1}
\langle\, k\,T(t)\,\rangle_{s}^{0}\,=\,
\frac{\int\,d^{3}p\,\Le H_{\bot}+H_{\|}\Ra\,f_{s0}(\vec{r},\vec{v},t)}
{\int\,d^{3}p\,f_{s0}(\vec{r},\vec{v},t)}\,=\,
\langle\, k \,T_{\bot}(t)\,\rangle_{s}^{0}+\langle \,k\, T_{\|}(t)\,\rangle_{s}^{0}
\eeq 
with $H$ from \eq{DFunc41}:
\beq\label{Tt2}
H_{\bot}\,+\,H_{\|}\,=\,m\,c^{2}\,\frac{\gamma}{\beta}\,\frac{1}{2}\,\Le\,p_{r}^{2}\,+\,p_{\theta}^{2}\,\Ra\,+\,
\,m\,c^{2}\,p_{z}.
\eeq
First of all, consider the $\langle\, k \,T_{\bot}(t)\,\rangle_{s}^{0}$ value. We interesting in calculation of the temperature change, i.e. we will calculate
\beq\label{Tt3}
\Delta\,\langle\,k\,T_{\bot}(t)\,\rangle\,=\,k\,\Delta\,\langle\,T_{\bot}(t)\,\rangle\,=\,
\langle\,k\,T_{\bot}(0)\,\rangle_{s}^{0}\,-\,
\langle\,k\,T_{\bot}(t)\,\rangle_{s}^{0}\,,
\eeq
therefore we will calculate \eq{Tt1} taking $t\,=\,0$ limit in the end of the calculations.
We have for the \eq{Tt1} integral:
\beq\label{Tt4}
\frac{\langle\,k\,T_{\bot}(t)\,\rangle_{s}^{0}\,}{m\,c^2}\,=\,N^{-1}\,\frac{\gamma}{2\,\beta}\,
\int\,d^{3}p\,\Le\,p_{r}^{2}\,+\,p_{\theta}^{2} \,\Ra\,f_{s0}(\vec{r},\,\vec{v},t)\,.
\eeq
Using results of \eq{RR5} and \eq{TT3} we obtain:
\beqar\label{Tt5}
\frac{\langle\,k\,T_{\bot}(t)\,\rangle_{s}^{0}\,}{m\,c^2}\,& = &\,
\frac{1}{2}\,
\Le
\frac{k T_{\bot}}{m c^2}-\frac{\psi(r)}{m c^2}\Ra
\Le 1 + \frac{1 + l_{0}^{2}\omega_{r}^{2}t^{2}/c^{2}}{ 1 + t^{2} \alpha/\rho}\Ra\,+\,
\frac{\gamma}{2\beta}\,
\Le\langle\,v_{r}\,\rangle_{s}^{0}\Ra^{2}\Le 1 + t^{2}\frac{l_{0}^{2}\omega_{r}^{2}}{c^{2}}\Ra\,-\,\nonumber\\
& - & 2\,r\,t\,\frac{\gamma}{2\beta}\,\frac{l_{0}^{2}\omega_{r}^{2}}{c^{2}}\,\langle\,v_{r}\,\rangle_{s}^{0}\,+\,
r^{2}\,\frac{\gamma}{2\beta}\,\frac{l_{0}^{2}\omega_{r}^{2}}{c^{2}}\,.
\eeqar 
In terms of \eq{VC2}-\eq{VC21} definitions we rewrite \eq{Tt5} as :
\beqar\label{Tt6}
\frac{\langle\,k\,T_{\bot}(t)\,\rangle_{s}^{0}\,}{m\,c^2}\,& = &\,
\frac{1}{2}\,
\Le
\frac{k T_{\bot}}{m c^2}-\frac{\gamma r^{2}}{2\beta}\,B\,\Le 1 - \frac{t^{2} B}{1 + t^{2} B }   \Ra\Ra
\Le 1 + \frac{1 + t^{2} A^{2}}{ 1 + t^{2} B}\Ra\,+\,
\frac{\gamma}{2\beta}\,
\Le\langle\,v_{r}\,\rangle_{s}^{0}\Ra^{2}\Le 1 + t^{2} A^{2}\Ra\,-\,\nonumber\\
& - & 2\,r\,t\,\frac{\gamma}{2\beta}\,A^{2}\,\langle\,v_{r}\,\rangle_{s}^{0}\,+\,
r^{2}\,\frac{\gamma}{2\beta}\,A^{2}\,.
\eeqar 
Therefore we obtain:
\beq\label{Tt7}
\frac{\langle\,k\,T_{\bot}(0)\,\rangle_{s}^{0}\,}{m\,c^2}\, = \,
\Le
\frac{k T_{\bot}}{m c^2}-\frac{\gamma r^{2}}{2\beta}\,B\,\Ra\,+
r^{2}\,\frac{\gamma}{2\beta}\,A^{2}\,,
\eeq 
where only linear on $\Lambda_{0}$ terms were preserved in the answer.
Inserting obtained expressions in \eq{Tt3} we obtain:
\beqar\label{Tt8}
\frac{\Delta\,\langle\,k\,T_{\bot}(t)\,\rangle}{m c^{2}}\,& = &\,\frac{1}{2}
\frac{k T_{\bot}}{m c^2}\frac{t^{2}\Le B-A^2\Ra}{1 + t^{2} B}\,-\,
\frac{\gamma r^{2}}{4\beta}\,B\,t^{2}\,\Le 
\frac{2 B}{ 1 + t^{2} B}\,+\,\frac{B - A^{2}}{ \Le 1 + t^{2} B\Ra^{2}}
\Ra\,-\,\nonumber\\
&-&
\frac{\gamma}{2\beta}\,\Le \Le\langle\,v_{r}\,\rangle_{s}^{01}\Ra^{2}\,+\,2\,
\langle\,v_{r}\,\rangle_{s}^{01}\,\langle\,v_{r}\,\rangle_{s}^{02}\,\Ra\,
\Le 1 + t^{2} A^{2}\Ra\,+\,
2\,r\,t\,\frac{\gamma}{2\beta}\,A^{2}\,\langle\,v_{r}\,\rangle_{s}^{01}\,.
\eeqar
We can simplify expression \eq{Tt8} considering only leading on $\gamma$ parameter terms:
\beq\label{Tt9}
\frac{\Delta\,\langle\,k\,T_{\bot}(t)\,\rangle}{m c^{2}}\,\approx \,\frac{1}{2}
\frac{k T_{\bot}}{m c^2}\frac{t^{2}\Le B-A^2\Ra}{1 + t^{2} B}\,
\eeq
or
\beq\label{Tt10}
\Delta_{\bot}\,=\,\frac{\Delta\,\langle\,T_{\bot}(t)\,\rangle}{T_{\bot}}\,\approx \,\frac{1}{2}
\frac{t^{2}\Le B-A^2\Ra}{1 + t^{2} B}\,,
\eeq
which is valid in the case of ultra-relativistic expansion of the hot spot of interest.

 Second contribution into \eq{Tt1} integral is given by the following expression:
\beq\label{Tt11} 
\frac{\langle\,k\,T_{\|}(t)\,\rangle_{s}^{0}\,}{m\,c^2}\,=\,N^{-1}\,\frac{\gamma}{2\,\beta}\,
\int\,d^{3}p\,p_{z}^{2}\,f_{s0}(\vec{r},\,\vec{v},t)\,.
\eeq
Using \eq{ZVel4} result we obtain:
\beq\label{Tt12}
\frac{\langle\,k\,T_{\|}(t)\,\rangle_{s}^{0}\,}{m\,c^2}\,=\,
\frac{\beta}{\gamma}\,+\,\Lambda_{0}\,\frac{\alpha}{\rho}\frac{r\,t\,}{\Le 1\,+\,t^2\,\alpha/\rho\Ra}\,
\eeq
and similarly to \eq{Tt9} we can define:
\beq\label{Tt13}
\frac{\Delta\,\langle\,k\,T_{\|}(t)\,\rangle}{m c^{2}}\,=\,-\,
\Lambda_{0}\,\frac{\alpha}{\rho}\frac{t\,r}{\Le 1\,+\,t^2\,\alpha/\rho\Ra}\,=\,-\,
\Lambda_{0}\,\frac{t\,B}{ 1\,+\,t^2\,B}\,,
\eeq
where we take $r\,=\,1$ as in the previous calculations.
Rewriting \eq{Tt13} as 
\beq\label{Tt14}
\Delta_{\|}\,=\,\frac{\Delta\,\langle\,T_{\|}(t)\,\rangle}{T_{\bot}}\,
=\,-\,
\Lambda_{0}\,\frac{m c^{2}}{k\,T_{\bot}}\,
\frac{\alpha}{\rho}\frac{t\,r}{\Le 1\,+\,t^2\,\alpha/\rho\Ra}\,=\,-\,
\Lambda_{0}\,\frac{m c^{2}}{k\,T_{\bot}}\,\frac{t\,B}{ 1\,+\,t^2\,B}\,,
\eeq
we obtain finally
\beq\label{Tt15}
\Delta\,=\,\Delta_{\bot}\,+\,\Delta_{\|}\,=\,
\frac{1}{2}
\frac{t^{2}\Le B\,-\,A^2\Ra}{1 +\, t^{2}\, B}\,-\,
\Lambda_{0}\,\frac{t\,B\,D\,}{ 1\,+\,t^2\,B}\,,
\eeq
where 
\beq\label{Tt17Add3}
D\,=\,m c^{2}\,/\,k\,T_{\bot}\,>\,1\,. 
\eeq
Further we will consider only the case of an adiabatic expansion of the fluid volume when  
$\Lambda_{0}$ is negative. 

 Now, designating $\langle\,T(t\,=\,1)\,\rangle_{s}^{0}\,=\,T_{c}$ and 
$\langle\,T(t\,=\,0)\,\rangle_{s}^{0}\,=\,T_{0}\,=\,T_{\bot}$ we get from \eq{Tt15}
following expression:
\beq\label{Tt17}
\frac{T(t\,=\,0)\,-\,T_{c}}{T(t=0)}\,=\,\frac{T_{0}\,-\,T_{c}}{T_{0}}\,=\,
1\,-\,\frac{T_{c}}{T_{0}}\,=\,\frac{1}{2}
\frac{B\,-\,A^2}{1\, +\,B}\,+\,
|\Lambda_{0}|\,\frac{\,B\,D\,}{ 1\,+\,B}\,,
\eeq
which allows calculate $D$ parameter on the base of known $\Lambda_{0},\,A,\,B$ parameters and  $T_{c}\,/\,T_{0}$ ratio, which we also define as
\beq\label{Tt18}
\Delta_{1}\,=\,\,1\,-\,\frac{T_{c}}{T_{0}}\,.
\eeq
Fixing value of $T_{0}$ we fix the value of $\Delta_{1}$. Further, basing on results of \cite{Lattice},  we determine the maximum upper limit of initial temperature as 
$T_{0}^{Max}\approx\,25\,T_{c}$ that
gives $\Delta_{1}^{Max}\,=\,0.96$ and determines maximum temperature range of interest as $T_{c}\,<\,T(t)\,<\,25\,T_{c}\,.$ 

 The requested $t\,=\,t(T/T_{c})$ dependence we will obtain solving \eq{Tt15} together with \eq{Tt17}:
\beq
1\,-\,\frac{T}{T_{0}}\,=\,1\,-\,\frac{T}{T_{c}}\,\frac{T_{c}}{T_{0}}\,=\,
\frac{1}{2}
\frac{t^{2}\Le B\,-\,A^2\Ra}{1 +\, t^{2}\, B}\,+\,
|\Lambda_{0}|\,\frac{t\,B\,D\,}{ 1\,+\,t^2\,B}\,.
\eeq 
Inserting solution of this quadratic equation into \eq{VC35}-\eq{VC32} for the transport coefficients we will obtain 
these coefficient as function of $T\,/\,T_{c}$ ratio. Therefore, it is interesting to 
apply obtained formulas to the quark-gluon plasma $\eta\,/\,s$ ratio calculated in \cite{Lattice,Lattice1}. 
In order to perform these calculations, we note that in our approach entropy is constant. Therefore, 
taking $s\,=\,1$, we have for the ratio of interest $\eta\,/\,s\,=\,\eta $ in dimensionless variables of the problem.

  The behavior of the viscosity coefficients \eq{VC35}-\eq{VC32} as functions of time (temperature) is different for the each coefficient and depends on a few parameters. We assume, that among coefficients \eq{VC35}-\eq{VC32} the only one
is  not equal zero or not very small at $t\,=\,0$, whereas all other coefficients are small. In our framework, only \eq{VC33} and \eq{VC31} coefficients can be large enough at $t\,=\,0$ decreasing or increasing  during the evolution
toward some values till $t\,=\,1$. The largeness of one of them at $t\,=\,0$ requires a smallness  of another and vise versa. Also, we will consider a different temperature ranges in the calculations. We will perform the calculations for
$T_{0}\,=\,5\,T_{c},\,\Delta_{1}\,=\,0.8$ and for $T_{0}\,=\,25\,T_{c},\,\Delta_{1}\,=\,0.96$ temperature ranges, both of them correspond to the $t\,=\,0\,-\,1\,$ time range.

\begin{figure}[thbp]
\begin{center}
\psfig{file=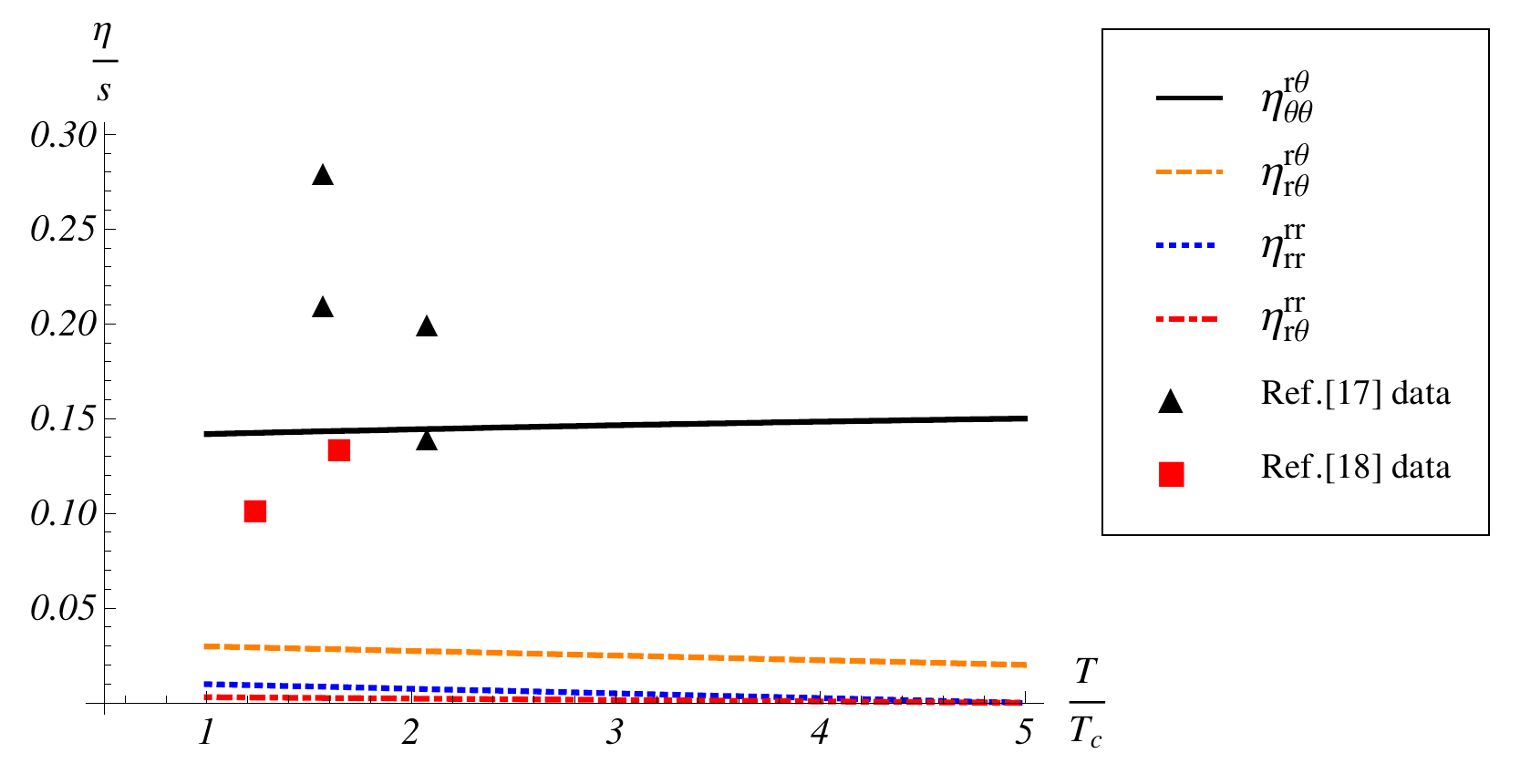 ,width=130mm} 
%\psfig{file=Pos_Lambda_r_t_t_t_Coeff_B0_01_L0_5.eps ,width=60mm} & 
%\psfig{file=Pos_Lambda_r_t_t_t_Coeff_B0_L0_5.eps,width=60mm}\\
% &  \\ 
%\fig{Vis_rttt_1}-c & \fig{Vis_rttt_1}-d \\
% &  \\

\end{center}
\caption{\it The $\eta\,/\,s$ ratio for the different viscosity coefficients \eq{VC32}-\eq{VC35} at $\Delta_{1}\,=\,0.8$, $A\,=\,0.12$, $B\,=\,0.03$ and $\Lambda_{0}\,=\,0.01$. Results of the lattice calculations for this ratio are taken from \cite{Lattice,Lattice1}.}
\label{Ratio11}
\end{figure}

\begin{figure}[thbp]
\begin{center} 
\psfig{file=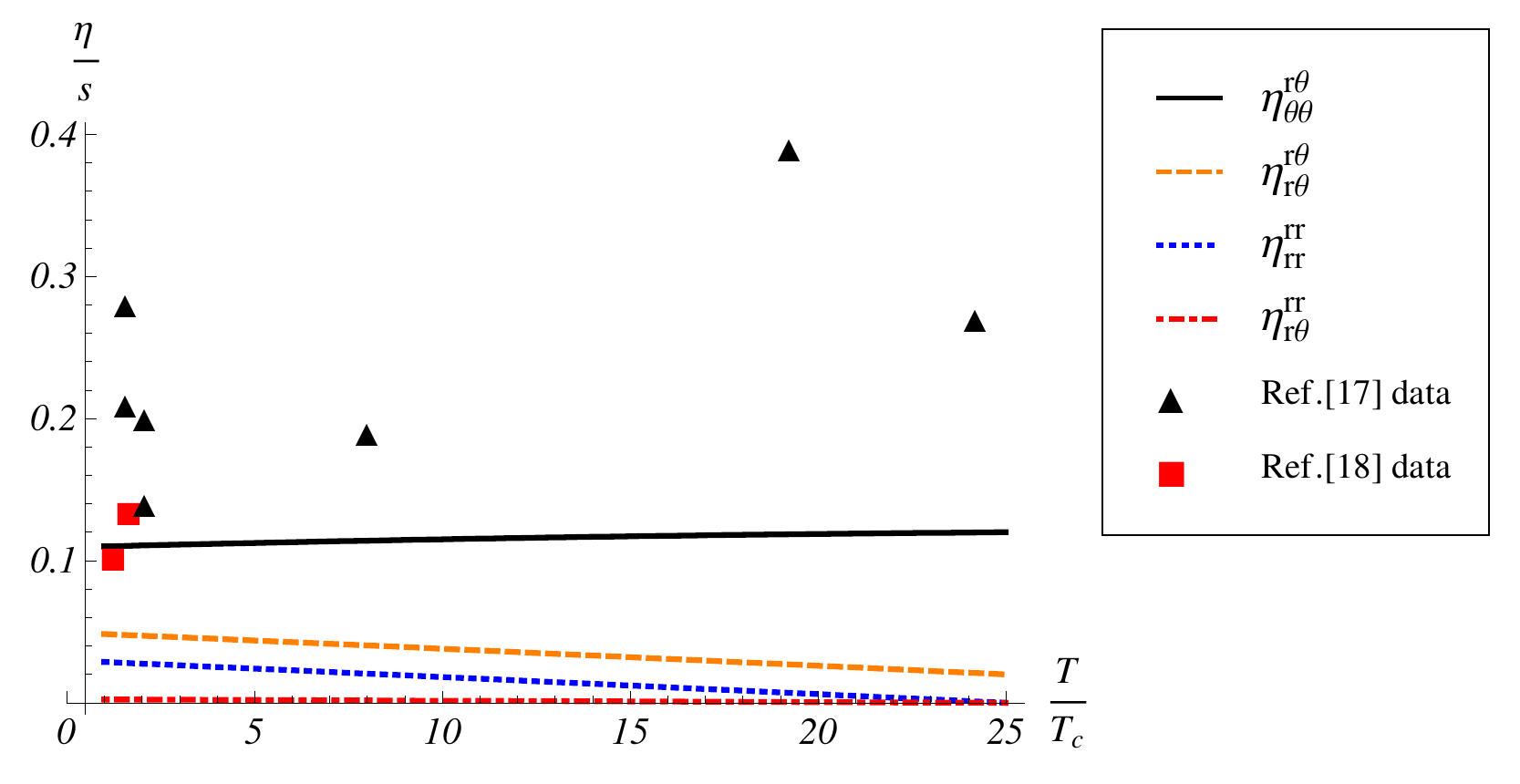,width=130mm}
%\psfig{file=Pos_Lambda_r_t_t_t_Coeff_B0_01_L0_5.eps ,width=60mm} & 
%\psfig{file=Pos_Lambda_r_t_t_t_Coeff_B0_L0_5.eps,width=60mm}\\
% &  \\ 
%\fig{Vis_rttt_1}-c & \fig{Vis_rttt_1}-d \\
% &  \\
\end{center}
\caption{\it The $\eta\,/\,s$ ratio for the different viscosity coefficients \eq{VC32}-\eq{VC35} at   $\Delta_{1}\,=\,0.96$, $A\,=\,0.12$, $B\,=\,0.03$ and $\Lambda_{0}\,=\,0.01$. Results of the lattice calculations for this ratio are taken from \cite{Lattice,Lattice1}.}
\label{Ratio12}
\end{figure}

  There are following different possibilities for the temperature dependence of the viscosity coefficients which we discussed above. The first one is when $\eta_{r\theta}^{\theta\theta}$ viscosity coefficient large and all other are small:
\beq\label{Tt21}  
A\,\sim\,10^{-1}\,;\,\,A\,\gg\,B,\,\Lambda_{0}\,;\,\,B\,\simeq\,\Lambda_{0}\,.   
\eeq  
Taking into account \eq{MagF}, \eq{VC21} and assuming that
\beq\label{ParComb1}
\omega_{r}\,\ll\,c\,/\,l_{0}\,,
\eeq
we obtain for this combination of parameters
\beq\label{ParComb2}
\frac{|q|\,B_{z0}}{l_{0}}\,\ll\,m\,c^{2}\,. 
\eeq
This scenario one can call a variant with small $B_{z0}$ field.
The results of the calculations are presented in \fig{Ratio11}-\fig{Ratio12}
for $\Delta_{1}\,=\,0.8\,,T_{0}\,=\,5\,T_{c}$ and 
$\Delta_{1}\,=\,0.96\,,T_{0}\,=\,25\,T_{c}$ values correspondingly. Due the uncertainties presented in the  model
we did not fit the lattice $\eta\,/\,s\,$ data in respect to our parameters  obtaining instead different values of the parameters for different values of
$\Delta_{1}$. The problem of more precise fixing of the parameters we discuss in the conclusion of the paper.
We note, that in both cases of \fig{Ratio11}-\fig{Ratio12}, parameter \eq{Tt17Add3} is
$\,D\,\sim\,10^{3}\,$, that is in accordance with our assumption. 

\begin{figure}[thbp]
\begin{center}
\psfig{file=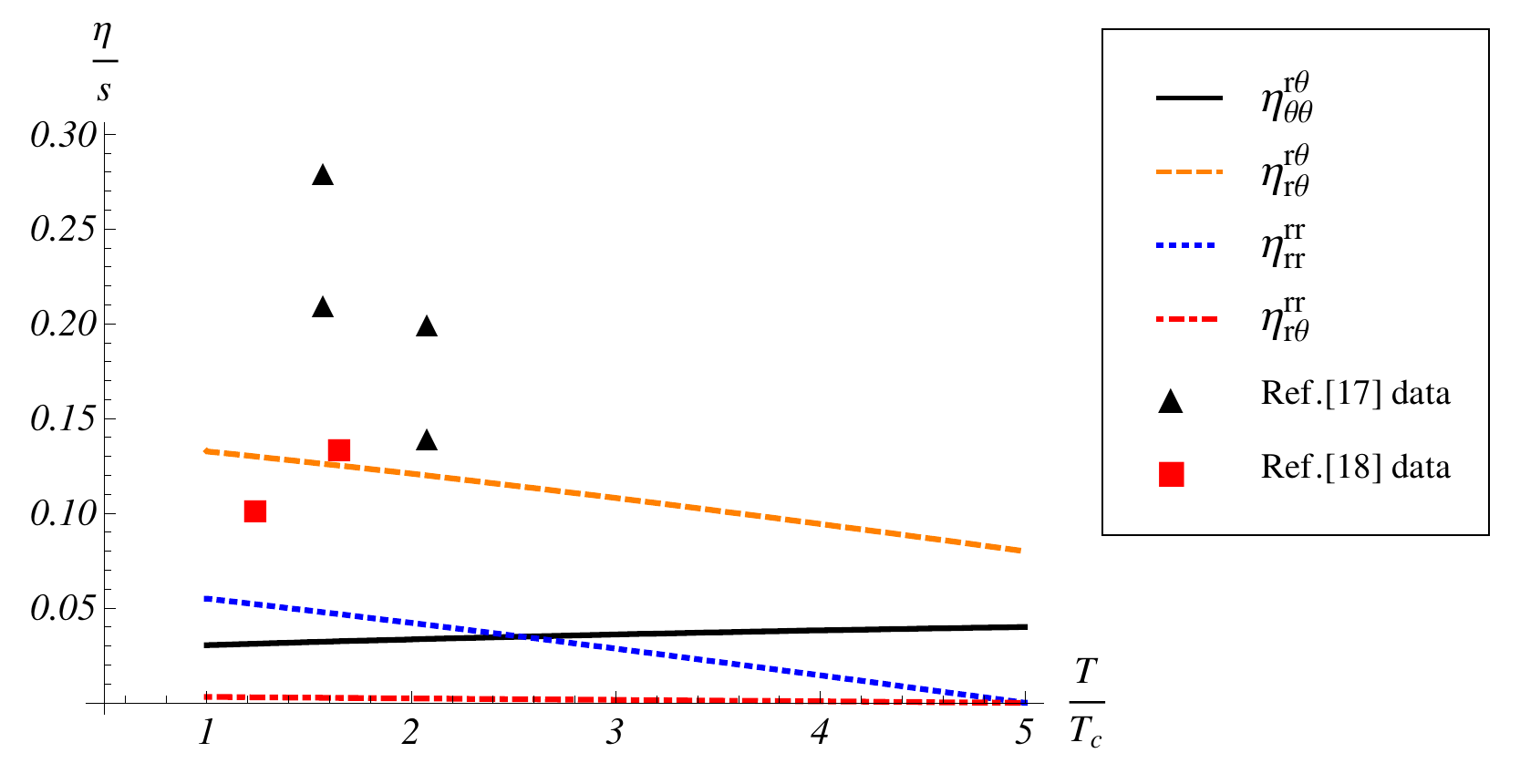 ,width=130mm} 
%\psfig{file=Pos_Lambda_r_t_t_t_Coeff_B0_01_L0_5.eps ,width=60mm} & 
%\psfig{file=Pos_Lambda_r_t_t_t_Coeff_B0_L0_5.eps,width=60mm}\\
% &  \\ 
%\fig{Vis_rttt_1}-c & \fig{Vis_rttt_1}-d \\
% &  \\

\end{center}
\caption{\it The $\eta\,/\,s$ ratio for the different viscosity coefficients \eq{VC32}-\eq{VC35} at $\Delta_{1}\,=\,0.8$, $A\,=\,0.04$, $B\,=\,0.06$ and $\Lambda_{0}\,=\,0.04$. Results of the lattice calculations for this ratio are taken from \cite{Lattice,Lattice1}.}
\label{Ratio21}
\end{figure}

\begin{figure}[thbp]
\begin{center} 
\psfig{file=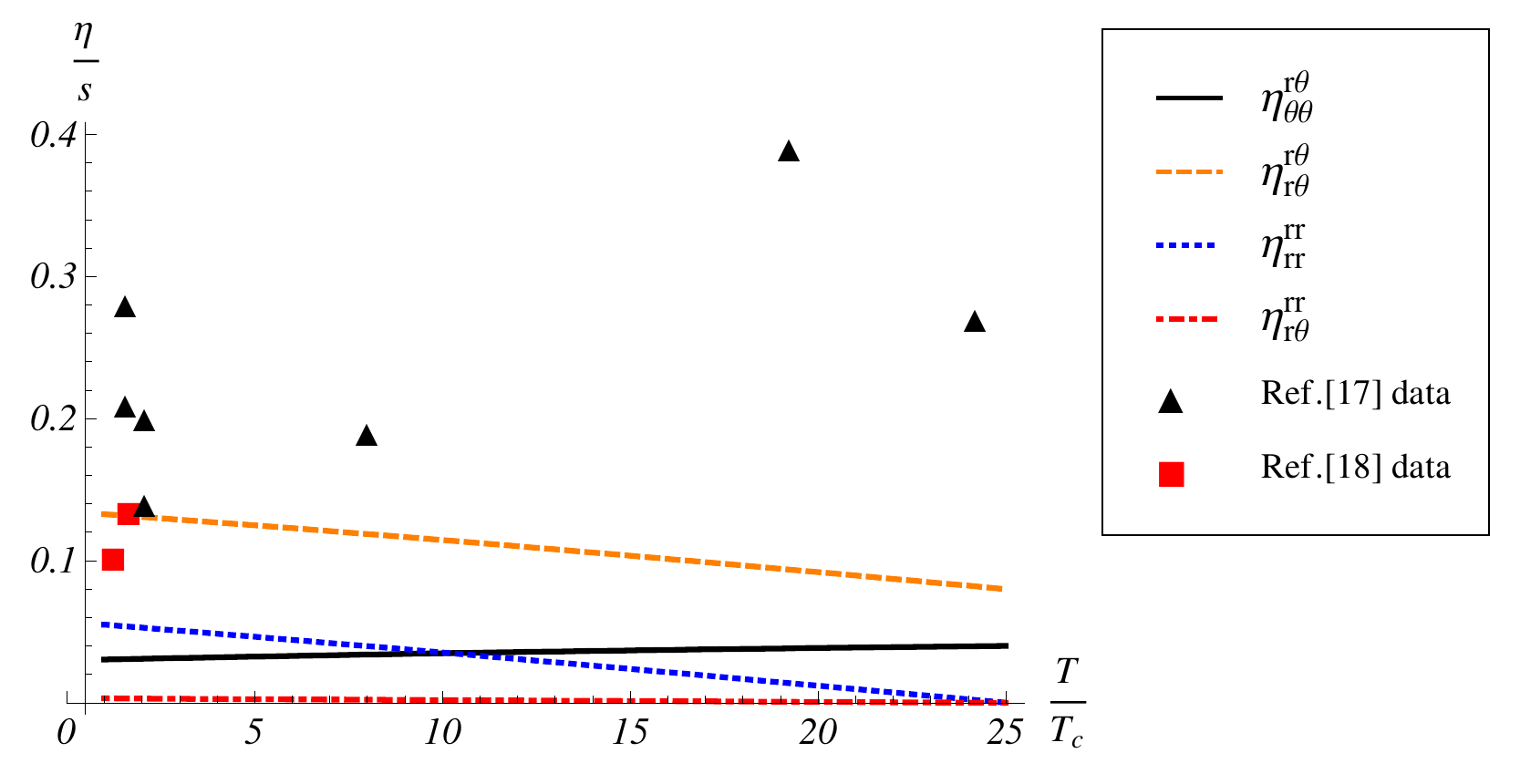,width=130mm}
%\psfig{file=Pos_Lambda_r_t_t_t_Coeff_B0_01_L0_5.eps ,width=60mm} & 
%\psfig{file=Pos_Lambda_r_t_t_t_Coeff_B0_L0_5.eps,width=60mm}\\
% &  \\ 
%\fig{Vis_rttt_1}-c & \fig{Vis_rttt_1}-d \\
% &  \\
\end{center}
\caption{\it The $\eta\,/\,s$ ratio for the different viscosity coefficients \eq{VC32}-\eq{VC35} at   $\Delta_{1}\,=\,0.96$, $A\,=\,0.04$, $B\,=\,0.06$ and $\Lambda_{0}\,=\,0.04$. Results of the lattice calculations for this ratio are taken from \cite{Lattice,Lattice1}.}
\label{Ratio22}
\end{figure}

 Another possible restrictions on the problem's parameters which satisfy our requests are the following:
\beq\label{Tt23}  
\,A\,\sim\,B\,\sim\,\Lambda_{0}\,\sim\,10^{-2}\,.   
\eeq
It gives
\beq\label{ParComb3}
\frac{|q|\,B_{z0}}{l_{0}}\,\sim\,m\,c^{2}\,
\eeq
and therefore it can be named as  variant with large $B_{z0}$ field.
In this case the viscosity coefficient \eq{VC31} is large at $t\,=\,0$ whereas all other coefficients remains small during the evolution. The results of the calculations with these parameter's restrictions are presented in \fig{Ratio21}-\fig{Ratio22}
for $\Delta_{1}\,=\,0.8\,,T_{0}^{Mac}\,=\,5\,T_{c}$ and 
$\Delta_{1}\,=\,0.96\,,T_{0}^{Mac}\,=\,25\,T_{c}$ values correspondingly. For these values of parameters we obtain  $D\sim\,10^{2}$.

\begin{figure}[thbp]
\begin{center}
\psfig{file=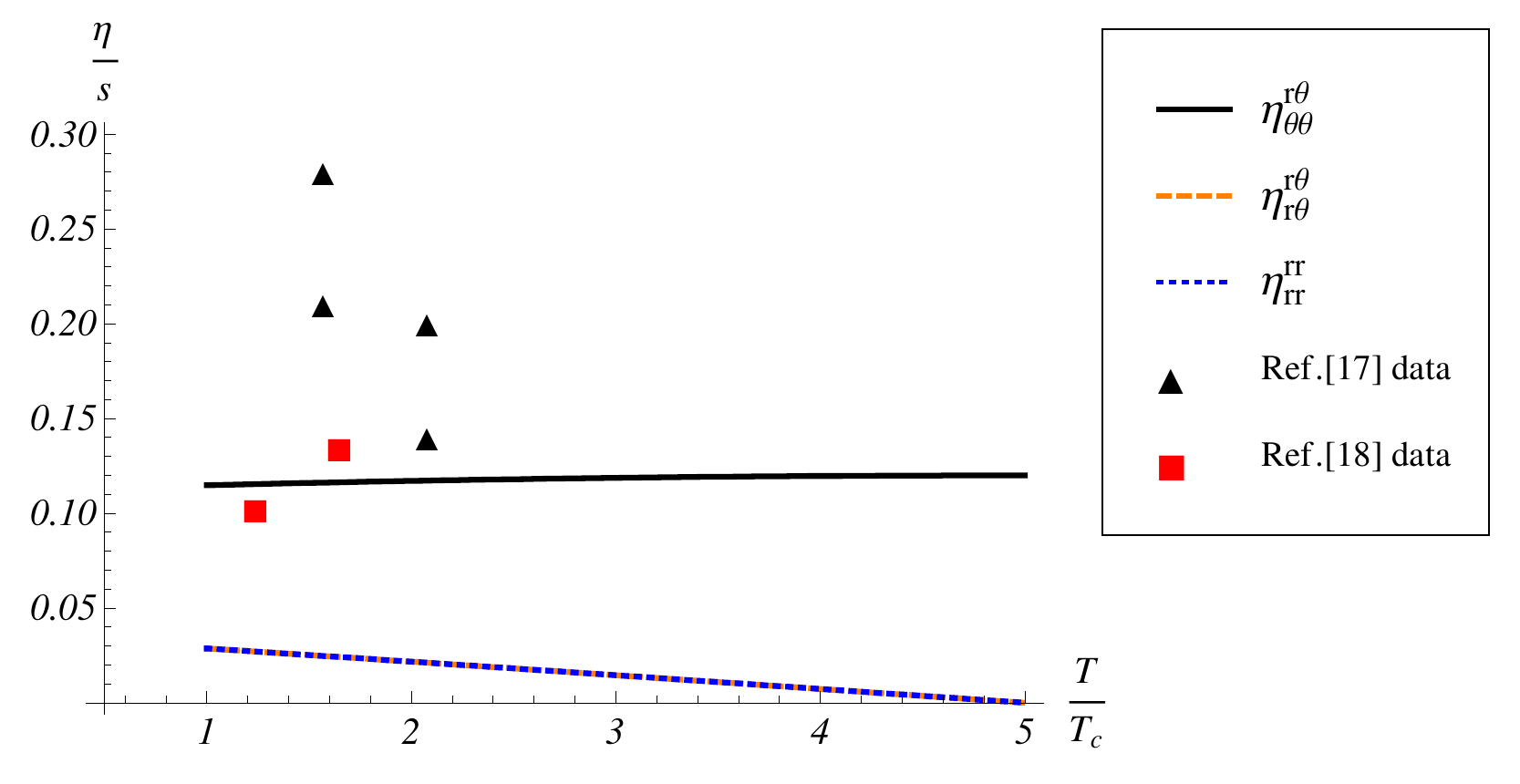 ,width=130mm} 
%\psfig{file=Pos_Lambda_r_t_t_t_Coeff_B0_01_L0_5.eps ,width=60mm} & 
%\psfig{file=Pos_Lambda_r_t_t_t_Coeff_B0_L0_5.eps,width=60mm}\\
% &  \\ 
%\fig{Vis_rttt_1}-c & \fig{Vis_rttt_1}-d \\
% &  \\

\end{center}
\caption{\it The $\eta\,/\,s$ ratio for the different viscosity coefficients \eq{VC32}-\eq{VC35} at $\Delta_{1}\,=\,0.8$, $A\,=\,0.12$, $B\,=\,0.03$ and $\Lambda_{0}\,=\,0$. Results of the lattice calculations for this ratio are taken from \cite{Lattice,Lattice1}.}
\label{Ratio31_1}
\end{figure}

\begin{figure}[thbp]
\begin{center} 
\psfig{file=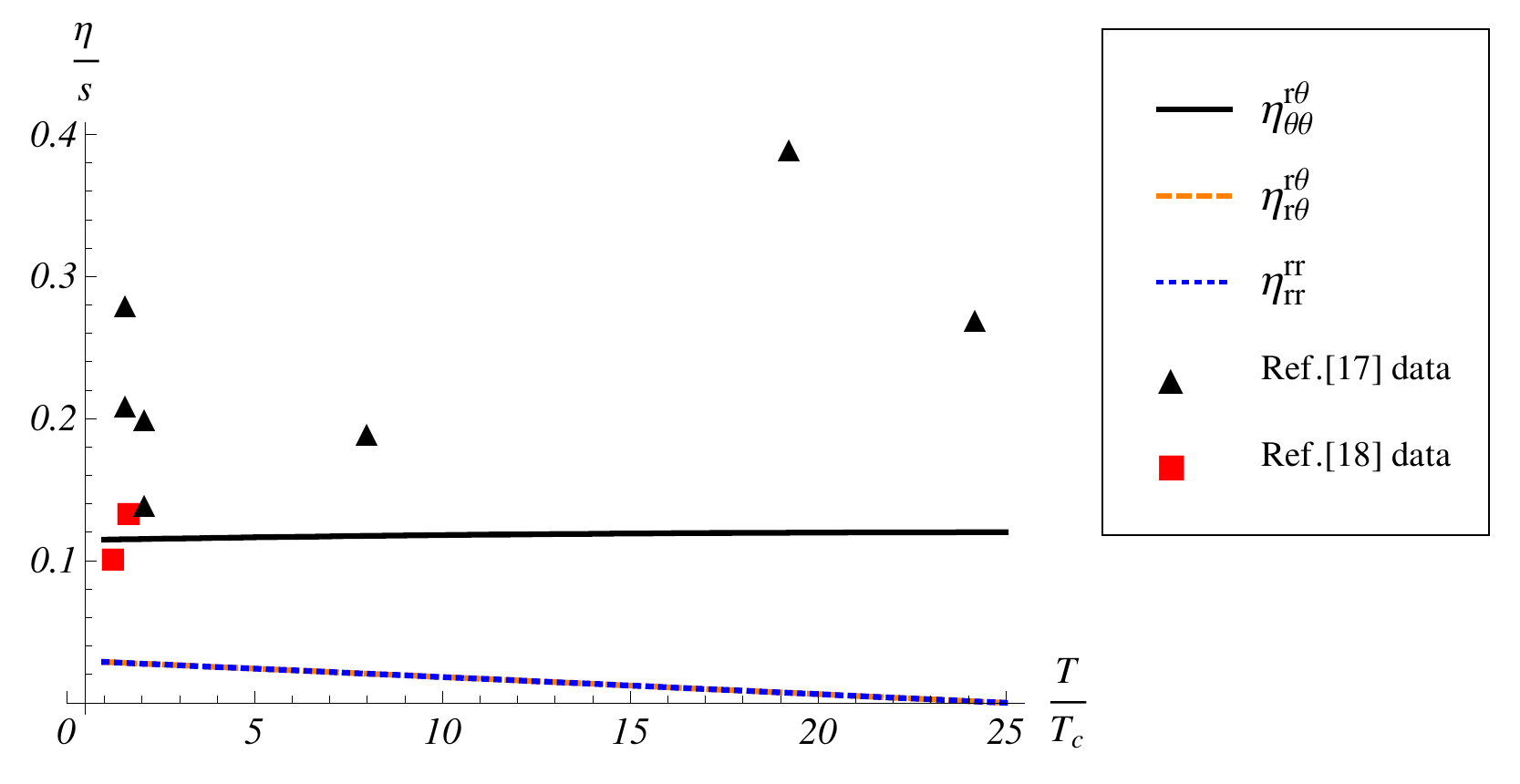,width=130mm}
%\psfig{file=Pos_Lambda_r_t_t_t_Coeff_B0_01_L0_5.eps ,width=60mm} & 
%\psfig{file=Pos_Lambda_r_t_t_t_Coeff_B0_L0_5.eps,width=60mm}\\
% &  \\ 
%\fig{Vis_rttt_1}-c & \fig{Vis_rttt_1}-d \\
% &  \\
\end{center}
\caption{\it The $\eta\,/\,s$ ratio for the different viscosity coefficients \eq{VC32}-\eq{VC35} at   $\Delta_{1}\,=\,0.96$, $A\,=\,0.12$, $B\,=\,0.03$ and $\Lambda_{0}\,=\,0$. Results of the lattice calculations for this ratio are taken from \cite{Lattice,Lattice1}.}
\label{Ratio32_1}
\end{figure}

\begin{figure}[thbp]
\begin{center}
\psfig{file=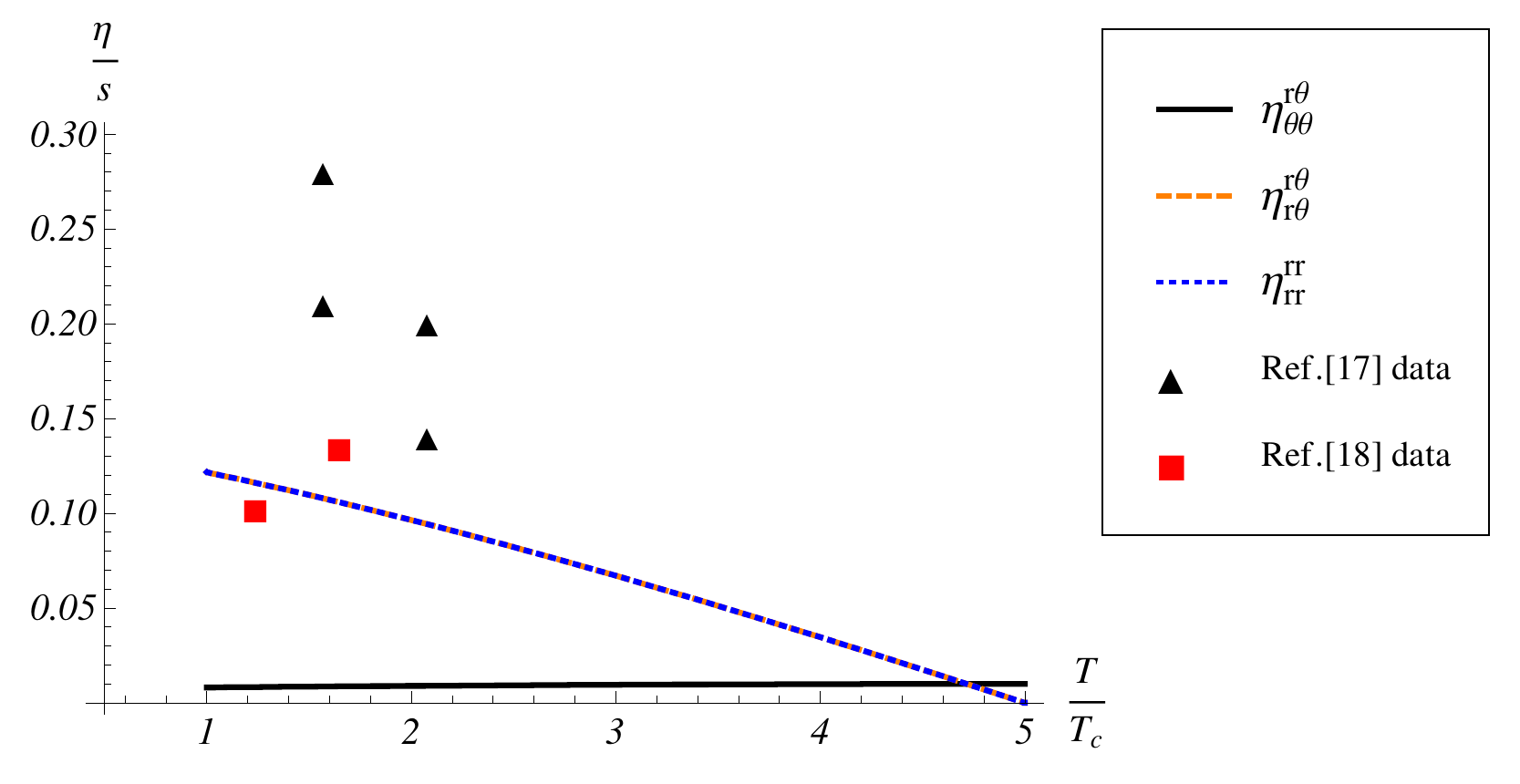 ,width=130mm} 
%\psfig{file=Pos_Lambda_r_t_t_t_Coeff_B0_01_L0_5.eps ,width=60mm} & 
%\psfig{file=Pos_Lambda_r_t_t_t_Coeff_B0_L0_5.eps,width=60mm}\\
% &  \\ 
%\fig{Vis_rttt_1}-c & \fig{Vis_rttt_1}-d \\
% &  \\

\end{center}
\caption{\it The $\eta\,/\,s$ ratio for the different viscosity coefficients \eq{VC32}-\eq{VC35} at $\Delta_{1}\,=\,0.8$, $A\,=\,0.07$, $B\,=\,0.4$ and $\Lambda_{0}\,=\,0$. Results of the lattice calculations for this ratio are taken from \cite{Lattice,Lattice1}.}
\label{Ratio31_2}
\end{figure}

\begin{figure}[thbp]
\begin{center} 
\psfig{file=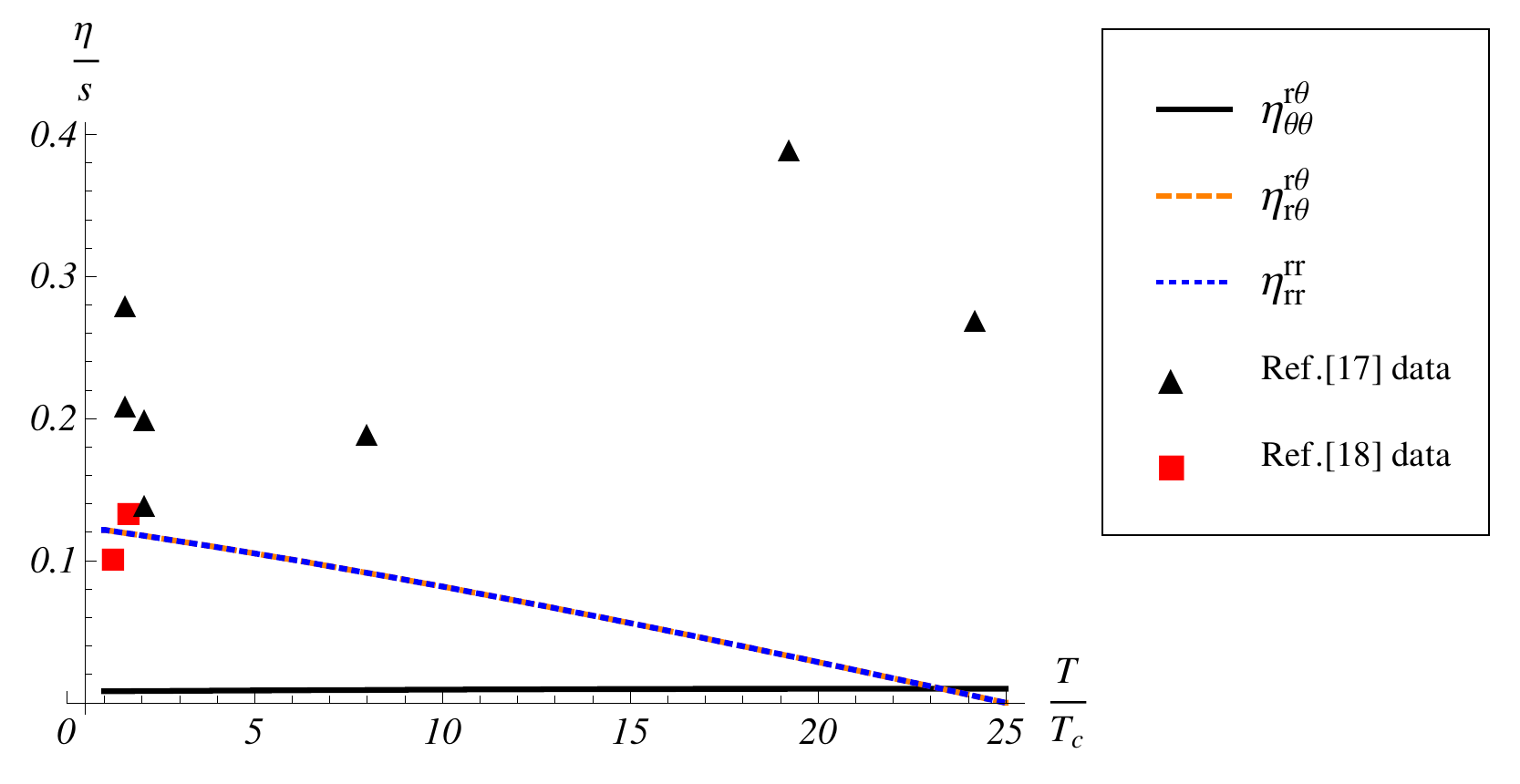,width=130mm}
%\psfig{file=Pos_Lambda_r_t_t_t_Coeff_B0_01_L0_5.eps ,width=60mm} & 
%\psfig{file=Pos_Lambda_r_t_t_t_Coeff_B0_L0_5.eps,width=60mm}\\
% &  \\ 
%\fig{Vis_rttt_1}-c & \fig{Vis_rttt_1}-d \\
% &  \\
\end{center}
\caption{\it The $\eta\,/\,s$ ratio for the different viscosity coefficients \eq{VC32}-\eq{VC35} at   $\Delta_{1}\,=\,0.96$, $A\,=\,0.07$, $B\,=\,0.4$ and $\Lambda_{0}\,=\,0$. Results of the lattice calculations for this ratio are taken from \cite{Lattice,Lattice1}.}
\label{Ratio32_2}
\end{figure}

 The case when $\Lambda_{0}\,=\,0$, i.e. when the hot spot is neutral, we have
again two different scenarios. the first one is similar  to \eq{Tt21}, whereas the second oen is given by
\beq\label{Tt21Add4}  
B\,\sim\,10^{-1}\,;\,\,B\,\gg\,A\,.   
\eeq   
The plots for these cases are presented in \fig{Ratio31_1}-\fig{Ratio32_1} and \fig{Ratio31_2}-\fig{Ratio32_2}
correspondingly
for $\Delta_{1}\,=\,0.8\,,T_{0}^{Mac}\,=\,5\,T_{c}$ and 
$\Delta_{1}\,=\,0.96\,,T_{0}^{Mac}\,=\,25\,T_{c}$. 
Parameter \eq{Tt17Add3} in this case does not affect on the plots, also
we note that at $\,\Lambda_{0}\,=\,0$ we have $\eta^{rr}_{rr}\,=\,\eta^{r\theta}_{r\theta}$ at $\Lambda_{0}\,=\,0$.

  It is worth emphasizing, that we compared results of our calculations with the \cite{Lattice,Lattice1} data in order to show the importance of new degrees of freedom introduced into the model. We did not try fix the parameters precisely by the data fitting, therefore, 
obtained numerical values of the parameters serve as approximate restrictions which we will use in the future investigations of quark-gluon plasma phenomenological models.

\section{Conclusion}

\,\,\,\,\,\, In this paper we considered a model for adiabatic expansion of a small, dense, charged and rotating 
hot spot
which was perturbed by an external electromagnetic field at the moment of time prior to the subsequent hydrodynamical evolution.
The main motivation for the model's construction was an investigation of a possibility for the small shear viscosity  to entropy ratio in the system described by Vlasov's equation. We demonstrated, that an introduction of the additional degrees of freedom in the system allows to obtain requested ratio for the expanding small volume of an ideal fluid element.

  Solving perturbatively Vlasov's equation \eq{Vl9} we obtained the
non-equilibrium distribution function with dependence on the particular, inhomogeneous initial conditions 
\eq{Vl10} and \eq{DFun2}. Using this function for the calculation of the momentum flux tensor
we also obtained a family of the viscosity coefficients \eq{VC1}\,-\,\eq{VC13} whose values depend on the model's parameters,
that we consider as a main result of the manuscript.
There are  two such new parameters introduced in the model, initial angular velocity of the hot spot and 
ratio of it's interaction with the external field to the kinetic energy at the initial moment of time. Taking these parameters to zero, we see 
from \eq{VC1}\,-\,\eq{VC13} that at the leading order the viscosity coefficients are equal to zero. Therefore, we obtained that 
the value of viscosity to entropy ratio is controlled by these parameters, which allow to vary this ratio correspondingly to the initial conditions of the problem, see results of Section 5, where 
different scenarios for the value of 
viscosity coefficients are described.  We underline, that the shear viscosity of the bulk of QGP is result of the averaging of the  independent processes of expansion of the different fluctuations immersed in some 
smooth and 
less dense surround,  see \cite{New1,New2,New3,New4,New5}. The calculation of that
averaged shear viscosity coefficients is a separate task, which we will consider in the subsequent publications.

 In our calculations  we consider the electromagnetic interactions between the particles, which are similar
in the first approximation to the weakly interacting (asymptotically free) quarks and gluons inside a very small and dense initial fluctuation of the matter in the high energy scattering. Despite the fact that at high energy the interaction between relativistic particles may be very complicated, see for example \cite{BFKL,Bond}, our calculations can serve as a simple model for 
studying properties of the Quark-Gluon Plasma (QGP).
We see from the Fig.\ref{Ratio11}\,-\,Fig.\ref{Ratio32_2}, that in our framework, the smallness of the $\eta\,/\,s$ ratio for QGP means some restrictions on the properties of this initial fluctuation and it's interaction with the 
surround.
Additional restrictions in the problem are arising due the fact that we demand the smallness of all but one shear viscosity coefficients. This condition gives some additional information about microscopic structure of the system of interest.
It is important to note, that even in the case of the neutral fluid element, when 
$\Lambda_{0}\,=\,0$, the approach allows to describe possible ranges of the $\eta\,/\,s$ ratio from \cite{Lattice, Lattice1}, see Fig.\ref{Ratio31_1}\,-\,Fig.\ref{Ratio32_2}. 
 
 For the case of QGP, adopting the $\eta\,/\,s$ ratio  from \cite{Lattice,Lattice1}, we obtained that the possible ranges of the values of the parameters  are
$A\,=\,l_{0}\,\omega_{r}\,/\,c\,\sim\,0.01\,-\,0.2$, $B\,=\,l_{0}^{2}\,\omega_{r}\,\omega_{c}\,/\,c^2\,-\,
l_{0}^{2}\,\omega_{c}^{2}\,/\,c^2\,\sim\,0.01\,-\,0.4$, $\Lambda_{0}\,\sim\,0\,-\,0.04$. We see, that the value of $\Lambda_{0}$ is small in any combination, whereas the value of $\omega_{r}$ can vary rather widely. In order to fix the $\omega_{r}$ value some additional data is required therefore. Perhaps, an application of the model 
to the calculations of the multiplicities of the particles produced at high-energy scattering , see for example \cite{BNL}, can resolve this problem.

 Additional perturbative corrections to our results are provided by the contributions in the next order at $\Lambda$ parameter, see  
\eq{Pert1}, the present calculations were obtained in the first order to $\Lambda_{0}$ and zeroth order to $\Lambda$ 
perturbative parameters. Therefore, obtained results must be supplemented by the contributions to the next order in 
$\Lambda$ in the r.h.s. of \eq{Vl9}. This, in turn, requires to solve \eq{Pot8}\,-\,\eq{Pot9}
with the currents determined by mean velocities of \eq{RadVel44} and \eq{AzVel22}. We will present  these calculations in the subsequent publication.

  Finally we conclude, that our model can be useful for the clarification of the microscopic dynamics of the interactions of asymptotically free 
quarks and gluons inside a QCD hot spot as well as in explaining  mechanism of the shear viscosity smallness in the processes of the ideal fluid element expansion.
We believe also, that the microscopic theory of 
the hydrodynamical expansion of the charged hot spot will provide connection
between the data, obtained in high-energy collisions of protons and nuclei in the LHC and RHIC experiments \cite{Fluid,Fluid1,Fluid2,BNL}, and microscopic fields inside the collision region as well. 

%%%%%%%%%%%%%%%%%%%%%%%%%%%%%%%%%%%%%%%%%%%%%%%%%%%%%%%%%%%%%%%%%%%%%%%%%%%%%%

%%%%%%%%%%%%%%%%%%%%%%%%%%%%%%%%%%%%%%%%%%

\end{document}